\documentclass[a4paper]{article} 
\usepackage[fleqn]{amsmath}
\usepackage{amsfonts,amssymb}
\usepackage{pifont}
\usepackage{mathrsfs}
\usepackage{graphicx}
\usepackage{savesym} 
\savesymbol{AND} 
\usepackage{algorithmic}
\usepackage{algorithm}
\usepackage{accents}
\usepackage{bm}
\usepackage{multirow}
\usepackage{paralist}
\usepackage{caption}
\usepackage{subcaption}
\usepackage{enumitem}
\usepackage[stable]{footmisc}
\usepackage[%
    all,
    defaultlines=3 
]{nowidow}

\newcounter{Thmnbr}
\def\theenumi{{\arabic{enumi}}.}     
\def\theenumii{\bff{\roman{enumii}}.}       
     \def\theenumiii{\alph{enumiii}}
       
\def\p@enumii{\theenumi}
\def\p@enumiii{\theenumi(\theenumii)}
\def\p@enumiv{\p@enumiii\theenumiii}

\newtheorem{lemm}{Lemma}[section]
\newtheorem{theo}[lemm]{Theorem}
\newtheorem{propo}[lemm]{Proposition}
\newtheorem{assump}[lemm]{Assumptions}

\newtheorem{aim}[lemm]{Aims}

\newtheorem{lemms}{Lemma}[subsection]
\newtheorem{theos}[lemms]{Theorem}

\newtheorem{manualtheoreminner}{Lemma}
\newenvironment{manualtheorem}[1]{%
  \renewcommand\themanualtheoreminner{#1}%
  \manualtheoreminner
}{\endmanualtheoreminner}

\newtheorem{defi}{Definition}[section]
\newtheorem{rema}{Remark}[section]

\newcommand{\findemo}{{\hfill \ding{113}}}

\newcommand{\debdemo}{{\noindent {\bf  Proof}.\  }}

\newcommand{\R}{{\mathop{\rm I\mkern -3.5mu R}}}
\newcommand{\N}{{\mathop{\rm I\mkern -3.5mu N}}}

\newcommand{\Rn}{{{\R}^n}}
\newcommand{\Rm}{{{\R}^m}}

\newcommand{\Rq}{{{\R}^q}}

\newcommand{\Rnn}{{{\R}^{n\times n}}}

\newcommand{\Rqq}{{{\R}^{q\times q}}}

\newcommand{\Rnq}{{{\R}^{n\times q}}}
\newcommand{\Rnm}{{{\R}^{n\times m}}}

\newcommand{\Rnl}{{{\R}^{n\times l}}}

\newcommand{\ith}{$^{\textrm{th}}$\ }

\newcommand{\ea}{{\it et al.\ }} 
\newcommand{\cf}{{\it cf.\ }} 

\newcommand{\ssi}{{\emph{if and only if\ }}}

\newcommand{\ie}{{\it i.e.}} 
\newcommand{\tq}{\emph{s.t. }}
\newcommand{\wrt}{\emph{w.r.t. }}
\newcommand{\resp}{\emph{resp}}
\newcommand{\aka}{\emph{a.k.a. }}

\let\d\displaystyle
\let\nn\nonumber

\newcommand{\nid}{{\noindent}}

\DeclareMathOperator*{\diag}{Diag}

\DeclareMathOperator*{\bdiag}{Bdiag}
\DeclareMathOperator*{\tr}{tr} 

\DeclareMathOperator*{\rank}{rank}

\DeclareMathOperator*{\dif}{d\!}
\DeclareMathOperator*{\card}{Card}
\DeclareMathOperator*{\vol}{vol} 
\DeclareMathOperator*{\ssal}{ssal} 
\DeclareMathOperator*{\Vol}{\textstyle\vol_\dag\!} 

\newcommand{\norm}[1]{\left\Vert#1\right\Vert}

\newcommand{\abs}[1]{{\ensuremath{\left| #1 \right|}}}
\newcommand{\bff}[1]{\mbox{\boldmath$ #1 $}} 
\newcommand{\eqd}{{\ensuremath{\, =: \,}}}
\newcommand{\eqg}{{\ensuremath{\, := \,}}}
\newcommand{\eq}{\ensuremath{\Leftrightarrow}}
\newcommand{\guil}[1]{{``#1''}}

\newcommand{\Det}[1]{{\left\vert#1\right\vert}_{\dag}} 

\newcommand{\BC}{{\mathcal{B}}}
\newcommand{\CC}{{\mathcal{C}}}
\newcommand{\DC}{{\mathcal{D}}}
\newcommand{\EC}{{\mathcal{E}}}

\newcommand{\GC}{{\mathcal{G}}}
\newcommand{\HC}{{\mathcal{H}}}

\newcommand{\KC}{{\mathcal{K}}}

\newcommand{\NC}{{\mathcal{N}}}
\newcommand{\OC}{{\mathcal{O}}}

\newcommand{\PC}{{\mathcal{P}}}
\newcommand{\SC}{{\mathcal{S}}}

\newcommand{\ZC}{{\mathcal{Z}}}

\newcommand{\Fd}{{\mathscr{F}}}

\def\u{\bff{u}}
\def\vb{\bff{v}}

\def\x{\bff{x}}
\def\y{\bff{y}}
\def\zb{\bff{z}}

\def\taub{\bff{\tau}}

\def\r{\bff{r}}

\def\byb{\bar{\y}}

\def\bdel{\bar{\bff{\delta}}}

\def\bybk{\bar{\y}_{k}}

\def\Akk{A_{k-1}}

\def\Rk{R_{k}}

\def\taubk{\bff{\tau}_k}
\def\taubkk{\bff{\tau}_{k-1}}

\def\rki{\bff{r}_{{k}_i}}
\def\rkit{\bff{r}_{{k}_i}^T}

\def\yk{\bff{y}_{k}} 
\def\delk{\bff{\delta}_{k}} 
\def\vk{\bff{v}_{k}}

\def\xz{\bff{\hat{x}_{0}}} 
\def\xza{\bff{x_{0}}} 
\def\xk{\hat{\bff{x}}_{k}} 
\def\xkk{\hat{\bff{x}}_{k/k-1}} 

\def\dxk{\tilde{\bff{x}}_{k}} 
\def\dxkkk{\tilde{\bff{x}}_{k-1}} 
\def\xka{\bff{x}_{k}} 
\def\xkka{\bff{x}_{k-1}} 
\def\bxk{\breve{\bff{x}}_{k}}
\def\bxkk{\breve{\bff{x}}_{k-1}}
\def\zk{\bff{z}_k}

\def\zbz{\bff{z}_0}

\def\Vscr{\mathscr{V}}

\def\Pk{P_{k}}

\def\Pkk{P_{k/k-1}}

\def\Pkkk{P_{k-1}}

\def\sigz{\varsigma_{0}}

\def\sigk{\varsigma_{k}}
\def\sigkp{\varsigma_{k+1}}

\def\sigkkk{\varsigma_{k-1}}
\def\sig{\varsigma}

\def\ki{_{k_i}}
\def\kii{_{k_{i-1}}}
\def\kz{_{k_0}}
\def\kk{_{k/k-1}}
\def\kki{_{k/k-1_i}}

\def\kkk{_{k-1}}

\def\zero{{\mathbf{0}}} 

\def\kp{_{k_+}}
\def\kkp{_{{k-1}_+}}

\def\yki{\bff{y}_{k_i}}

\def\Rpi{{\R}^{p_i}}

\def\xki{\hat{\bff{x}}_{k_i}}

\def\xkiii{\hat{\bff{x}}_{k_{i-1}}}
\def\xkz{\hat{\bff{x}}_{k_{0}}}

\def\gamki{\gamma_{k_i}}

\def\yki{{y_{k_i}}}

\def\sigki{\varsigma_{k_i}}

\def\sigkiii{\varsigma_{k_{i-1}}}
\def\sigkz{\varsigma_{k_0}}

\def\sigkkki{\varsigma_{k-1_i}}

\def\delki{{\delta_{k_i}}}
\def\delkis{{\delta_{k_i}^{2}}}

\def\omki{\omega_{k_i}}

\def\sigkj{\varsigma_{k_j}^2}

\DeclareMathOperator*{\mink}{\oplus}

\makeatletter
\DeclareRobustCommand\sfrac[1]{\@ifnextchar/{\@sfrac{#1}}%
                                            {\@sfrac{#1}/}}
\def\@sfrac#1/#2{\leavevmode\kern.1em\raise.5ex
         \hbox{$\m@th\fontsize\sf@size\z@
                           \selectfont#1$}\kern-.1em
         /\kern-.15em\lower.25ex
          \hbox{$\m@th\fontsize\sf@size\z@
                            \selectfont#2$}}
\makeatother
\makeatletter
\newcommand{\mathl}{\@fleqntrue\@mathmargin0pt} 
\newcommand{\mathc}{\@fleqnfalse} 
\newcommand{\pushr}[1]{\ifmeasuring@#1\else\omit\hfill$\displaystyle#1$\fi\ignorespaces}
\newcommand{\pushl}[1]{\ifmeasuring@#1\else\omit$\displaystyle#1$\hfill\fi\ignorespaces}
\makeatother

\DeclareMathAlphabet{\mathpzc}{OT1}{pzc}{m}{it}                                                        
\allowdisplaybreaks

\makeatletter
\newcounter{ALC@tempcntr}
\newcommand{\LCOMMENT}[1]{%
    \setcounter{ALC@tempcntr}{\arabic{ALC@rem}}
    \setcounter{ALC@rem}{1}
    \item \{#1\}
    \setcounter{ALC@rem}{\arabic{ALC@tempcntr}}
}%

\begin{document}

\setlength{\abovedisplayskip}{0pt}
\setlength{\belowdisplayskip}{0pt}
\setlength{\abovedisplayshortskip}{0pt}
\setlength{\belowdisplayshortskip}{0pt}

\newlength{\mes}
\setlength{\mes}{5em}
\newcommand{\underb}[1]{\underaccent{\bar}{#1}}
\newcommand{\Rrho}{{{\R}^{\varrho}}}
\newcommand{\Rrhoq}{{{\R}^{\varrho\times q}}}
\newcommand{\Rqrho}{{{\R}^{\varrho\times q}}}
\newcommand{\Rrhorho}{{{\R}^{\varrho\times\varrho}}}
\newcommand{\Rrhon}{{{\R}^{\varrho\times n}}}
\def\Rpi{\R^\pi}
\def\Rnpi{\R^{n\times\pi}}
\def\Rnrho{\R^{n\times\varrho}}
\def\Rnrho{\R^{n\times\varrho}}
\def\txt{\textstyle}
\def\ii{_{i-1}}
\def\sqrtexp{^{\frac{1}{2}}}
\newcommand{\kkj}[1]{_{{k/k-1}_{#1}}}
\newcommand{\kkkj}[1]{_{{k-1}_{#1}}}
\def\kkp{_{k+1/k}}
\def\kp{_{k+1}}
\def\kpi{_{k+1_i}}
\newcommand{\kij}[1]{_{k_{#1}}}
\def\xbk{\xb_{k}}
\def\xkpa{\xb_{k+1}}
\def\xkp{\hat{\xb}_{k+1}}
\def\xkkp{\hat{\xb}_{k+1/k}}
\def\bxkkp{\bar{\xb}_{k+1/k}}
\def\bxkk{\bar{\xb}_{k/k-1}}
\def\Pkkp{P_{k+1/k}}
\def\tPk{\tilde{P}_{k}}
\def\tPkp{\tilde{P}_{k+1}}
\def\tPkkk{\tilde{P}_{k-1}}
\newcommand{\kkpj}[1]{_{{k+1/k}_{#1}}}
\def\ybk{\bm{y}_{k}}
\def\xibk{\bm{\xi}_k}
\def\xibkp{\bm{\xi}_{k+1}}
\def\xibkk{\bm{\xi}_{k-1}}
\def\xibh{\hat{\bm{\xi}}}
\def\xibhk{\hat{\bm{\xi}}_{k}}
\def\xibhkk{\hat{\bm{\xi}}_{k/k-1}}
\def\xibhkkk{\hat{\bm{\xi}}_{k-1}}
\def\xibhkp{\hat{\bm{\xi}}_{k+1}}
\def\etab{\bm{\eta}}
\def\etabk{\bm{\eta}_k}
\def\etabkk{\bm{\eta}_{k-1}}
\def\dbk{\bm{d}_k}
\def\dbt{\bm{d}^T}
\newcommand{\bk}[1]{{b}_{k_{#1}}}
\newcommand{\bki}{{b}_{k_{i}}}
\def\udel{\underb{\delta}}
\def\bdel{\bar{\delta}}
\def\udelkis{\underb{\delta}_{k_i}^{2}}
\def\udelkj{\underb{\delta}_{k_j}}
\def\uyi{\underb{y}_{i}}
\def\byi{\bar{y}_{i}}
\def\uyki{\underb{y}_{k_i}}
\def\byki{\bar{y}_{k_i}}
\def\uybk{\underb{\bm{y}}_{k}}
\def\bybk{\bar{\bm{y}}_{k}}
\def\bbybk{\bar{\bar{\bm{y}}}_{k}}
\def\bbyk{\bar{\bar{{y}}}_{k}}
\def\yb{\bm{y}}
\def\xb{\bm{x}}
\def\ab{\bm{a}}
\def\bb{\bm{b}}
\def\cb{\bm{c}}
\def\db{\bm{d}}
\def\gb{\bm{g}}
\def\lb{\bm{l}}
\def\mb{\bm{m}}
\def\nb{\bm{n}}
\def\qb{\bm{q}}
\def\rb{\bm{r}}
\def\brb{\bar{\rb}}
\def\urb{\underb{\rb}}
\def\trb{\tilde{\rb}}
\def\ub{\bm{u}}
\def\bdb{\bar{\db}}
\def\udb{\underb{\db}}
\def\tdb{\tilde{\db}}
\def\wb{\bm{w}}
\def\wbk{\bm{w}_k}
\def\wbkk{\bm{w}_{k-1}}
\def\bwbk{\bar{\bm{w}}_k}
\def\bwbkk{\bar{\bm{w}}_{k-1}}
\def\twbk{\tilde{\bm{w}}_k}
\def\twbkk{\tilde{\bm{w}}_{k-1}}
\def\ubk{\ub_{k}}
\def\mub{\bm{\mu}}
\def\mubk{\bm{\mu}_k}
\newcommand{\muki}[1]{{\mu}_{k_{#1}}}
\newcommand{\chiki}[1]{{\chi}_{k_{#1}}}
\def\omb{\bm{\omega}}
\def\bomb{\bar{\bm{\omega}}}
\def\ombk{\bm{\omega}_k}
\def\bombk{\bomb_k}
\def\nub{\bm{\nu}}
\def\nubk{\bm{\nu}_k}
\def\nubkk{\bm{\nu}_{k-1}}
\def\bnubk{\bar{\bm{\nu}}_k}
\def\bnuki{\bar{\nu}_{k_i}}
\def\nuki{\nu_{k_i}}

\def\bPsi{\bar{\Psi}}
\def\bPsik{\bar{\Psi}_k}
\def\psib{\bm{\psi}}
\def\phib{\bm{\varphi}}
\def\bpsib{\bar{\psib}}
\def\bpsibk{\bar{\psib}_k}
\def\bZk{\bar{\ZC}_k}
\def\bEC{\bar{\EC}}
\def\bEk{\bar{\EC}_k}
\def\bPk{\bar{P}_k}
\def\bPhik{\bar{\Phi}_k}
\def\bOmk{\bar{\Omega}_k}

\def\bsigk{\bar{\varsigma}_k}
\def\brsig{\breve{\varsigma}}
\def\bxk{\bar{\bm{x}}_k}

\def\brQ{\breve{Q}}
\def\brrho{\breve{\varrho}}

\def\ccb{\check{\cb}}
\def\bccb{\bar{\ccb}}
\def\bbccb{\bar{\bccb}}
\def\bcb{\bar{\cb}}
\def\bbcb{\bar{\bcb}}
\def\ib{\bm{i}}
\def\eb{\bm{e}}
\def\hb{\bm{h}}
\def\bxb{\bar{\xb}}
\def\bbxb{\bar{\bxb}}
\def\dxb{\tilde{\xb}}
\def\dxbk{\tilde{\xb}_k}
\def\dxbkk{\tilde{\xb}\kk}
\def\dbxbk{\tilde{\bxb}_k}
\newcommand{\dbxbki}[1]{\tilde{\bxb}_{k_{#1}}}
\newcommand{\dxbki}[1]{\tilde{\xb}_{k_{#1}}}
\newcommand{\dxbkkki}[1]{\tilde{\xb}_{k-1_{#1}}}
\def\by{\bar{y}}
\def\uy{\underb{y}}
\def\bby{\bar{\by}}
\def\fb{\bm{f}} 
\def\brxb{\breve{\x}} 
\def\bfb{\bar{\fb}} 
\def\brfb{\breve{\fb}} 
\def\brEC{\breve{\EC}} 
\def\brE{\breve{E}} 
\def\tfb{\bff{\tilde{f}}}
\def\bfk{\bar{\fb}}
\def\bbfb{\bar{\bfb}}
\def\bP{\bar{P}}
\def\bbP{\bar{\bP}}
\def\cP{\check{P}}
\def\bcP{\bar{\cP}}
\def\bbcP{\bar{\bcP}}
\def\csig{\check{\sigma}}
\def\uFk{\underb{F}_{k}}
\newcommand{\sigj}[1]{\varsigma_{#1}}
\newcommand{\sigkij}[1]{\varsigma_{k_{#1}}}
\newcommand{\brFk}{\breve{F}_{k}}
\newcommand{\fbki}{\bm{f}_{k_{i}}}
\newcommand{\fbk}[1]{\bm{f}_{k_{#1}}}
\newcommand{\bfbk}[1]{\bar{\bm{f}}_{k_{#1}}}
\newcommand{\tfbk}[1]{\tilde{\bm{f}}_{k_{#1}}}
\newcommand{\brfbk}[1]{\breve{\bm{f}}_{k_{#1}}}
\newcommand{\bfbki}{\bar{\bm{f}}_{k_{i}}}
\newcommand{\tfbki}{\tilde{\bm{f}}_{k_{i}}}
\newcommand{\brfbki}{\breve{\bm{f}}_{k_{i}}}
\newcommand{\cfbk}[1]{\check{\bm{f}}_{k_{#1}}}
\newcommand{\ufbk}[1]{\underb{\bm{f}}_{k_{#1}}}
\newcommand{\uyk}[1]{\underb{{y}}_{k_{#1}}}
\newcommand{\udek}[1]{\underb{\delta}_{k_{#1}}}
\newcommand{\udeki}{\underb{\delta}_{k_{i}}}
\newcommand{\bdek}[1]{\bar{\delta}_{k_{#1}}}
\newcommand{\bdeki}{\bar{\delta}_{k_{i}}}
\newcommand{\urho}{\underb{\rho}}
\newcommand{\brho}{\bar{\rho}}
\newcommand{\urhok}[1]{\underb{{\rho}}_{k_{#1}}}
\newcommand{\urhoki}{\underb{{\rho}}_{k_{i}}}
\newcommand{\brhok}[1]{\bar{{\rho}}_{k_{#1}}}
\newcommand{\brhoki}{\bar{{\rho}}_{k_{i}}}
\def\uybk{\underb{\y}_{k}}
\newcommand{\byk}[1]{\bar{{y}}_{k_{#1}}}
\newcommand{\brybk}{\breve{\y}_{k}}
\newcommand{\baki}[1]{\bar{a}_{k_{#1}}}
\newcommand{\bomi}[1]{\bar{\omega}_{#1}}
\newcommand{\bomki}[1]{\bar{\omega}_{k_{#1}}}
\newcommand{\ak}[1]{{a}_{k_{#1}}}
\newcommand{\aki}{{a}_{k_{i}}}
\newcommand{\caki}[1]{\check{a}_{k_{#1}}}
\newcommand{\cyk}[1]{\check{y}_{k_{#1}}}
\newcommand{\cybk}{\check{\yb}_{k}}
\newcommand{\bvk}[1]{\bar{\bm{v}}_{k_{#1}}}
\newcommand{\gbkj}[1]{{\bm{g}}_{k_{#1}}}
\newcommand{\bgbkj}[1]{\bar{\bm{g}}_{k_{#1}}}
\newcommand{\bdbki}[1]{\bar{\bm{d}}_{k_{#1}}}
\newcommand{\cdbki}[1]{\check{\bm{d}}_{k_{#1}}}
\newcommand{\dbki}{\bm{d}_{k_{i}}}
\newcommand{\dbkij}[1]{{\bm{d}}_{k_{#1}}}
\newcommand{\rbkij}[1]{{\bm{r}}_{k_{#1}}}
\newcommand{\rbki}{{\bm{r}}_{k_{i}}}
\newcommand{\vkj}[1]{{\bm{v}}_{k_{#1}}}
\newcommand{\zkj}[1]{{z}_{k_{#1}}}
\newcommand{\bzkj}[1]{\bar{z}_{k_{#1}}}
\newcommand{\ukj}[1]{{\bm{u}}_{k_{#1}}}
\newcommand{\alphak}[1]{\alpha_{k_{#1}}}
\newcommand{\betak}[1]{\beta_{k_{#1}}}
\newcommand{\gammak}[1]{\gamma_{k_{#1}}}
\newcommand{\alphaki}{\alpha_{k_{i}}}
\newcommand{\betaki}{\beta_{k_{i}}}
\newcommand{\gammaki}{\gamma_{k_{i}}}
\newcommand{\lamki}{\lambda_{k_{i}}}
\newcommand{\lamk}[1]{\lambda_{k_{#1}}}
\newcommand{\thetaki}{\theta_{k_{i}}}
\newcommand{\alphakinv}[1]{\alpha_{k_{#1}}^{-1}}
\newcommand{\balphak}[1]{\bar{\alpha}_{k_{#1}}}
\newcommand{\balphakinv}[1]{\bar{\alpha}_{k_{#1}}^{-1}}
\newcommand{\bF}{\breve{F}}
\newcommand{\bFk}{\bar{F}_k}
\newcommand{\brF}{\bar{F}}
\newcommand{\bbFk}{\bar{\bar{F}}_k}
\newcommand{\phibki}{{\bm{\varphi}}_{k_{i}}}
\newcommand{\phibk}[1]{{\bm{\varphi}}_{k_{#1}}}
\newcommand{\psibk}[1]{{\psib}_{k_{#1}}}
\newcommand{\bpsibki}[1]{{\bpsib}_{k_{#1}}}
\newcommand{\xbki}[1]{\hat{\xb}_{k_{#1}}}
\newcommand{\bpk}{\breve{p}_{k}}
\newcommand{\bqk}{\bar{q}_{k}}
\newcommand{\racine}[1]{{#1}^{\frac{1}{2}}}
\newcommand{\iracine}[1]{{#1}^{-\frac{1}{2}}}
\newcommand{\iracineT}[1]{{#1}^{-\frac{T}{2}}}
\newcommand{\racineT}[1]{{#1}^{\frac{T}{2}}}
\newcommand{\ppzc}{\mathpzc{P}}
\newcommand{\zpzc}{\mathpzc{Z}}
\newcommand{\gpzc}{\mathpzc{G}}
\newcommand{\range}{\mathpzc{R}}
\newcommand{\nul}{\mathpzc{N}}
\newcommand{\col}{\mathscr{Col}}
\newcommand{\kf}{\mathscr{K}}
\newcommand{\pscrk}{\mathscr{P}_k}
\newcommand{\zscrk}{\mathscr{Z}_k}
\newcommand{\gscrk}{\mathscr{G}_k}
\newcommand{\ugscrk}{\underb{\mathscr{G}}_k}
\newcommand{\bgscrk}{\bar{\mathscr{G}}_k}
\newcommand{\mscrk}{\mathscr{M}_k}
\newcommand{\hscr}[1]{\mathscr{H}_{#1}}
\newcommand{\hscrk}{\mathscr{H}_k}
\newcommand{\dscrk}{\mathscr{D}_k}
\newcommand{\bdscrk}{\bar{\mathscr{D}}_k}
\newcommand{\brdscrk}{\breve{\mathscr{D}}_k}
\newcommand{\cscr}{\mathscr{C}}
\newcommand{\cscrk}{{\mathscr{C}}_k}
\newcommand{\bGC}{\bar{\GC}}
\newcommand{\uGC}{\underb{\GC}}
\newcommand{\bIn}[1]{\bar{I}_{n,#1}}
\newcommand{\uIn}[1]{\underb{I}_{n,#1}} 

\def\bEk{\bar{E}_k}
\def\Ek{E_k}
\def\Vk{V_k}
\def\Vkp{V_{k+1}}
\def\Vo{V_{\bot}}
\def\Uk{U_k}
\def\Ukp{U_{k+1}}
\def\bVk{\bar{V}_k}
\def\bVkp{\bar{V}_{k+1}}
\def\Uo{U_{\bot}}
\def\bUk{\bar{U}_k}
\def\bUkp{\bar{U}_{k+1}}
\def\tFk{\tilde{F}_k}
\def\tfbk{\tilde{\fb}_k}
\def\tybk{\tilde{\y}_k}
\def\ubki{{\ub}_{k_i}}
\def\vbki{{\vb}_{k_i}}
\def\bubki{\bar{\ub}_{k_i}}
\def\bvbki{\bar{\vb}_{k_i}}

\def\Qinv{\Theta}
\def\Qki{Q}
\def\qqkkpj{\kappa}
\def\qqkj{\kappa}
\def\PPki{\Pi}
\def\spo{s}
\renewcommand{\alphaki}{\alpha_{i}}
\renewcommand{\betak}[1]{\beta_{#1}}
\renewcommand{\betaki}{\beta_i}
\renewcommand{\alphak}[1]{\alpha_{#1}}
\renewcommand{\gammaki}{\gamma_{i}}
\renewcommand{\gammak}[1]{\gamma_{#1}}
\renewcommand{\phibki}{\phib_{i}}
\renewcommand{\delki}{\delta_{i}}
\renewcommand{\delkis}{\delta_{i}^{2}}
\renewcommand{\lamki}{\eta_{i}}
\renewcommand{\thetaki}{\theta_{i}}
\renewcommand{\brhoki}{\bar{\rho}_{i}}
\renewcommand{\urhoki}{\underb{\rho}_{i}}
\renewcommand{\byki}{\bar{y}_{i}}
\renewcommand{\uyki}{\underb{y}_{i}}
\renewcommand{\sigkiii}{\sigma_{i-1}}
\renewcommand{\sigki}{\sigma_{i}}
\renewcommand{\sigkz}{\sigma_{0}}
\renewcommand{\xkiii}{z_{i-1}}
\renewcommand{\xki}{z_{i}}
\renewcommand{\xkz}{z_{0}}
\renewcommand{\bqk}{\tilde{q}}
\renewcommand{\bP}{P_\oplus}
\renewcommand{\bEC}{\EC_\oplus}
\renewcommand{\bcb}{\cb_\oplus}
\renewcommand{\sigkj}[1]{\sigma_{#1}}

\title{Bounded-error constrained state estimation of LTV systems in presence of sporadic measurements}
\author{Yasmina BECIS-AUBRY
\thanks{*The author is with Universit\'{e} d'Orl\'{e}ans, Laboratoire PRISME EA 4229 (Univ. Orl\'{e}ans - INSA CVL). 63 av. de Lattre de Tassigny, 18020 Bourges Cedex, FRANCE. Tel. +33 2 48 23 84 78 {\tt\small Yasmina.Becis@univ-orleans.fr}.}
}


\maketitle

\begin{abstract}
This contribution proposes a recursive set-membership method for the ellipsoidal state characterization for discrete-time linear time-varying models with additive unknown disturbances vectors, bounded by possibly degenerate zonotopes and polytopes, impacting respectively, the state evolution equation and the sporadic measurement vectors, which are expressed as linear inequality and equality constraints on the state vector. New algorithms are designed considering the unprecedented fact that, due to equality constraints, the shape matrix of the ellipsoid characterizing all possible values of the state vector is non invertible. The two main size minimizing criteria (volume and sum of squared axes lengths) are examined in the {\it time update} step and also in the {\it observation updating}, in addition to a third one, minimizing some error norm and ensuring the input-to-state stability of the estimation error.

The author's papers \cite{Bec:21} and \cite{Bec:23} were combined into this longer, more comprehensive version. It includes all the proofs and a few images and is meant to be a support for the reader. 
There is no introduction, no conclusion, and no application examples.

%

\end{abstract}


\setlength{\mathindent}{0pt}

\section{\bf Notations and definitions}\label{subsec_notations}
{%
\begin{enumerate}
\item The symbol $\eqg$ (\resp. $\eqd$) means that the {Left Hand Side} (\resp.   {RHS}) \emph{is defined to be equal to} the {Right Hand Side} (\resp.   {LHS}). 
Normal lowercase letters are used for scalars, capital letters for matrices, bold lowercase letters for vectors and calligraphic capital letters for sets.
$\R$, $\R^*$, $R_+$,  $R_+^*$ denote the sets of real, non-zero, nonnegative and positive numbers \resp. $\N$ and $\N^*$ are the sets of nonnegative and positive integers \resp.
$l,m,n,p,q\in\N$ designate vectors and matrices dimensions. The subscript $k\in\N$ is the discrete time step and $i,j\in\N^*$ are vector and matrix component indices. 
\item $x_i$ is the $i$\ith component of the vector $\x$.  $a_{ij}$ is the $i$\ith row and $j$\ith column  element of $A\eqg
[\ab_j]_{j=1}^m\in\Rnm$ and
$\ab_j\in\Rn$ is its $j$\ith column vector  (if $n=0$ or $m=0$, $A$ is an empty matrix).
\item\label{identity}$\bff{0_n}\in\R^n$ and $0_{n,m}\in\Rnm$ are vector and matrix of zeros; 
$I_n\eqg[\ib_{1}\ldots\ib_{n}]$ is the $n\times n$ identity matrix. 
\item $A^T$, $A^\dag$, $\rank(A)$, $\nul(A)$ and $\range (A)$ stand \resp. for the transpose, 
Moore-Penrose inverse, rank, kernel and range of the matrix $A$. If A is square, $\tr(A)$, $\abs{A}$ and $A^{-1}$, are its trace, determinant and inverse (if any) \resp.  
\item $\diag(x_i)_{i=1}^k$ is a diagonal matrix where  $x_1,\ldots,x_k$ are its diagonal elements.
%
%
 
%
\item A {Symmetric} matrix $M$ is {Positive Definite}, denoted by {SPD} or $M>0$ (\resp.   {Positive Semi-Definite} or non-negative definite, denoted by {SPSD} or $M\geq 0$) if and only if $\forall\x\in\Rn\text{--}\{\zero\}$, $\x^TMx>0$ (\resp.   $\x^TMx\geq~0$). This condition is met if and only if all its eigenvalues are real (because of its symmetry) and positive (\resp.   non-negative). The matrix inequality $M>N$ (\resp. $M\geq N$) means that  $M-N>0$ (\resp. $M-N\geq 0$).
\item 
 $\norm{\x}\eqg\norm{\x}_2\eqg\sqrt{\x^T\x}$ is the 2-norm of the vector $\x$ and  \linebreak 
  $\norm{A}\eqg\norm{A}_2\eqg\sup_{\bm{x}\neq\zero}\frac{\norm{A\xb}}{\norm{\xb}}$ 
 is the largest singular value of $A$. 
%
\item\label{unit_ball} 
$\BC^n_p\eqg\{\zb\in\R^n|\norm{\zb}_p\leq 1\}$ is a unit ball in $\R^n$ for the $p-$norm. 
$\BC^n_2$ and $\BC^n_\infty\eqg[-1,1]^n$ are the centered unit hypersphere and hypercube/box \resp.
%
\item\label{Minkowski} 
$\SC_1\oplus\SC_2\eqg\{\x\in\R^n|\x=\x_1+\x_2,\x_1\in\SC_1,\x_2\in\SC_2\}$ is the Minkowski sum of the sets $\SC_1,\SC_2\subset\R^n$ and 
%
$\mink_{i=1}^m\SC_i\eqg\SC_1\oplus\cdots\oplus\SC_m$.
\item\label{Ellipsoid} $\EC(\cb,P) \eqg \{x \in \R^n|\ (x-\cb)^T P^{-1}(x-\cb)\leq 1\}$ is an ellipsoid in $\R^n$, where $\cb\in \R^n$ is its center and $P\in\R^{n\times n}$ is an {\bf SPD} matrix that defines its shape, size and orientation in the $\R^n$ space. If $P$ is not invertible {\bf SPSD}, the ellipsoid is degenerate (\ie\ some of its axis lengths are zero) and is then defined as an affine transformation of matrix $M$, \tq $M^TM=P$, of the unit Euclidean ball $\BC_2^n$: $\EC(\cb,M^TM)=~\{\x \in \R^n|\ \x=\cb+M\zb, \zb\in\BC_2^p\}$.%
\item\label{Hyperplane} $\HC(\db ,a)\eqg\{\x \in \R^n|\x^T\db =a\}$ is a hyperplane in $\Rn$ of normal vector $\db \in\Rn$ and whose signed distance from the origin is $\frac{{a}}{\norm{\db }}$.
Let also ${\GC}(\db ,a)\eqg\{\x: \x^T\db \leq a\}$ be one of the two halfspaces into which the hyperplane divides the $\R^n$ space and ${\GC}(-\db ,-a)$  
 is the other one.  Now let $\DC(\db ,a)\eqg{\GC}(\db,1+a)\cap{\GC}(-\db,1-a)$, \ie, $\DC(\db ,a)\eqg\{\x\in\R^n: \abs{\x^T\db-a}\leq 1\}$
which is the strip of $\R^n$,  of width ${2}{\norm{\db }}^{-1}$, that can also be seen as a degenerate unbounded ellipsoid or zonotope centered at $\HC(\db ,a)$.
%
$\PC(C,\db )=\bigcap_{i=1}^m\GC(\cb_i,d_i)$ is a polyhedron.
\item\label{Zonotope} $\ZC(\cb,L)\eqg \{\x \in \R^n|\ \x=\cb+L\zb, \zb\in\BC_\infty^m\}$$=\mink_{j=1}^q\{t_j\bm{l}_{j},\abs{t_j}\leq 1\}\oplus\{\cb\}$ is a zonotope of center $\cb$, obtained by affine transformation, of shape matrix $L\in\R^{n\times m}$, of the unit box $\BC_\infty^m$, 
where $m$ can be smaller, equal to or greater than $n$. A zonotope is also a convex polyhedron with centrally symmetric faces in all dimensions. 
%
\item\label{Support function} The support function of a set $\SC\subset\Rn$ is $\rho_\SC~:\Rn\rightarrow\R$, \linebreak
$\d\x\mapsto\rho_\SC(\x)\eqg\sup_{\ub\in\SC}\u^T\x$.  $\HC\big(\x,\rho_\SC(\x)\big)$ is the supporting hyperplane of $\SC$ and $\SC\subset \GC\big(\x,\rho_\SC(\x)\big)$. 
$\d\rho_{\EC(\cb,P)}(\x)=\cb^T\x+\sqrt{\x^TP\x}$ \cf \cite{Che:94}.
\end{enumerate}
}

\section{\uppercase{Problem formulation}}\label{sec_prob_form}
%
Consider the following linear discrete time 
system 
\mathl
\begin{subequations}\label{system0}
\begin{align}
&&\xka  &= \Akk \xkka+B\kkk \taub\kkk+R\kkk\wbkk, \quad k\in\N^*\label{state_eq0}\\
\text{where }\ &&\x_{0}&\in\EC(\xz,\sigz P_0)\eqd\EC_0\subset\Rn\text{ and }\wbk\in\BC_{\infty}^m, \label{init_state_noise_bound}
\end{align}
\end{subequations}
where $\xka\in\Rn$, $\taubk\in\R^{l}$ and $\wbk\in\R^{m_k}$  
are \resp.\ the unknown state vector to be estimated, a known and bounded control vector and an unobservable bounded process noise vector with unknown statistical characteristics and which size $m\eqg {m_{k}}$ is possibly time-varying; 
$\EC(\xz, \sigz P_0)\eqd\EC_0$ is a known ellipsoid (\cf $\S$~\ref{subsec_notations}.\ref{Ellipsoid}),  where $\xz\in\Rn$ is the initial estimate of $\xka$ at $k=0$,
$P_0\in\Rnn$ is a SPD matrix, $\sigz\in\R_+^*$ is a scaling positive scalar (can be set to 1), the product $\sigz P_0$ is chosen as large as the confidence in $\xz$ is poor; $\BC_{\infty}^m$ is the unit ball for the $\infty-$norm in $\Rm$ (\cf $\S$~\ref{subsec_notations}.\ref{unit_ball}); $A_k\in\Rnn$ and  $B_k\in\Rnl$  are known state and input matrices, \resp. and $\Rk\in\Rnm$ is the generator matrix defining the shape of the zonotope bounding the unknown input vector
$R_k\wbk\in\ZC(\zero_n,R_k)$. 
Now consider the output equation for the system \eqref{system0}:\hspace{-4pt}
\begin{subequations}\label{bounds0}
\mathc
\begin{align}
&F_k^T\xka = \yk, \quad \yk\in\R^{p_k}\\
&\uyki\leq \yki\leq \byki,\ i\in\pscrk\eqg\{1,\ldots,{p_k}\},
\end{align}
\end{subequations}
where, the output matrix $F_k\eqg[\fbk{j}]_{j=1}^{p_k}\in\R^{n\times {p_k}}$ 
is time varying and so is the number of its columns, ${p}\eqg p_k\in\N$, 
which\footnote{Absolutely all variables appearing in the algorithms of this paper, except $n$, 
are time varying as attested by the subscript $k$. Yet, for an improved readability, it will be skipped on some of them, when no confusion can arise about the time step $k$.}  can be zero sometimes (in the absence of measurements).
Indeed, the measurements are available in varying amounts, at not all but only some sporadic, not a priori known, time steps $k$. Three cases can be exhaustively enumerated: 
%
1) for some $i\in\dscrk\subset\pscrk$, both (finite and distinct) bounds are available: $\uyki<\byki$; 
2) for some $i\in\gscrk\eqg(\ugscrk\cup\bgscrk)\subset\pscrk$, only one bound, either  $\byki$ (if $i\in\bgscrk$) or  $\uyki$  (if $i\in\ugscrk$)  is available, in this case, the other (unavailable) bound is considered as $\mp\infty$.
%
3) for some other $i\in\hscrk\subset\pscrk$, the bounds are equal:  $\uyki=\byki$. The sets $\dscrk$, $\bgscrk$, $\ugscrk$ and $\hscrk$ form a partition for $\pscrk$: $\pscrk=\dscrk\cup\bgscrk\cup\ugscrk\cup\hscrk$. 
The measurement inequalities \eqref{bounds0} can be rewritten as ($\rightarrow$ stands for \guil{tends to}):
\begin{subequations}\label{bounds}
\mathl
\begin{align}
\fbki^T\xka\leq\byki\text{ and } \uyki\rightarrow-\infty\eq \xka\in\bGC\ki\eqg\GC(\fbki,\byki),&\text{ if }i\in\bgscrk, \label{boundb_HSpace0}\\
 \fbki^T\xka\geq\uyki\text{ and } \byki\rightarrow+\infty\eq \xka\in\uGC\ki\eqg\GC(-\fbki,-\uyki),&\text{ if } i\in\ugscrk, \label{boundu_HSpace0}\\
\fbki^T\xka=\byki,\text{ and } \uyki=\byki\ \eq \xka\in\HC\ki\eqg\HC(\fbki,\byki), &\text{ if } i\in\hscrk,\label{bound_HyperP0}\\
\abs{\tfrac{1}{\gammaki}\fbki^T\xka-\yki}\leq1 \eq\xka\in\DC\ki\eqg\DC\Big(\tfrac{1}{\gammaki}\fbki,\yki\Big),\text{ otherwise }& (i\in\dscrk)\\
\text{where } \gammaki\eqg\tfrac{\byki-\uyki}{2}\text{ and } \yki\eqg\tfrac{\byki+\uyki}{2\gammaki};\label{bound_Strip0}
\end{align}
\end{subequations}
%
where $\GC$, $\HC$ and $\DC$ are a halfspace, a hyperplane and a strip resp. (\cf $\S$~\ref{subsec_notations}.\ref{Hyperplane}).
%
The linear inequality constraint on the state vector of the form \eqref{boundb_HSpace0} (resp. \eqref{boundu_HSpace0}) stands for a measurement corrupted by an error whose only upper (\resp. lower) bound is known; the linear equality constraint on the state vector \eqref{bound_HyperP0} represent a noiseless output; as for \eqref{bound_Strip0}, it  acts as a measurement that is affected by a bounded noise. All these three categories of outputs can obviously coexist, \ie, occur at the same time step $k$.
\begin{rema}
{The output equation \eqref{bounds0} can be  derived from the one with measurements vector: $\zk=G_k^T\xka+\vk$, subject to a noise vector belonging to a polyhedron:  $$\vk\in\PC([C_k\ -C_k],[\bar{\db}_k^T\ -\underb{\db}_k^T]^T )\subset\R^{2{p}},$$ where  $\bybk\eqg[\byk{1}\cdots\byk{p}]^T=~\!\bar{\db}_k-C_k^T\zk$, $\uybk\eqg~\![\uyk{1}\cdots\uyk{p}]^T=\underb{\db}_k-C_k^T\zk$ and $F_k = -G_kC_k$.
}
\end{rema}
\begin{assump}\label{assum}
From now on, we assume that
\begin{enumerate}
\item\label{assum_all_matrices_bounded} all known matrices and vectors intervening in \eqref{system0} and \eqref{bounds}, 
as well as 
 the SPD  $P_0$ and $\sigz\in\R_+^*$  
 are bounded;
\item\label{assum_all_matrices_nonzero} all the columns of all the matrices intervening in \eqref{system0} 
and those of $F_k$, if any, are nonzero; 
\item\label{assum_uF_full_column_rank}  the matrix $F_{{\mathscr{H}}}\eqg[\fbki]_{i\in{\hscrk}}$, intervening in \eqref{bound_HyperP0}, has full column  rank (thus avoiding contradictory constraints leading to an empty set);
\end{enumerate}
\end{assump}
\begin{aim}
We are intending here to design an estimator $\xk$ for the state  vector $\xka$  of the system \eqref{system0}-\eqref{bounds0}, such that, 
\begin{enumerate}
\item\label{requirement1}
a set (ellipsoid $\EC_k$ of center $\xk$) containing all possible values of the true state vector
$\xka$ is quantified, at each time step $k\in\N^*$ (standard requirement for a set-membership approach);
\item\label{requirement2} 
the state estimate vector $\xk$ is \emph{acceptable}, \ie, it belongs to all the sets defined in \eqref{bounds}.
%
\item\label{requirement3} 
under some conditions, the estimator $\xk$ is ISS, (Input-to-State Stable, \cf  Theorem \ref{lemm_ISS}). 
This is one of the distinguishing features of the algorithm designed here.
\end{enumerate}
\end{aim}
The other distinguishing feature is that, unlike the other set-membership techniques, such as those using exclusively intervals, zonotopes or polytopes, the one detailed here delivers an optimal (\wrt some chosen criteria) set, without any conservatism. 
Since the only measured information about the true state vector $\xka$ consists in its belonging to the sets defined in \eqref{bounds},  there is no better estimate than the one that belongs to these sets. But such an estimator is not unique and is not necessarily stable so the most suitable one will be
chosen among the set of all possible estimators by optimizing a given cost-function. 

Let $\EC_{k}\eqg\EC(\xk,\sigk P_{k})$ be the ellipsoid containing all possible values of the true state vector
$\xka$. 
Please note that the singular values 
 of the shape matrix $\sigk P_{k}$ 
correspond to the semi-lengths of its axes, whose directions are
defined by the associated--orthogonal since $P_{k}$ is symmetric--eigenvectors.
The parameter $\sigk$ is used to model the possibly non-monotonic part of the shape matrix of the ellipsoid $\EC_k$, during the measurement correction stage, since the matrix $P_k$ is decreasing then. It can be seen as the upper bound on a squared weighted estimation error norm, $(\xka-\xk)^TP_k^\dag(\xka-\xk)$.
In what follows, we have to determine the progression law for the
ellipsoid $\EC_{k}$ (and thence for the state
estimate vector $\xk$) such that the aims \bff{i}.--\bff{iii}. are
fulfilled. 
\section{ Time update (prediction stage)\ }\label{sec_time_update}
%
In the two first paragraphs of this subsection, useful tools are established in view of the development of the prediction algorithm in $\S$ \ref{subsec_algo_time_up}.
%
\subsection{ Minkowski sum of an ellipsoid and a line segment}\label{subsec_sum-ellips-segment}
%
The  lemma below gives the parameterized family of ellipsoids, $\bEC(\mu)$, that contains the Minkowski sum of the ellipsoid $A\EC\eqg\{\y\in\Rn|\y=A\x,\ \x\in\EC\}$,
on one hand and the segment $\ZC(\zero_n,\rb\rb^T)$, on the other.
\begin{lemm}\label{lem_ell_zono_sum}
Let $\cb,\ub,\rb\in\Rn$, with $\rb\neq\zero_n$, $A\in\Rnn$ and $P\in\Rnn$ SPSD. \\ 
For any $\xb\in\EC\eqg\EC(\cb,\sig P)$, $\wb\in\ZC(\ub,\rb)$ and for any $\mu\in\R_+^*$, 
\begin{subequations}\label{Ebar}
\mathl
\begin{align}
A\xb+\wb &\in A\EC(\cb,P)\oplus\EC(\ub,\rb\rb^T)
\subset\bEC(\mu)\eqg\EC\big(\bcb,\sig\bP(\mu)\big),
\\
\text{where }\qquad\qquad\bcb&\eqg A\cb+\ub, \label{cbar}\\
\bP(\mu)&\eqg(1+\mu){Q}+\tfrac{1+\mu}{\sig\mu}\rb\rb^{T}\text{, where } {Q}\eqg APA^T.\label{Pbar}
\end{align}
\end{subequations}
\end{lemm}
{\debdemo 
\cf Appendix \ref{Appendix_lem_ell_zono_sum}. 
\findemo 
}

$\mu$ is a positive scalar parameter, chosen in such a way as to minimize the size of $\bEC(\mu)$, as detailed  in the next paragraph.
\subsection{ Optimal values for the parameter $\mub$}\label{subsec_mu-opt}
%
Now, the most telling two measures of the size of an ellipsoid, \ie the volume and the SSAL (sum of the squared axes lengths) will be minimized. Since the eigenvalues of $\sig\bP(\mu)$ are the squared semi-axes lengths of $\bEC(\mu)$, the former is proportional to their product, \ie, to $\abs{\sig \bP(\mu)}$ and the latter is equal to $\sig\tr\big(\bP(\mu)\big)$.
\subsubsection{ Pseudo-volume minimization}\label{subsubsec_mu-opt-vol}
The equality constraints on the state vector, introduced by the measurements $i~\!\in\hscrk$ and resulting in the intersection of the state ellipsoid $\EC_k$ with hyperplanes (studied in $\S$~\ref{subsec_hyp}), causes the ellipsoid's shape matrix $\Pk$ to loose rank during the correction stage,  ensuing in its dimension reduction by zeroing some axes lengths and, therefore, bestowing this ellipsoid a zero volume.  
Thereupon, we shall introduce a generalized volume, the \emph{pseudo-volume} of an ellipsoid, when its usual volume can be zero.
Let us first recall that if $P\in~\!\Rnn$ is SPD, then the usual volume of an ellipsoid is $\vol\Big(\EC(\cb,P)\Big)\eqd \dfrac{2\pi^{\frac{n}{2}}}{n\Gamma(\frac{n}{2})}\abs{P}$, where $\abs{P}$ is the usual determinant of $P$ and $\Gamma$ denotes the $\Gamma-$function \cite{Wil:09}.
\begin{defi}\label{Def_pseudo-volume}
 For any SPSD matrix $P$ and any $\cb\in\Rn$, the pseudo-volume of the ellipsoid $\EC\big(\cb,P)$ is proportional to the determinant of $P$:
 \mathc
 \begin{align}
 \Vol\big(\EC(\cb,P)\big)&\eqd 
\vol(\BC^q_2)\Det{P},  \text{ where }\vol(\BC^q_2)=\tfrac{\pi^{\frac{q}{2}}}{\Gamma(\frac{q}{2}+1)},\label{eq_pseudo-volume}
 \end{align} 
where $q\eqg \rank(P)$ and $\Det{P}\eqd \d\lim_{t\rightarrow 0}\dfrac{\abs{P+tI_n}}{t^{n-q}}$ is the pseudo-determinant of the matrix $P$, \ie, the product of all its nonzero singular values.
The pseudo-volume of $\EC(\cb,P)$ is nothing else than the volume of the projection of $\EC(\cb,P)$ onto $\range(P)$.
\end{defi}
\begin{propo}\label{propo_pseudo-determinant} Let ${Q}_+\eqg b\big(Q+a\rb\rb^{T})$, $a,\: b\in\R_+^*$; then
\begin{subequations}\label{eq_pseudo-inv-rec}
\mathl
\begin{align}
\textit{i.}\qquad\qquad{q_+\eqg\rank\big({Q_+}\big)}&=
\begin{cases}
q\eqg\rank(Q),&\text{ if } \vb= \zero_n,\\
q+1,&\text{ otherwise.}
\end{cases}\label{eq_rank-rec}
\end{align}
\begin{align}
\textit{ii.}\qquad\qquad\Det{{Q}_+}&=
\begin{cases}
b^{q}\Det{{Q}}\big(1+a\rb^T\ub\big),&\text{if }\vb=\zero_n;\\
b^{q+1}\Det{{Q}}a\vb^T\vb,&\text{otherwise;}
\end{cases}\label{eq_pseudo-det_rec}
\end{align}
\mathl
\begin{align}
\text{where }\qquad\ub\eqg Q^\dag\rb\qquad\text{ and }\qquad
\vb\eqg (I_n-QQ^\dag)\rb. \label{eq_u_v_def}
\end{align}
\end{subequations}
\end{propo}
{\debdemo 
\cf Appendix \ref{Appendix_propo_pseudo-determinant}. 
\findemo 
}

\begin{theo}\label{theo_mu-vol} 
$\bEC(\mu)$ defined in \eqref{Ebar} has the minimum pseudo-volume if  $$\mu=~\!\mu_{\text{v}}\eqg\arg\min_{\mu\in\R_+^*}\Vol(\bEC(\mu)):$$ \nopagebreak
\mathl
\begin{subequations}\label{mu-vol_}
\begin{align}
&\mu_{\text{v}}\eqg
\begin{cases}
\tfrac{1}{2{q}}\sqrt{({q}-1)^{2}{h}^{2}+4{q}{h}}-\tfrac{{q}-1}{2{q}}{h},&\text{ if }\vb= \zero_n,\\
\tfrac{1}{q}, &\text{ otherwise; }
\end{cases}\label{mu-vol}\\
&\text{where $\ub$ and $\vb$ are defined in \eqref{eq_u_v_def}, }{q}\eqg\rank(Q)
\text{ and }
{h}\eqg\sig^{-1}\rb^{T}\ub.
\label{eq_q_h_def}
\end{align}
\end{subequations}
\end{theo}
{\debdemo 
\cf Appendix \ref{Appendix_theo_mu-vol}.
\findemo 
}

Noticing that the minimization of the volume of $\bEC(\mu)$ requires the computation of the pseudo-inverse, the result hereafter will serve to express 
$\bP(\mu)^\dag$ by means of 
$Q^\dag$, allowing to deduce it recursively, without the need to compute it anew at each step. 
\mathl
\begin{propo}\label{propo_pseudo-inv} 
%
Let ${Q}_+\eqg b\big(Q+a\rb\rb^{T})$, $a\in\R_+^*$; then
\begin{subequations}
\begin{align}
{Q_+^\dag}&=\tfrac{1}{b}\big(Q^\dag+{\Delta}\big), \text{ with }
{\Delta}\eqg
\begin{cases}
\tfrac{1}{\norm{\rb}^2}\Big(\!\tfrac{c}{\norm{\rb}^2}\vb\vb^T-\ub\vb^T-\vb\ub^T\!\Big),&\!\!\!\text{if }\vb\neq\zero_n;\\
-\tfrac{1}{c}\ub\ub^T,&\!\!\!\text{otherwise;}
\end{cases}\label{Delta_pseudo-inv-rec}
\end{align}
\end{subequations}
where $c\eqg\tfrac{1+a\rb^TQ^\dag\rb}{a}$ and $\ub$ and $\vb$ are defined in \eqref{eq_u_v_def}. 
\end{propo}
{\debdemo 
\eqref{eq_pseudo-inv-rec} can be obtained by applying Thm 1 and Thm 3 of \cite{Mey:73}. 
\findemo
}
\subsubsection{SSAL minimization}\label{subsubsec_mu-opt-tr}
As for the minimization of the sum of the squared axes lengths of the ellipsoid, \aka the trace criterion, it is given directly by the following theorem ensuing from the literature, where $\ssal\big(\EC(\cb,\varsigma P)\big)\eqg \varsigma\tr(P)$.
\begin{theo}[\cite{Mak:96b}]\label{theo_mu-tr_Maksarov}  $\bEC(\mu)$ defined in \eqref{Ebar} has the minimum  SSAL 
if $\mu=\mu_{\text{s}}$: \nopagebreak
\mathc
\begin{align}\label{mu-tr}
\mu_{\text{s}}&\eqg\arg\min_{\mu\in\R_+^*}\ssal\big(\bEC(\mu)\big)=\arg\min_{\mu\in\R_+^*}\tr\big(\bP(\mu)\big)=\sqrt{\dfrac{\rb^{T} \rb}{\sig\tr(APA^T)}}.
\end{align}
%
\end{theo}
\renewcommand{\muki}[1]{\mu_{#1}}
\subsection{The time update algorithm}\label{subsec_algo_time_up}
Let $\EC\kkp\eqg\EC(\xkkp,\sigk\Pkkp)$ be the ellipsoid containing the
\guil{reachable set} 
of every possible value of $\xka\in\EC_k\eqg\EC(\xk,\sigk P_k)$ that
evolves according to the plant dynamics eq. \eqref{state_eq0}, subject to \eqref{init_state_noise_bound}. The following theorem gives the parametrized family of ellipsoids $\EC\kkp$ (of parameter $\mub$) that contains 
 $A_k\EC_k\oplus\ZC(\zero_n,{R}_k)$.
\begin{theo}[Prediction stage]\label{theo_time_update}
If $\xka\in\EC_k\eqg\EC(\xk,\sigk\Pk)$ and $\xka$ obeys to \eqref{system0}, then $\forall\mub\eqg(\mu_1,\ldots,\mu_m)^T\in]0,+\infty[^m$,%
\begin{subequations}\label{predic_eq}
\mathl
\begin{align*}
\xkpa\in\EC(\xkkp,\sigk P\kkp)\eqd\EC\kkp\eqg\EC\kkpj{m}
\end{align*}
\mathl
\begin{align}
\text{where }\quad\EC\kkpj{i}&\eqg \EC(\xkkp,\sigk\Qki_{i})\supseteq \EC\kkpj{i-1}\oplus\ZC(\zero_n,\rbki\rbki^T),\\
\text{where }\quad\xkkp&\eqg A_k\xk+B_k\taubk,\label{x_pred}\\
\Pkkp&\eqg \Qki_{m},\label{P_pred}\\
 \Qki_{0}&\eqg A_k\Pk A_k^{T};\label{P_pred0}\\
\Qki_{i}&\eqg(1+\mu_i)\left(\Qki_{i-1} +\tfrac{1}{\mu_i\sigk}\rbki\rbki^{T}\right), \forall i\in\{1,\ldots,m\};\label{P_pred1}
\end{align}
\end{subequations}
$\rbki$ (the $i$\ith column of $\Rk$) 
being the generator vector of the zonotope
containing all possible values of the process noise $ R_k\wbk$.
\end{theo}
{\debdemo 
\cf Appendix \ref{Appendix_theo_time_update}.
\findemo 
}

Now, the results of $\S$ \ref{subsec_mu-opt} are employed in order to express the optimal predicted ellipsoid $\EC\kk$ according to the volume and trace criterion respectively.
\subsubsection{Pseudo-volume minimization}
\begin{theo}
\label{theo_mu-vol_k} 
$\EC\kkpj{i}$ (\cf Thm \ref{theo_time_update})
has the minimum pseudo-volume \ssi $\d\mu_i\eqg\arg\min_{\mu_i\in\R_+^*}\Det{\Qki_{i}}=\mu_{\text{v}}$, given by \eqref{mu-vol},  
where
\nopagebreak
\mathl
\begin{subequations}\label{mu-vol_k}
\mathl
\begin{align}
q&\eqg \qqkkpj_{i-1}, \ 
h\eqg\sigk^{-1}\rbki^{T} \ub, \ 
\ub\eqg \Qinv_{i-1}\rbki, \ 
\vb\eqg \rbki - \Qki_{i-1} \ub,\label{eq_u-v-mu-vol-pred} \\ 
\qqkkpj_{i}&
\eqg
\begin{cases}
\qqkkpj_{i-1},&\text{ if }\vb = \zero_n,\\
\qqkkpj_{i-1}+1&\text{ otherwise;}
\end{cases}
\label{eq_q-mu-vol-pred}\\
{\qqkkpj_{0}}&\eqg\begin{cases}
{q}_k,&\text{ if }\rank(A_k)=n; \\
\rank(\Qki_{0}),&\text{ otherwise;}
\end{cases}\label{eq_q-mu-vol-pred}\\
\intertext{where ${q}_{k}\eqg\rank(P_k)$ (\cf \eqref{eq_correc-q}); $\qqkkpj_{i}\!\eqg\rank(\Qki_{i-1})$, $q\kkp\eqg\qqkkpj_{m}$ and $\Qinv_{i}\eqg \Qki_{i}^\dag$:} 
\Qinv_{i}&\eqg\tfrac{1}{1+\mu_{i}}\big(\Qinv_{i-1}+  {\Delta_{i}}\big), \ i\in\{1,\ldots,m\}, \label{eq_bQ-mu-vol-pred}\\ 
\Qinv_{0} &\eqg \Qki_{0}^\dag= \big(A_kP_kA_k^T\big)^\dag,
\label{Q_pseudo-inverse}\\
\text{with }\Delta_{i}&=\Delta \text{ given in \eqref{Delta_pseudo-inv-rec}, where }\rb\eqg\rbki\text{ and }c\eqg{\sigk\mu_{i}+\rbki^T\ub}.\label{Delta_k_def}
\end{align}
\end{subequations}
\end{theo}
{\debdemo 
Direct consequence of Thm \ref{theo_mu-vol} and Proposition \ref{propo_pseudo-inv}, where $a\eqg\tfrac{1}{\mu_i\sigk}$ and $b\eqg 1+\mu_i$.
\findemo 
}

Algorithm \ref{Algo_Predic_VolMin} resumes the last two theorems and computes $\EC\kkp$ from $\EC_k$, where 
$\x\leftarrow \xk$, $P\leftarrow P_k$, $\varsigma\leftarrow \sigk$, $q\leftarrow q_k$, $A\leftarrow A_k$, $B\leftarrow B_k$, $\rb_j\leftarrow \rbkij{j},j=1,m$, 
 $\taub\leftarrow \taubk$ and $m\leftarrow m_k$.
{%
\begin{algorithm}[H]
\caption{Computation of the minimal pseudo-volume predicted ellipsoid}
\begin{algorithmic}[1]\label{Algo_Predic_VolMin}
\REQUIRE $\x$, $P$, $\varsigma$, $q$, $A$, $B$, $\rb_{1}\cdots\rb_{m}$, $\taub$, $m$
\ENSURE $\x$, $P$, $q$
\STATE $ {Q}\eqg AP A^{T}$;  \COMMENT{\cf\eqref{P_pred0}}
\STATE $ \Qinv \eqg{Q}^\dag$; \COMMENT{\cf\eqref{Q_pseudo-inverse}}
\STATE\label{Algo_Predic_VolMin_q0} $q\eqg\rank({Q})$;
\COMMENT{this line can be skipped if $\rank({A_k})= n$; \cf\eqref{eq_q-mu-vol-pred}} 
\FOR{$i=1,\cdots,m$} 
\STATE $\ub\eqg \Qinv\rb_i$;
$\vb\eqg \rb_i - Q \ub$; \COMMENT{\cf\eqref{eq_u-v-mu-vol-pred}}
\IF{$\vb=\zero_n$}
\STATE $h\eqg\varsigma\rb_i^{T}\ub$; $\mu=\tfrac{1}{2{q}}\sqrt{({q}-1)^{2}{h}^{2}+4{q}{h}}-\tfrac{{q}-1}{2{q}}{h}$; \COMMENT{\cf\eqref{eq_u-v-mu-vol-pred}, \eqref{mu-vol} \resp.}
\STATE $ {\Delta}\eqg-\tfrac{1}{\varsigma\mu+\rb_i^T\ub}\ub\ub^T$; \COMMENT{\cf\eqref{Delta_k_def} and \eqref{Delta_pseudo-inv-rec}}
\ELSE
\STATE $\mu=\tfrac{1}{{q}}$; \COMMENT{\cf\eqref{mu-vol}}
\STATE\label{Algo_Predic_VolMin_q}  $q\leftarrow q+1$; \COMMENT{\cf \eqref{eq_q-mu-vol-pred}}
\STATE $V\eqg \ub\vb^T$;
\STATE $ {\Delta}\eqg\tfrac{1}{\norm{\rb_i}^2}\bigg(\tfrac{{\varsigma\mu+\rb_i^T\ub}}{\norm{\rb_i}^2}\vb\vb^T-V-V^T\bigg)$; \COMMENT{\cf\eqref{Delta_k_def} and \eqref{Delta_pseudo-inv-rec}}
\ENDIF
\COMMENT{\cf\eqref{mu-vol}}
\STATE ${Q}\eqg(1+\mu)\big({Q}+\tfrac{1}{\mu\varsigma}\rb_i\rb_i^{T}\big)$;  \COMMENT{\cf\eqref{P_pred1}}
\STATE $\Qinv\eqg(1+\mu)^{-1}\big(\Qinv+  {\Delta}\big)$; \COMMENT{\cf\eqref{eq_bQ-mu-vol-pred}}
\ENDFOR
\STATE ${\x}\gets A\x+B\taub$ \COMMENT{\cf\eqref{x_pred}};
${P}\gets {Q}$ \COMMENT{\cf\eqref{P_pred}}; 
\end{algorithmic}
\end{algorithm}
}
\begin{rema}\label{rema_trace-vor-choice-mu}
It is worth noting that the volume minimization problem \linebreak $\d\arg\min_{\mu_i}\Det{\Qki_{i}}$ has an explicit solution here.  If the unknown input vector was bounded by an ellipsoid, as was the case in \cite{Mak:96b,Dur:01,Bec:08,She:18}, rather than by an $\infty-$norm bounded set, such as a zonotope, $\d\mu_{\text{v}}$ would be the unique positive root of an $n-$order polynomial. Nevertheless, considering that the pseudo-inverse of a $n\times n$ matrix is needed at each time step $k$, in line with \eqref{Q_pseudo-inverse}, 
the trace minimization is more appealing, at least from the computational point of view.
\end{rema}
\subsubsection{Sum of the squared axes' lengths minimization}
\begin{theo}
\label{theo_mu-tr} $\EC\kkpj{i}$ defined in Thm \ref{theo_time_update} has the minimum  SSAL, \ssi 
$\d\mub\eqg\arg\min_{\mu_i\in\R_+^*}\ssal(\EC\kkp)=\mub_{\text{s}}$,  
\mathl
\begin{align}\label{mu-tr_k}
\text{where } \mu_{{\text{s}}_{i}}&\eqg\arg\min_{\mu_i\in\R_+^*}\tr(\Qki_{i})=\sqrt{\dfrac{\rbki^T \rbki}{\sigk\tr(\Qki_{i-1})}},  i\in\{1,\cdots,m\}, 
\end{align}
%
and the recursive formula \eqref{P_pred}-\eqref{P_pred1} becomes: 
\begin{subequations}\label{predic_eq_glo}
\mathl
\begin{align}
P\kkp&=\big(1+\tfrac{\varpi}{{\vartheta}}\big)\big(\Qki_{0}+\tfrac{{\vartheta}}{\sigk}{M}_{k}\big),\label{P_pred_tr}\\
{\text{where } }\qquad
 {\vartheta}&\eqg \sqrt{\sigk\tr(\Qki_{0})}\text{ and $\Qki_{0}$ given in \eqref{P_pred0},}\label{mu0_def}
  \end{align}
 \begin{align}
{\varpi}\eqg\sum_{i=1}^m {\norm{\rbki}} 
\text{ and }{M}_{k}\eqg\sum_{i=1}^m\tfrac{1}{{\norm{\rbki}}}\rbki\rbki^T
=R_k{\diag\Big(\tfrac{1}{{\norm{\rbki}}}\Big)_{i=1}^m}R_k^T.\label{bR_def}
\end{align}
\end{subequations}
\end{theo}
{\debdemo 
\cf Appendix \ref{Appendix_theo_mu-tr}. 
\findemo 
}

Algorithm \ref{Algo_Predic_TrMin} computes the value of $\mub_{\text{s}}$ given in Thm \ref{theo_mu-tr} .
{%
\begin{algorithm}
\caption{Computation of the minimal SSAL predicted ellipsoid}
\begin{algorithmic}[1]\label{Algo_Predic_TrMin}
\REQUIRE $\x$, $P$, $\varsigma$, $A$, $B$, $\rb_{1}\cdots\rb_{m}$, $\taub$, $m$ 
\ENSURE $\x$, $P$
\STATE $ {Q}\eqg APA^{T}$; \COMMENT{\cf\eqref{P_pred0}}
\STATE $ {\vartheta}\eqg \sqrt{\varsigma\tr(Q)}$; \COMMENT{\cf\eqref{mu0_def}}
\STATE ${\varpi} = 0$;
\STATE ${M}=0_{n\times n}$;
\FOR{$i=1,\cdots,m$} 
\STATE $\epsilon\eqg{\norm{\rb_i}}$;
\STATE ${\varpi} \leftarrow {\varpi} + \epsilon$;
\STATE  ${M}\leftarrow {M}+\frac{1}{\epsilon}\rb_i \rb_i^T$;
\ENDFOR
\STATE $\x\gets A\x+B\taub$ \COMMENT{\eqref{x_pred}}; $ P\gets \big(1+\tfrac{{\varpi}}{{\vartheta}}\big)\big(Q+\tfrac{{\vartheta}}{\varsigma}{M}\big)$\COMMENT{\eqref{P_pred_tr}};\\ 
%
\end{algorithmic}
\end{algorithm}
}

\begin{rema}
When minimizing the SSAL of $\EC\kk$, there is no need to compute the $m_k$ intermediate values of neither $\Qki_{i}$, nor $ \mu_{{\text{s}}_{i}}$ given by the recursive formulae \eqref{P_pred}-\eqref{P_pred1} and \eqref{mu-tr_k}. $\Pkk$ can be computed directly using \eqref{predic_eq_glo} instead. Notice also that all ${{\norm{\rbki}}}$ are nonzero 
thanks to the assumption \ref{assum_all_matrices_nonzero}
\end{rema}
\begin{rema}
It is possible to minimize the weighted sum of the squared axes lengths of  $\EC\kkp$: $\tr(C\Pkkp C^T)$, for any $C\in\R^{n_C\times n}$, $n_C\in\N^*$. In this case, instead of \eqref{mu-tr_k}, the optimal value for $\mub$ would be $\mub_{k_{\text{s}}}\eqg\mub_{\text{s}}\eqg[\mu_{{\text{s}}_{1}},\ldots,\mu_{{\text{s}}_{m}}]$, where 
\begin{subequations}
\mathc
\begin{align}\label{mu-tr-C}
\mu_{{\text{s}}_{i}}&\eqg\sqrt{\frac{\rkit C^TC\rki}{\sigk\tr(C \Qki_{i} C^T)}},\  i\in\{1,\cdots,m\},
\end{align}
and the equations \eqref{bR_def} would be replaced by
\begin{align}
{\varpi}&\eqg\sum_{i=1}^m  {\norm{C\rbki}} \text{ and }
{M}_{k}\eqg\sum_{i=1}^m{\norm{C\rbki}} ^{-1}C\rbki\rbki^TC^T.\label{bR_def_C}
\end{align}
\end{subequations}
\end{rema}

Given the ellipsoid  $\EC_k$ at the time step $k$, Thm \ref{theo_time_update} provides the predicted ellipsoid $\EC\kkp$, whose center is given by \eqref{x_pred} and whose shape matrix is given, up to the factor $\sigk$, by the recursive formulae \eqref{P_pred}-\eqref{P_pred1}, which depends on $\mub$. Thms \ref{theo_mu-vol_k} and \ref{theo_mu-tr}  offer the optimal values for this parameter according to two cost functions, whose choice is left to the reader. 

%
Discussions on how to choose the parameters $\mu_{i}$ to achieve more optimality conditions for the ellipsoid bounding the reachable set for continuous-time systems are proposed in \cite{Che:05} and \cite{Kur:00}. 
See also survey books \cite{Che:94}, \cite{Kur:97} for a more complete overview. 
%
\section{ Measurement update (correction stage)\ }\label{sec_measurement_update}
%
The dynamic state evolution equation \eqref{system0} allowed to compute the predicted ellipsoid $\EC\kk$ which contains all possible values of the state vector $\xka$ taking into account all the measurements up to time step $k-1$ if any. Now, let us recall the other sets containing $\xka$, obtained from the measurements:
\mathc%
\begin{align}
\eqref{bounds}  &\eq\xka\in\bigcap_{i\in\gscrk}\GC\ki\cap\bigcap_{i\in\dscrk}\DC\ki\cap\bigcap_{i\in\hscrk}\HC\ki,\text{ if } {p}_{k}\neq 0.\label{set_XC_P_eq}
\end{align}
${p}\eqg {p_k}$ is the number of measurements at time step $k$. It is interesting to notice that the intersection of half-spaces can be considered as a possibly unbounded polyhedron and that the intersection of strips is a zonotope:  
\begin{align}
\bigcap_{i\in\gscrk}\GC\ki&\eqd\PC_k\eqg\PC\big([\fbki]_{i\in\gscrk},[\yki]_{i\in\gscrk}\big),\\
\bigcap_{i\in\dscrk}\DC\ki&\eqd\ZC_k\eqg\ZC^{\HC}\big([\fbki]_{i\in\dscrk},[\yki]_{i\in\gscrk}\big).
\end{align}
The correction stage consists in performing the intersection between $\EC\kk$ and the set \eqref{set_XC_P_eq}, allowing to find $\EC_k\supset\SC_k$ in light of the current measurements, where
\mathc
\begin{align}
\SC_k&\eqg\Big(\big(\EC\kk\cap\bigcap_{i\in\gscrk}\GC\ki\big)\cap\bigcap_{i\in\dscrk}\DC\ki\Big)\cap\bigcap_{i\in\hscrk}\HC\ki\nn\\
&=\Big(\big(\EC\kk\cap\PC_k\big)\cap\ZC_k\Big)\cap\bigcap_{i\in\hscrk}\HC\ki.
\label{SC_set} 
\end{align}
It will be shown that this intersection is the one between $\EC\kk$ and the possibly degenerate (if $\hscrk\neq\emptyset$) zonotope.
It does not result in an ellipsoid in general and has to be circumscribed by such a set, which is the subject of the upcoming paragraphs.
We shall begin by working on the intersection $\d\EC\kk\cap\GC\ki$ in $\S$~\ref{subsec_poly}.
Secondly, we'll be dealing with the intersection between an ellipsoid and a  strip in order to carry out the set obtained in $\S$~\ref{subsec_poly} and intersecting it with $\d\bigcap_{i\in\dscrk}\DC\ki$;  
$\S$~\ref{subsec_strip} provides the optimal ellipsoid overbounding this intersection.
Thirdly, the intersection of an ellipsoid with a hyperplane will be presented in $\S$~\ref{subsec_hyp}, in order to handle the intersection of the previously obtained ellipsoid with $\d\bigcap_{i\in\hscrk}\HC\ki$.
Finally, all these results will be compiled in a unique state estimation algorithm in $\S$~\ref{subsec_algo}.
\subsection{ Intersection of an ellipsoid with a halfspace}\label{subsec_poly}
The intersection between the ellipsoid $\EC\kk$ obtained in $\S$~\ref{sec_time_update} and the polyhedron $\PC_k$ 
can be reformulated as the intersection of $\EC\kk$ and a series of strips $\DC\ki$. 
To grasp this idea, take any closed convex set  $\SC$ and a hyperplane $\HC$ intersecting it. The intersection of $\SC$ with a halfspace $\GC$ delimited by $\HC$ is nothing else that its intersection with the strip formed between $\HC$ and a support hyperplane of $\SC$, parallel to $\HC$ and contained in $\GC$. Now, if $\HC$ doesn't intersect $\SC$, the latter is either a subset of $\GC$ or lies outside of it, and if $\HC$ is tangent to $\SC$ (being its support hyperplane), then $\SC$ is either again a subset of $\GC$  or it has only one point in common with it. 
In the case where $\SC$ is an ellipsoid and the intersecting halfspace corresponds to the constraint \eqref{boundb_HSpace0}, the theorem below provides the parameters of the intersecting strip. 
To obtain the intersection of an ellipsoid with the halfspace given by the constraint \eqref{boundu_HSpace0}, it suffices to replace $\fb$ by $-\fb$ and  $\by$ by $\uy$:
\begin{theo}[ellipsoid-halfspace intersec.]\label{theo_ellips_HSpace_inter} Let 
$\cb\in~\Rn$, $\fb\in\Rn\text{--}\{\zero_n\}$, $P\in\Rnn$ SPSD, $\varsigma\in\R_+^*$ and $\by\in\R$.
\mathc
\begin{subequations}
\begin{align}
\intertext{{If } $\by<-\urho$,\hfill{\normalfont{(case 1)}}}
&&\hspace{-5mm}\EC(\cb,\varsigma P)\cap\GC(\fb,\by)&=\emptyset;\label{cas1a}\\
\intertext{else if $\by\geq\brho$,\hfill{\normalfont{(case 2)}}}
&&\hspace{-5mm}\EC(\cb,\varsigma P)\cap\GC(\fb,\by)&=\EC(\cb,\varsigma P);\label{cas1b}\\
\intertext{else if 
$\by=-\urho$,\hfill{\normalfont{(case 3)}}}
&&\hspace{-5mm}\EC(\cb,\varsigma P)\cap\GC(\fb,\by)&=\EC(\cb,\varsigma P)\cap\HC(\fb,-\urho)
=\{\cb-\varsigma^{\frac{1}{2}}(\fb^TP\fb)^{-\frac{1}{2}} P\fb\};\hspace{-5mm}\label{cas3_HSpace}\\
\intertext{{else  (}$-\urho<\by<\brho$), \hfill\normalfont{(case 4)}}
&&\hspace{-5mm}\EC(\cb,\varsigma P)\cap\GC(\fb,\by)&=\EC(\cb,\varsigma P)\cap\DC(\tfrac{1}{\gamma}\fb,{y}),\label{case4_Hspace}\\
\intertext{where}
&&\gamma&\eqg \tfrac{1}{2}(\by+\urho)
\text{ and }{y}\eqg \tfrac{1}{2\gamma}(\by-\urho)\label{gamma_y_HSpace_def}\\ 
&&\urho&\eqg\rho_{\EC(\cb,\varsigma P)}(-\fb)=-\cb^T\fb+\sqrt{\varsigma\fb^TP\fb}\text{ (\cf $\S$~\ref{subsec_notations} \ref{Support function})}\hspace{-5mm}\label{urho_def}\\
&&\brho&\eqg\rho_{\EC(\cb,\varsigma P)}(\fb)=\cb^T\fb+\sqrt{\varsigma\fb^TP\fb}. \label{brho_def}
\end{align}
\end{subequations}
\end{theo}
\debdemo \cf Appendix \ref{Appendix_theo_ellips_HSpace_inter}. 
\findemo

\begin{figure}[h]
\captionsetup[subfigure]{labelformat=empty}
\begin{subfigure}[b]{.5\textwidth}
  \centering
  \includegraphics[width=\linewidth]{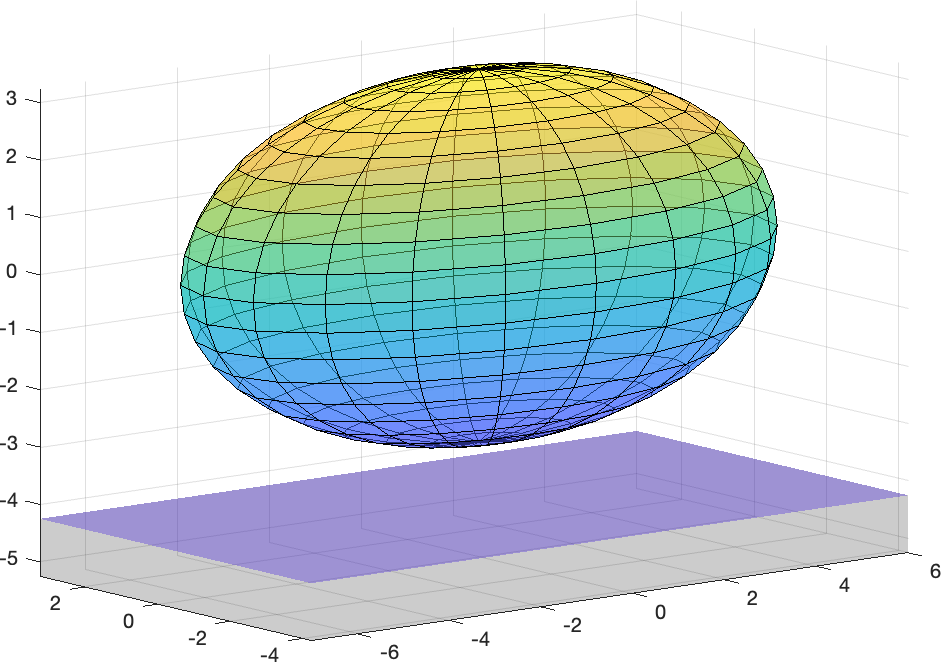}  
  \caption{Case 1: $\by-\cb^T\fb<-\sqrt{\varsigma\fb^TP\fb}$ }
\end{subfigure}
\hfill
\begin{subfigure}[b]{.5\textwidth}
  \centering
  \includegraphics[width=\linewidth]{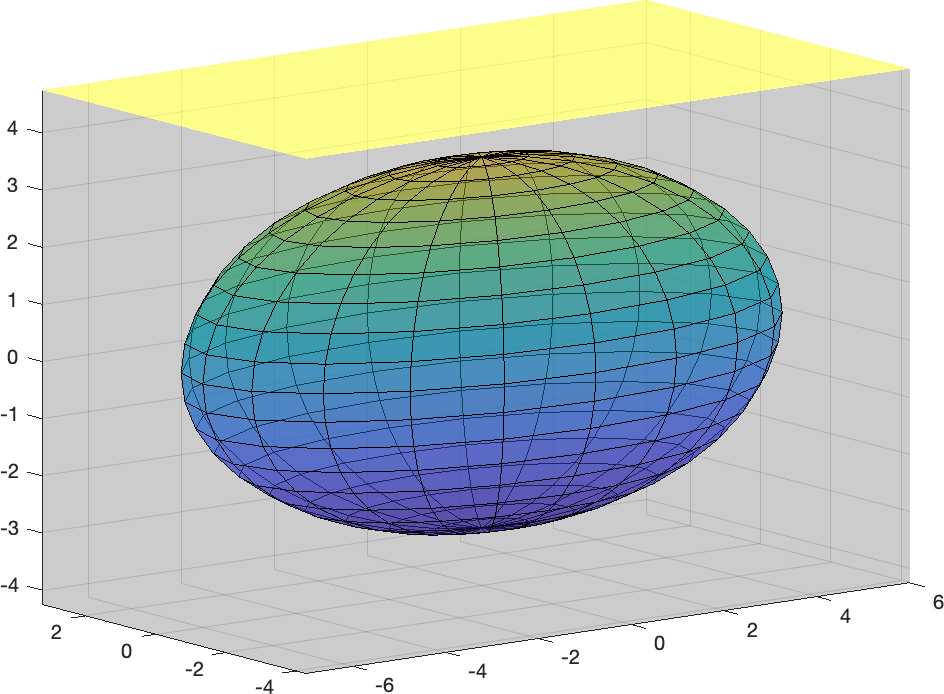}  
  \caption{Case 2: $\by-\cb^T\fb>\sqrt{\varsigma\fb^TP\fb}$} 
\end{subfigure}

\begin{subfigure}[b]{.5\textwidth}
  \centering
  \includegraphics[width=\linewidth]{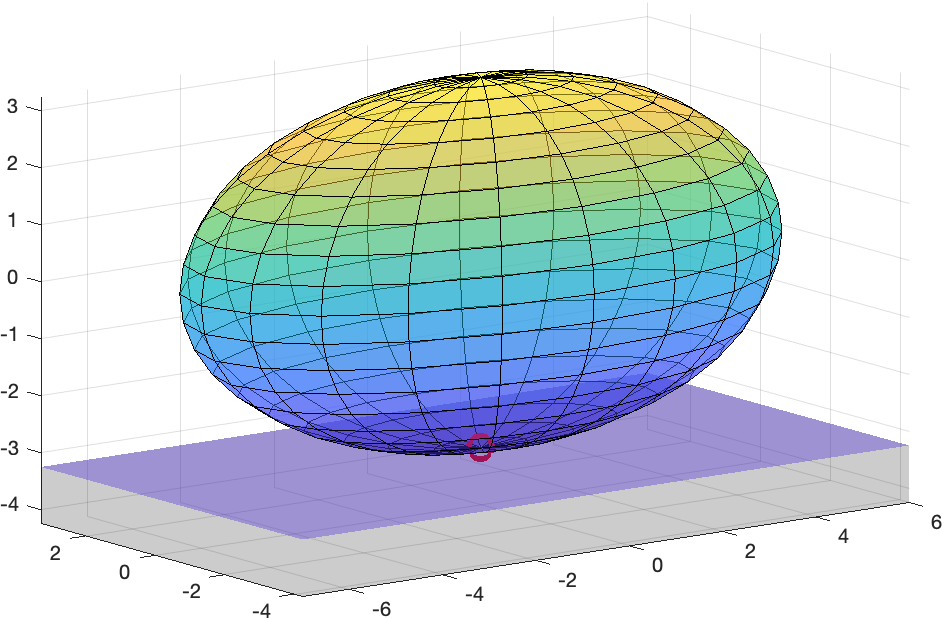}  
  \caption{Case 3: $\by-\cb^T\fb=-\sqrt{\varsigma\fb^TP\fb}$} 
\end{subfigure}
\hfill
\begin{subfigure}[b]{.5\textwidth}
  \centering
  \includegraphics[width=\linewidth]{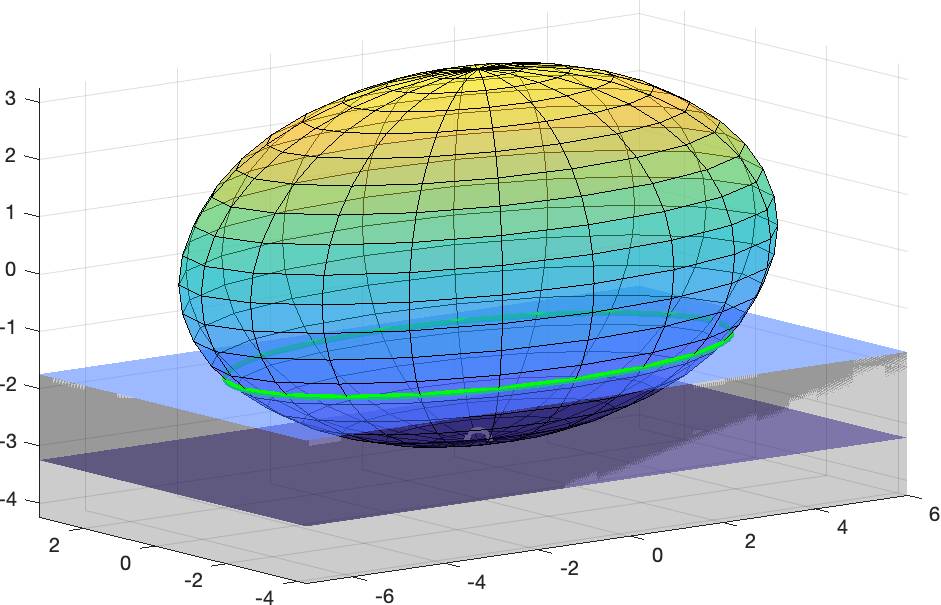} 
  \caption{Case 4: 
  $\abs{\by-\cb^T\fb}<\sqrt{\varsigma\fb^TP\fb}$} 
 \label{ell_hs_inters_fig}
\end{subfigure}
\caption{Illustration of the four cases of Theorem \ref{theo_ellips_HSpace_inter}  for $n=3$}
\label{fig_ellips_HSpace_inter}
\end{figure}

The figure \ref{fig_ellips_HSpace_inter} illustrates the above theorem in the case where $n=3$ ($\xka\in\R^3$). 
The halfspace is depicted by the shaded region and its upper boundary is the colored hyperplane.
\begin{description}
\item[Case 1:] when $\by-\cb^T\fb<-\sqrt{\varsigma\fb^TP\fb}$, the ellipsoid $\EC(\cb,\varsigma P)$ is outside the halfspace $\GC(\fb,\by)$ and their intersection is thus empty.
\item[Case 2:] when $\by-\cb^T\fb>\sqrt{\varsigma\fb^TP\fb}$, the ellipsoid is entirely contained in the halfspace and their intersection is nothing else than the ellipsoid $\EC(\cb,\varsigma P)$ itself.
\item[Case 3:] when $\by-\cb^T\fb=-\sqrt{\varsigma\fb^TP\fb}$, the ellipsoid is tangent to the halfspace and the intersection is reduced to one single point given by \eqref{cas3_HSpace} represented by a red \guil{o} on the figure \ref{fig_ellips_HSpace_inter}- (Case 3).
\item[Case 4:] when $\abs{\by-\cb^T\fb}<\sqrt{\varsigma\fb^TP\fb}$, the intersection between the ellipsoid, $\EC(\cb,\varsigma P)$  and the halfspace $\GC(\fb,\by)$ is the same as the intersection of this ellipsoid with a strip $\DC(\frac{1}{\gamma}\fb,y)$ (dark shaded area), bounded on one hand by the hyperplane $\HC(\fb,\by)$ (cyan), the boundary of $\GC(\fb,\by)$, and on the other hand, by the support hyperplane, $\HC(-\fb,-\urho)$ (violet) of the ellipsoid. 
\end{description}
%
\subsection{ Ellipsoid bounding the intersection of an ellipsoid with a strip} \label{subsec_strip} 
In the previous paragraph, we showed that the incorporation of the measurements $i\in\gscrk$ result, as for those $i\in\dscrk$, from the intersection of the predicted ellipsoid with a zonotope, formulated as an intersection of several strips. We need now to overbound this intersection by an ellipsoid. 
To begin with, the theorem below presents a family of parametrized ellipsoids (of parameter $\beta$) that contain an ellipsoidal layer, coming out of the intersection of $\EC(\cb,\varsigma P)$ with the strip 
$\DC(\fb,y)$, which can be considered--interestingly enough--as an ellipsoid unbounded in the direction orthogonal to $\fb$.
\begin{theo}[ellips./strip inters.]\label{theo_ellips_strip_inter}
Let $\cb\in\Rn$,  $\varsigma\in\R_+^*$, $y\in\R$, $P\in\Rnn$ SPSD and $\fb\in\Rn\text{--}\{\zero_n\}$,
%
 \mathc
\begin{subequations}\label{lemm_ellips_strip_inter_eq}
\begin{align}
\intertext{if $y-1>\brho\vee y+1<-\urho$,\hfill\normalfont{(case 1)}}
\hspace{-6pt}\DC(\fb,y)\cap\EC(\cb ,\varsigma P)&=\emptyset;\\
\intertext{else if $y+1\geq\brho \wedge y-1\leq -\urho$,\hfill\normalfont{(case 2)}}
\hspace{-6pt}\DC(\fb,y)\cap\EC(\cb ,\varsigma P)&=\EC(\cb ,\varsigma P);\\
\intertext{else if $y=-\urho-1$,\hfill\normalfont{(case 3.a)}}
\hspace{-6pt}\DC(\fb,y)\cap\EC(\cb ,\varsigma P)&=\HC(\fb,-\urho)\cap\EC(\cb,\varsigma P)=\{\cb-\varsigma^{\frac{1}{2}}(\fb^TP\fb)^{-\frac{1}{2}}P\fb\};\!\!\!\!\!\label{ellips_str_tang.up}\\
\intertext{else if $y=\brho+1$,\hfill\normalfont{(case 3.b)}}
\hspace{-6pt}\DC(\fb,y)\cap\EC(\cb ,\varsigma P)&=\HC(\fb,\brho)\cap\EC(\cb,\varsigma P)=\{\cb+\varsigma^{\frac{1}{2}}(\fb^TP\fb)^{-\frac{1}{2}}P\fb\};\label{ellips_str_tang.down}
\end{align}
\end{subequations}
\vspace{-\baselineskip}
\begin{subequations}\label{Ellips_Strip_Inter}
\begin{align}
\intertext{else $(-\urho< y+1<\brho\vee-\urho< y-1<\brho), \ \forall\beta\in]0,1[$,\hfill\normalfont{(case 4)}}
\hspace{-8mm}\DC(\fb,y)\cap\EC(\cb ,\varsigma P)=
\DC(\breve{\fb},\breve{y})\cap\EC(\cb ,\varsigma P)
\subset\EC\big({\cb }_{\DC}(\beta),{\varsigma}_{\DC}(\beta){P}_{\DC}(\beta)\big)\eqd{\EC}_{\DC}(\beta),\!\! \label{ellips_strip_inter}
\end{align}
\mathl
\begin{align}
\text{where }
\breve{\fb}\eqg\tfrac{1}{\gamma}\fb \text{ and } \breve{y}\eqg\tfrac{1}{2\gamma}(\by+\uy)=\tfrac{1}{\gamma}(\fb^T\cb+\delta), \label{breve_f_y_def}
\end{align}
\mathc
\begin{align}
{P}_{\DC}(\beta)& \eqg  P- {\alpha\beta} P\fb\fb^T P=P- {\alpha\beta} \phib\phib^T,\quad      \label{P_ellips_strip_inter}\\
{\cb}_{\DC}(\beta)   & \eqg  \cb +{\alpha\beta}\delta P\fb=\cb +{\alpha\beta}\delta\phib,  \label{c_ellips_strip_inter}\\
{\varsigma}_{\DC}(\beta) & \eqg  \varsigma+{\alpha\beta}\left(\tfrac{\gamma^2}{1-\beta}-\delta ^2\right),\label{sig_ellips_strip_inter}\\ 
\alpha&\eqg \big(\fb^TP\fb\big)^{-1}= \big(\fb^T\phib\big)^{-1}\text{ and } \phib\eqg P\fb, \label{alpha_def}\\
\delta&\eqg \tfrac{1}{2}(\by+\uy)-\fb^T\cb=\tfrac{1}{2}(\by+\uy-\brho+\urho),\label{delta_def}\\
    \gamma&\eqg \tfrac{1}{2}(\by-\uy),\label{gamma_def}\\ 
    \by&\eqg\min(y+1,\brho) \text{ and } \uy\eqg\max(y-1,-\urho),
 \end{align}
  \end{subequations}
and  $\urho$ and $\brho$ are defined in \eqref{urho_def}--\eqref{brho_def}.
\end{theo}
\debdemo \cf Appendix \ref{Appendix_theo_ellips_strip_inter}. 
\findemo
\begin{figure}[ht]
\captionsetup[subfigure]{labelformat=empty}
\hfill
\begin{subfigure}[b]{.32\linewidth}
  \centering
  \includegraphics[width=\linewidth]{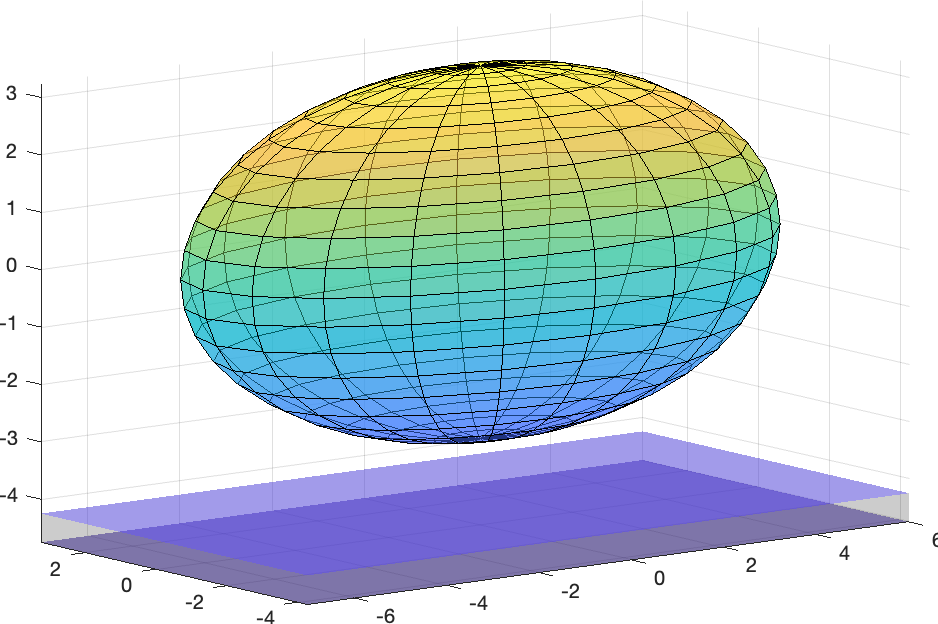}  
  \subcaption{{\tiny (1a): ${y-\cb^T\fb}<-\sqrt{\varsigma\fb^TP\fb} -1$}}
\end{subfigure}
\begin{subfigure}[b]{.32\linewidth}
  \centering
  \includegraphics[width=\linewidth]{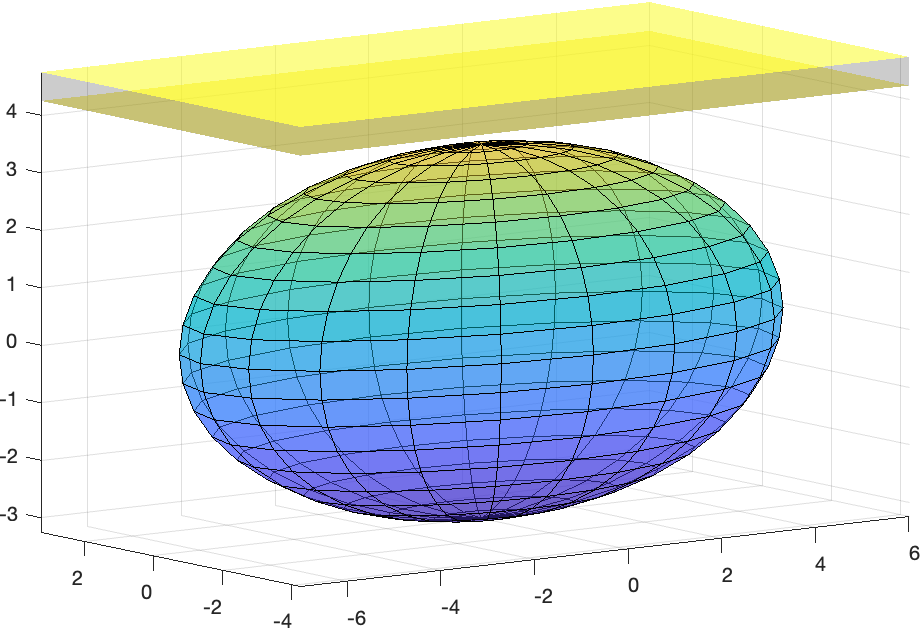}  
  \subcaption{{\tiny (1b): ${y-\cb^T\fb}>\sqrt{\varsigma\fb^TP\fb} +1$}}
\end{subfigure}
\hfill \newline

\hfill
\begin{subfigure}[b]{.32\linewidth}
  \centering
  \includegraphics[width=\linewidth]{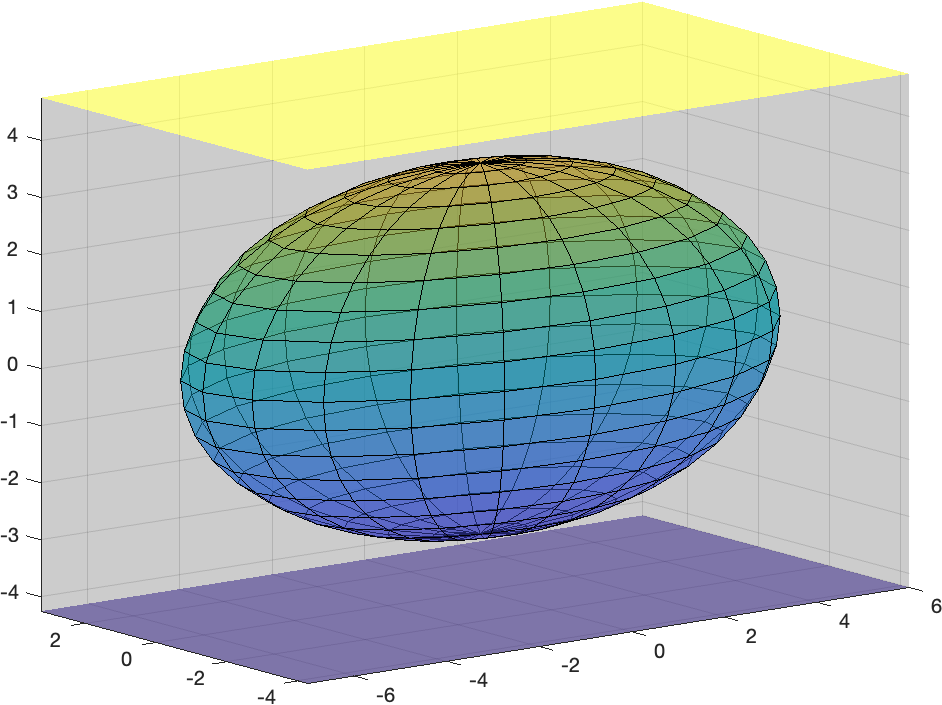}  
  \caption{{\tiny (2): $\abs{y-\cb^T\fb}<1-\sqrt{\varsigma\fb^TP\fb} $}}
\end{subfigure}
\hfill
\begin{subfigure}[b]{.32\linewidth}
  \includegraphics[width=\linewidth]{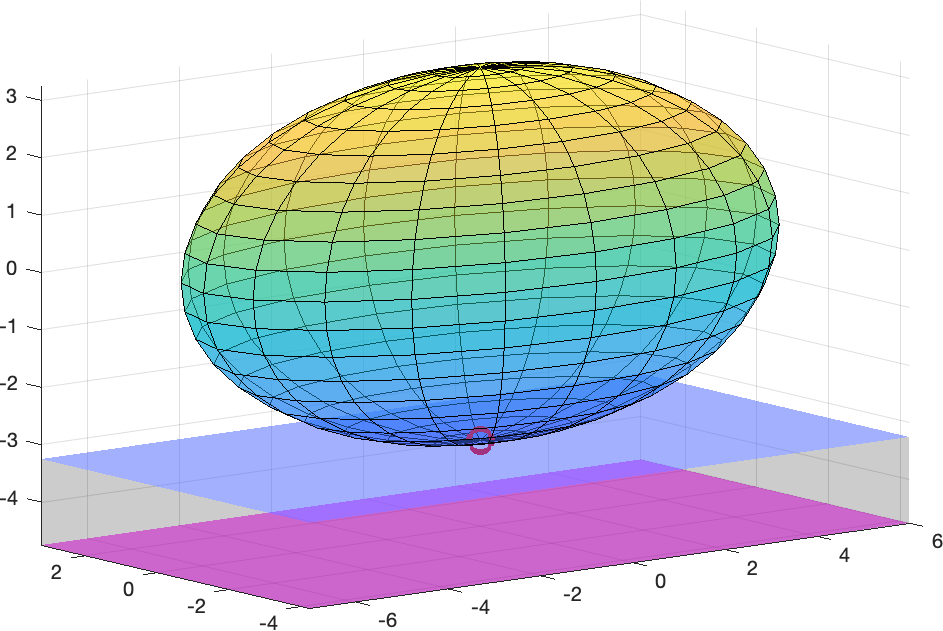}  
  \caption{{\tiny (3a): ${y-\cb^T\fb}=-\sqrt{\varsigma\fb^TP\fb} -1$}}
 \end{subfigure}
 \hfill
\begin{subfigure}[b]{.32\linewidth}
   \includegraphics[width=\linewidth]{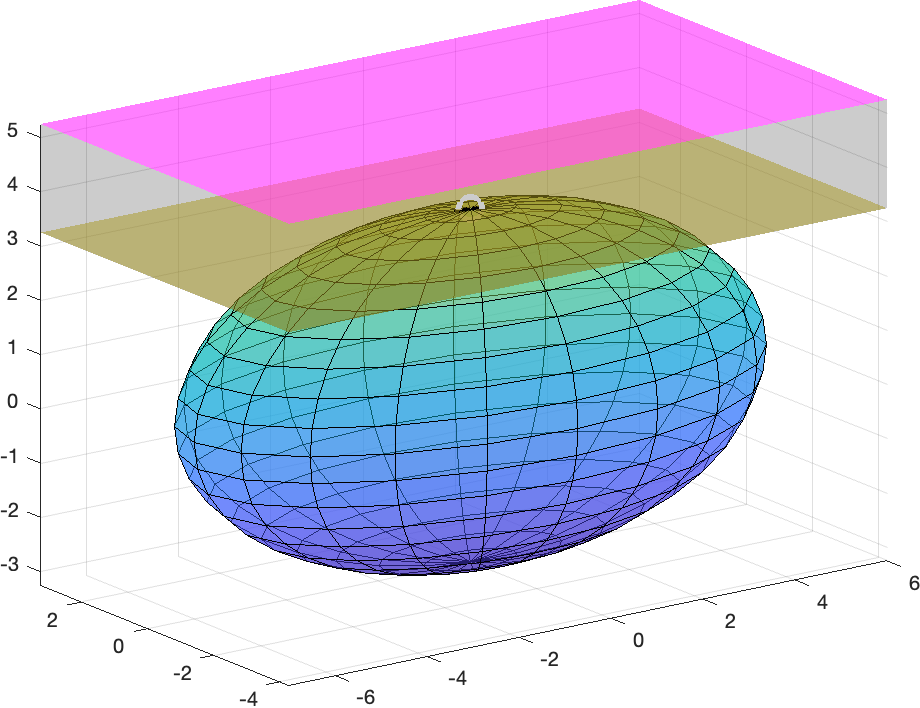}  
  \caption{{\tiny (3b): ${y-\cb^T\fb}=\sqrt{\varsigma\fb^TP\fb} +1$}}
\end{subfigure}

\begin{subfigure}[b]{.325\linewidth}
  \centering
  \includegraphics[width=\linewidth, height=0.14\textheight]{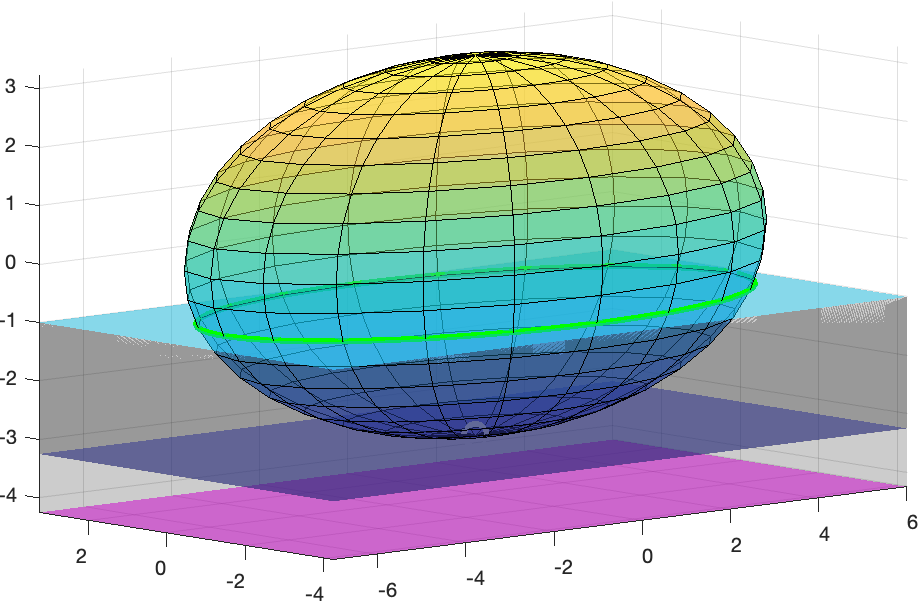} 
  \caption{{\tiny (4a): $\abs{y+1-\cb^T\fb}<\sqrt{\varsigma\fb^TP\fb} $ \\ and\quad$\abs{y-1-\cb^T\fb}>\sqrt{\varsigma\fb^TP\fb} $}}
    \end{subfigure}
\begin{subfigure}[b]{.325\linewidth}
  \includegraphics[width=\linewidth, height=0.14\textheight]{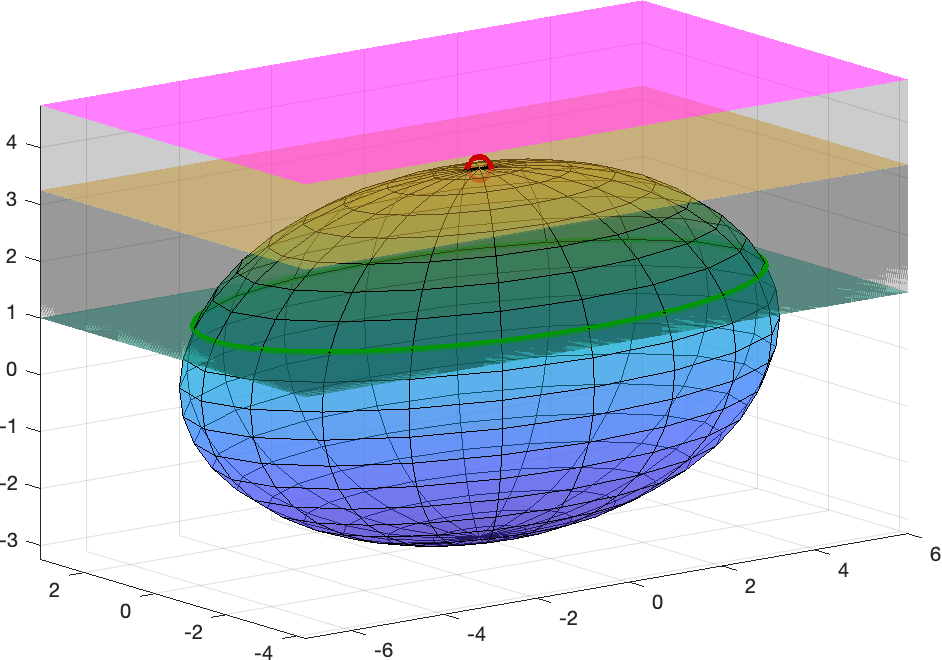} 
  \caption{{\tiny (4b): $\abs{y+1-\cb^T\fb}>\sqrt{\varsigma\fb^TP\fb} $\\ and\quad $\abs{y-1-\cb^T\fb}<\sqrt{\varsigma\fb^TP\fb} $}}
\end{subfigure}
\begin{subfigure}[b]{.325\linewidth}
  \includegraphics[width=\linewidth, height=0.14\textheight]{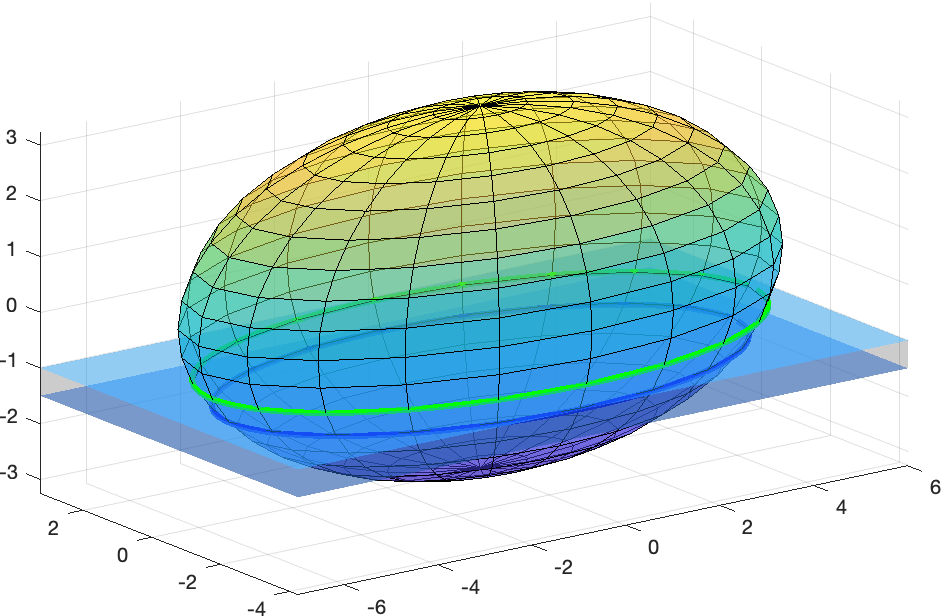} 
  \caption{{\tiny (4c): $\abs{y+1-\cb^T\fb}<\sqrt{\varsigma\fb^TP\fb} $\\ and\quad $\abs{y-1-\cb^T\fb}<\sqrt{\varsigma\fb^TP\fb} $}}
\end{subfigure}
\caption{Illustration of Theorem \ref{theo_ellips_strip_inter}  for $n=3$}
\label{fig_ellips_Str_inter}
\end{figure}

The figure \ref{fig_ellips_Str_inter} illustrates the above theorem in the case where $n=3$. 
The strip is depicted by the shaded region and its boundaries are the colored hyperplanes.
\begin{description}
\item[Case 1:] when { $\abs{y-\cb^T\fb}>\sqrt{\varsigma\fb^TP\fb} +1$}, the strip is either below (\cf fig. \ref{fig_ellips_Str_inter}.1a) or above (\cf fig. \ref{fig_ellips_Str_inter}.1b) with empty intersection. 
\item[Case 2:]  when $\abs{y-\cb^T\fb}<1-\sqrt{\varsigma\fb^TP\fb} $, the ellipsoid is entirely contained in the strip (\cf fig. \ref{fig_ellips_Str_inter}.2) and the intersection is the ellipsoid $\EC(\cb,\varsigma P)$ itself.
\item[Case 3:] when { $\abs{y-\cb^T\fb}=\sqrt{\varsigma\fb^TP\fb} +1$}, the ellipsoid is tangent to the strip and the intersection is reduced to a single point given by \eqref{ellips_str_tang.up} (represented by a red \guil{o} on fig. \ref{fig_ellips_Str_inter}.3a) or \eqref{ellips_str_tang.down} (white \guil{o} on fig. \ref{fig_ellips_Str_inter}.3.b).
\item[Case 4:]  when $\abs{y+1-\cb^T\fb}<\sqrt{\varsigma\fb^TP\fb}$ or $\abs{y-1-\cb^T\fb}<\sqrt{\varsigma\fb^TP\fb} $. This case can be decomposed in three sub-cases (not appearing in the theorem): 
\begin{description}
\item[4.a:] if $\abs{y+1-\cb^T\fb}<\sqrt{\varsigma\fb^TP\fb} $ and $\abs{y-1-\cb^T\fb}>\sqrt{\varsigma\fb^TP\fb} $, meaning that the hyperplane $\HC(\fb,y+1)$ representing the upper boundary (cyan) of the strip (the strip is depicted by all the shaded region in fig. \ref{fig_ellips_Str_inter}.4a) intersects the ellipsoid while the lower one $\HC(\fb,y-1)$ (magenta) does not; the strip is then reduced in width (to the new dark shaded strip) replacing the latter hyperplane by a support hyperplane (violet) of the ellipsoid, $\HC(\fb,\uy)$, where $\uy\eqg \cb^T\fb-\sqrt{\varsigma\fb^TP\fb}$.
\item[4.b:] if $\abs{y+1-\cb^T\fb}>\sqrt{\varsigma\fb^TP\fb} $ and $\abs{y-1-\cb^T\fb}<\sqrt{\varsigma\fb^TP\fb} $, meaning that the hyperplane $\HC(\fb,y-1)$ (green) representing the lower boundary of the strip is intersecting the ellipsoid while the upper one $\HC(\fb,y+1)$ (magenta) is not; the strip is then reduced in width replacing the latter by $\HC(\fb,\by)$ (yellow) where $\by\eqg \cb^T\fb+\sqrt{\varsigma\fb^TP\fb}$.
\item[4.c:] if $\abs{y+1-\cb^T\fb}<\sqrt{\varsigma\fb^TP\fb} $ and $\abs{y-1-\cb^T\fb}<\sqrt{\varsigma\fb^TP\fb}$, meaning that both the hyperplanes $\HC(\fb,y-1)$ and $\HC(\fb,y+1)$ representing \resp. the lower and upper boundaries of the strip are intersecting the ellipsoid.
\end{description}
\begin{figure}[ht]
\captionsetup[subfigure]{labelformat=empty}
\begin{subfigure}[b]{.325\linewidth}
  \centering
  \includegraphics[width=\linewidth, height=0.14\textheight]{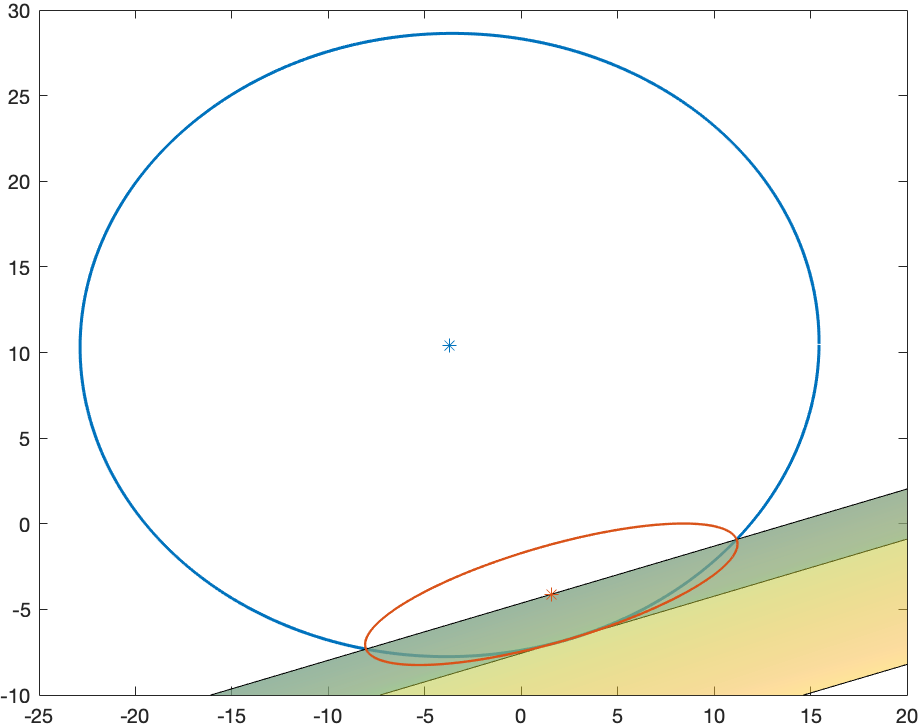} 
  \caption{{\tiny (4a): $\abs{y+1-\cb^T\fb}<\sqrt{\varsigma\fb^TP\fb} $ \\ and\quad$\abs{y-1-\cb^T\fb}>\sqrt{\varsigma\fb^TP\fb} $}}
    \end{subfigure}
\begin{subfigure}[b]{.325\linewidth}
  \includegraphics[width=\linewidth, height=0.14\textheight]{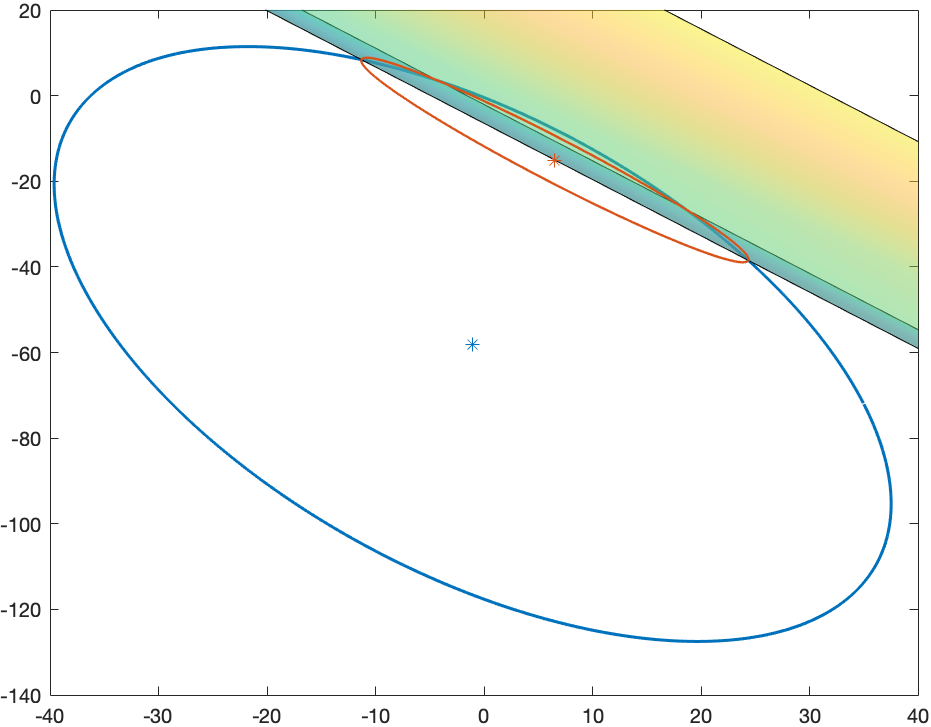} 
  \caption{{\tiny (4b): $\abs{y+1-\cb^T\fb}>\sqrt{\varsigma\fb^TP\fb} $\\ and\quad $\abs{y-1-\cb^T\fb}<\sqrt{\varsigma\fb^TP\fb} $}}
\end{subfigure}
\begin{subfigure}[b]{.325\linewidth}
  \includegraphics[width=\linewidth, height=0.14\textheight]{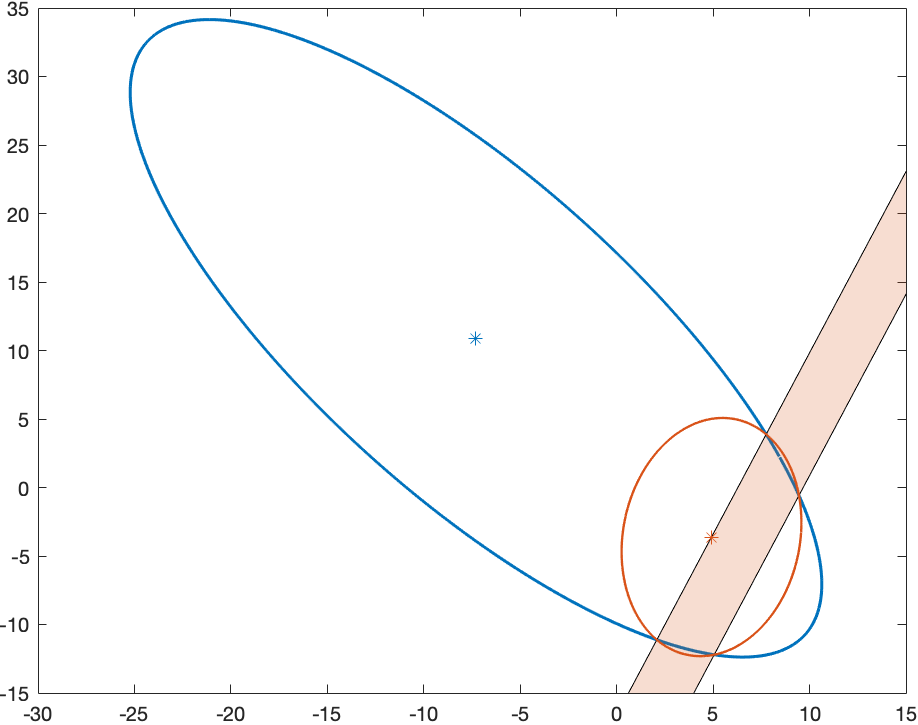} 
  \caption{{\tiny (4c): $\abs{y+1-\cb^T\fb}<\sqrt{\varsigma\fb^TP\fb} $\\ and\quad $\abs{y-1-\cb^T\fb}<\sqrt{\varsigma\fb^TP\fb} $}}
\end{subfigure}
\caption{Illustration of the case 4 Theorem \ref{theo_ellips_strip_inter}  for $n=2$}
\label{fig_ellips_Str_inter2d}
\end{figure}
After the strip is reduced, if necessary (case 4.a and 4.b), the intersection between the new strip and the ellipsoid is overbounded by a family of parametrized, function of $\beta$, ellipsoids. The red ellipsoids depicted in \ref{fig_ellips_Str_inter2d}.4a, \ref{fig_ellips_Str_inter2d}.4b and \ref{fig_ellips_Str_inter2d}.4c are only one possible occurrence of this family of ellipsoids for a particular value of $\beta$ given in \eqref{beta_opt_sig_min_}.
\end{description}

\subsection{ Optimal values for the  parameter $\beta$}\label{sec_beta_opt}
\indent In this section, the optimal value of the weighting parameter $\beta$ with respect to a
judiciously chosen criterion will be derived. 

Three optimization criteria have been used in literature for the parameter identification. The main two involve set measures on the ellipsoid and were first proposed by Fogel and Huang in their well-known pioneer paper  \cite{Fog:82}, in the parameter identification framework. They gave two optimal values for the weighting parameter $\omega\eqg\tfrac{\alpha\beta}{1-\beta}$: the first minimizing the determinant of $\tilde{\varsigma}(\omega)\tilde{P}(\omega)$ (where $\varsigma=1$) in Lemma \ref{lemm_Tan:97} and the second, its trace, 
thus optimizing  the volume and the sum,  \resp., of the squared semi-axes lengths of the ellipsoid $\EC\big(\tfrac{\omega}{\omega+\alpha}\big)$, defined in \eqref{Ellips_Strip_Inter}.  
Five years later, Dasgupta and Huang, in \cite{Das:87}, designed a modified least-squares parameter identification algorithm with a forgetting factor, where they used a weighting parameter $0\leq\nu\leq 1$ (the forgetting factor being $1-\nu$), which could be roughly related to the one used  in \cite{Fog:82} by $\nu=\tfrac{\omega}{\omega+1}$ and they introduced a new optimization cost-function for  $\nu$ based on the minimization of $\tilde{\varsigma}$.
Nayeri, Deller, Liu \ea, actively studied all these aspects during the nineties \cite{Nay:93,Nay:94,Liu:94,Nay:97,Del:00}, proposing a set-membership stochastic approximation identification algorithm and, in \cite{Del:94}, a unified framework of the general class of optimal bounding ellipsoid (OBE) for all the methods previously cited, on the basis of the weighted least squares identification method. They focused on the volume minimization criterion, for evident reasons.

Even if the convergence of such algorithms in the particular case of unconstrained set-membership parameter identification framework\footnote{The parameter identification can be seen, in a reductive manner, as a particular case of the state estimation where the estimated vector is supposed to be constant, \ie, $A_k=I_n$, $B_k=0_{n, l}$ and $R_k=0_{n, m}$ in \eqref{state_eq0}.} was proven for the three addressed criteria, each having its own interesting properties, when it comes to the set-membership state estimation, it is definitely an open issue. 

%
\subsubsection{Minimization of the worst case weighted estimation error}
The optimal value of $\beta$ developed in this paragraph is obtained by minimizing some quadratic measure of the estimation error vector\footnote{represented by its candidate Lyapunov function $\breve{\Vscr}_{\beta}(\x)$.}
in the worst noise case, embodied by ${\varsigma}_{\DC}$, in the manner of \cite{Bec:08,She:18}, inspired by some identification algorithms \cite{Das:87,Tan:97,Sun:01}.
 \begin{theos}\label{theo_beta_sig_min} If the case~4 of Thm \ref{theo_ellips_strip_inter} is met, where ${\EC}_{\DC}(\beta)$  is given by \eqref{Ellips_Strip_Inter}, 
then ${\varsigma}_{\DC}(\beta)$ defined in \eqref{sig_ellips_strip_inter} satisfies
\begin{subequations}\label{beta_opt_sig_min_}
\mathl
\begin{align}
{\varsigma}_{\DC}(\beta)&=\!\max_{\bm{x}\in\DC(\breve{\fb},\breve{y})\cap\EC(\cb,\varsigma P)}\!\!{\Vscr}_{{P}_{\DC}(\beta)}\big(\x-{\cb}_{\DC}(\beta)\big),
\text{ where }{\Vscr}_{{P}}(\xb)\eqg\xb^T{P}^\dag\xb, 
\label{Lyap_def_sig0}
\end{align}
\begin{align}
{\Vscr}_{{P}_{\DC}(\beta)}\big(\x-{\cb}_{\DC}(\beta)\big)-{\Vscr}_{{P}}(\x-\cb)&\leq{\varsigma}_{\DC}(\beta)-\varsigma= \tfrac{\alpha\beta\gamma^2}{1-\beta}-\alpha\beta{\delta^2}.\label{Lyap_sig_inequality}
\end{align}
and its minimum is achieved at
\mathc
\begin{align}\label{beta_opt_sig_min}
\beta_{\varsigma}& \eqg\arg \min_{\beta\in]0,1[} {\varsigma}_{\DC}(\beta)=
  \begin{cases}
  1-\gamma\abs{\delta}^{-1},  &\text{if } \abs{\delta }>\gamma,\\
      0, &\text{otherwise};
      \end{cases}
\end{align}
where $\breve{\fb}$, $\breve{y}$, $\alpha$ and $\delta$ are defined in \eqref{breve_f_y_def}, \eqref{alpha_def} and \eqref{delta_def}. 
\mathl
\begin{align}
\text{Furthermore,  if } \abs{\delta }>\gamma, \text{ then }{\varsigma}_{\DC}(\beta_{\varsigma})<\varsigma
\text{ and }{\varsigma}_{\DC}(\beta_{\varsigma})-\varsigma=-\alpha\beta_{\varsigma}^2{\delta}^2<0,\label{sig_decrease}\\
\Vol\big({\EC}_{\DC}(\beta_{\varsigma})\big)<\Vol\big(\EC(\cb,\varsigma P)\big)
\text{ and }\ssal\big({\EC}_{\DC}(\beta_{\varsigma})\big)<\ssal\big(\EC(\cb,\varsigma P)\big).\label{sig_vol_ssal_decrease}
\end{align}
\end{subequations}
\end{theos}
{\debdemo 
\cf Appendix \ref{Appendix_theo_beta_sig_min}.
\findemo }

The value of $\beta_{\varsigma}$ is resumed in Algorithm \ref{Algo_beta_SigMin}.
{%
\begin{algorithm}[H]
\caption{Optimal value of parameter $\beta$ minimizing $\breve\varsigma(\beta)$}
\begin{algorithmic}[1]\label{Algo_beta_SigMin}
\REQUIRE $\gamma$, $\delta$ \COMMENT{defined in \eqref{Ellips_Strip_Inter}}
\ENSURE $\beta_{\varsigma}$
\IF{$ \abs{\delta }>\gamma$ }
\STATE $\beta_{\varsigma}\eqg 1-\gamma\abs{\delta}^{-1}$; \COMMENT{\cf \eqref{beta_opt_sig_min}};
\ELSE
\STATE $\beta_{\varsigma}\eqg 0$;  \COMMENT{\cf \eqref{beta_opt_sig_min}};
\ENDIF
\end{algorithmic}
\end{algorithm}
}

\begin{rema} 
The representation of the output noise vector's bounding set as an intersection of strips, rather than as an ellipsoid, 
enables this optimization problem to have an analytical solution.
\end{rema}
\begin{minipage}[c]{0.45\linewidth}
As for fig. \ref{fig_ellips_Str_inter} (case 4), the figure \ref{ell_strip_inters_fig} on the right shows the (red/small) ellipsoid ${\EC}_{\DC}(\beta_{\varsigma})$ containing the intersection of the (blue/big) ellipsoid $\EC(\cb ,\varsigma P)$ with the (colored) strip $\DC(\fb,y)$, for $\beta_{\varsigma}$ given by \eqref{beta_opt_sig_min} and $n=2$. It is worth noting that with this value of $\beta_{\varsigma}$, the center ${\cb}_{\DC}(\beta_{\varsigma})$, of ${\EC}_{\DC}(\beta_{\varsigma})$ is the 
projection of $\cb$ (the center of  $\EC(\cb ,\varsigma P)$), on the hyperplan representing the nearest strip boundary, in the direction $P\fb$. 
\end{minipage}
\hfill
\begin{minipage}[c]{0.5\linewidth}
\begin{center}
\includegraphics[width=\linewidth]{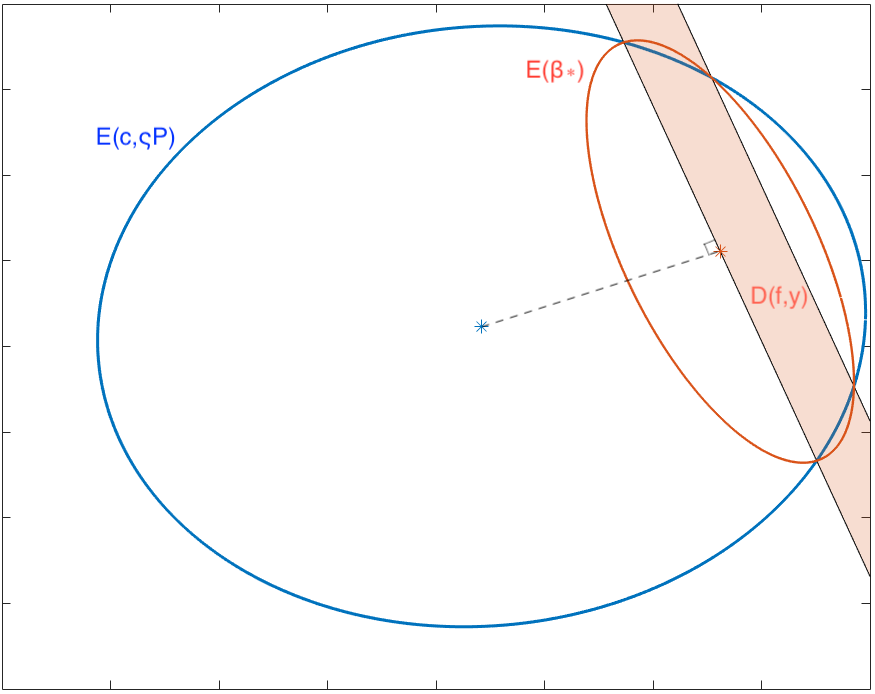}
\captionof{figure}{${\EC}_{\DC}(\beta_{\varsigma})\supset \EC(\cb ,\varsigma P)\cap \DC(\fb,y)$} 
\label{ell_strip_inters_fig}
\end{center}
\end{minipage}
\subsubsection{Minimization of the ellipsoid's volume}
As already mentioned in Section~\ref{sec_time_update}, the ellipsoid can possibility have zero axes lengths and, therefore, a zero volume and a noninvertible shape matrix. To our knowledge, this issue was not addressed in the set-membership estimation algorithms of the literature. 

The optimal value for the weighting parameter $\beta$ 
intervening in \eqref{Ellips_Strip_Inter}, that minimizes the pseudo-volume (\cf Definition \ref{Def_pseudo-volume}) of the ellipsoid ${\EC}_{\DC}(\beta)$, when its usual volume can be zero, is derived in what follows.
%
\begin{theos}\label{theo_beta_vol_min}
\begin{subequations}\label{eq_thm_beta_vol_min}
If the ellipsoid ${\EC}_{\DC}(\beta)$ given by \eqref{Ellips_Strip_Inter} of Thm  \ref{theo_ellips_strip_inter}, for which  ${q}\eqg \rank(P)>1$, then ${\EC}_{\DC}(\beta_{\text{v}})$ has the minimum (pseudo-)volume, where
$$\beta_{\text{v}}\eqg\arg \min_{\beta\in]0,1[}\Vol\big({\EC}_{\DC}(\beta)\big)$$ 
 is the unique solution of the quadratic equation $ a_2\beta^2+a_1\beta+a_0=0$, where
\mathl
\mathc
\begin{align}
a_0&\eqg{q}\alpha(\gamma^2-\delta^2)-\varsigma,\label{beta_vol_c}\\
a_1&\eqg(2{q}+1)\alpha\delta^2+\varsigma-\gamma^2\alpha,  \label{beta_vol_b}\\
a_2&\eqg-({q}+1)\alpha\delta^2, \label{beta_vol_a}
\end{align}
if $a_0<0$; otherwise $\beta_{\text{v}}= 0$.
\mathl
\begin{align}
\text{Furthermore, if $a_0<0$, 
then }\Vol\big({\EC}_{\DC}(\beta_{\text{v}})\big)<\Vol\big(\EC(\cb,\varsigma P)\big).\label{vol_decrease}
\end{align}
\end{subequations}
\end{theos}
{\debdemo 
\cf Appendix \ref{Appendix_theo_beta_vol_min}.
\findemo
}
{%
\begin{algorithm}[H]
\caption{Optimal value of parameter $\beta$ minimizing the volume of $\breve\EC(\beta)$}
\begin{algorithmic}[1]\label{Algo_beta_VolMin}
\REQUIRE $q$, $\alpha$, $\gamma$, $\delta$, $\varsigma$ \COMMENT{defined in \eqref{Ellips_Strip_Inter}}
\ENSURE $\beta_{\textit{v}}$
\STATE $a_0\eqg {q}\alpha(\gamma^2-\delta^2)-\varsigma$; \COMMENT{\cf \eqref{beta_vol_c}};
\IF{$a_0\geq0$ 
}
\STATE $\beta_{\textit{v}}\eqg 0$;  
\ELSE
\STATE $a_1\eqg (2{q}+1)\alpha\delta^2+\varsigma-\gamma^2\alpha$;  \COMMENT{\cf \eqref{beta_vol_b}};
\STATE $a_2\eqg-({q}+1)\alpha\delta^2$; \COMMENT{\cf \eqref{beta_vol_a}};
\STATE $\beta_{\textit{v}}\eqg \dfrac{-a_1+\sqrt{a_1^2-4a_0a_2}}{2a_2}$; 
\ENDIF
\end{algorithmic}
\end{algorithm}
}

\subsubsection{Minimization of the sum of the squared axes' lengths}
\begin{theos}\label{theo_beta_tr_min}
If the ellipsoid ${\EC}_{\DC}(\beta)$ is given by \eqref{Ellips_Strip_Inter} of Thm  \ref{theo_ellips_strip_inter}, then ${\EC}_{\DC}(\beta_{\text{s}})$ has the minimum SSAL, where
\begin{subequations}\label{eq-beta-opt-tr}
$$\beta_{\text{s}} \eqg\arg \min_{\beta\in]0,1[}\tr\big({\varsigma}_{\DC}(\beta){P}_{\DC}(\beta)\big)$$
is the unique positive solution to the cubic equation
\mathc
\begin{align}
&b_3\beta^3+b_2\beta^2+b_1\beta+b_0\eqg0,\label{eq_cubic}\\
\intertext{if $b_0<0$; $\beta_{\text{s}}=0$ otherwise; where 
}
b_0&\eqg\nu\varsigma-\tau(\delta^2-\gamma^2), \ \tau\eqg\tr(P)\text{ and }\nu\eqg\phib^T\phib=\fb^TP^2\fb,
\label{beta_tr_b0}\\
b_1&\eqg2\big(\tau\delta^2-\nu\varsigma+\alpha\nu(\delta^2-\gamma^2)\big),  \label{beta_tr_b1}\\
b_2&\eqg\nu\varsigma-\tau\delta^2+\alpha\nu(\gamma^2-4\delta^2), \label{beta_tr_b2}\\
b_3&\eqg2\alpha\nu\delta^2. \quad\label{beta_tr_b3}
\end{align}
\mathl
\begin{align}
\text{Furthermore, if $b_0<0$, then }
\ssal\big({\EC}_{\DC}(\beta_{\text{s}})\big)<\ssal\big(\EC(\cb,\varsigma P)\big).\label{ssal_decrease}
\end{align}
\end{subequations}\end{theos}
{\debdemo 
\cf Appendix \ref{Appendix_theo_beta_tr_min}.
\findemo
}

The computation of $\beta_{\text{s}}$ is elaborated in Algorithm \ref{Algo_beta_TrMin}. \vspace{-4pt}
{%
\begin{algorithm}[H]
\caption{Optimal value of the parameter $\beta$ minimizing the SSAL of $\breve\EC(\beta)$}
\begin{algorithmic}[1]\label{Algo_beta_TrMin}
\REQUIRE $P$, $\alpha$, $\gamma$, $\delta$, $\varsigma$, $\phib$ \COMMENT{defined in \eqref{Ellips_Strip_Inter}}
\ENSURE $\beta_{\text{s}}$
\STATE $\tau\eqg\tr(P)$; $\nu\eqg\phib^T\phib$;
\STATE $b_0\eqg\nu\varsigma-\tau(\delta^2-\gamma^2)$;
\IF{$b_0>=0$,}
\STATE $\beta_{\text{s}}\eqg 0$;
\ELSE
\STATE $b_1\eqg 2\big(\tau\delta^2-\nu\varsigma+\alpha\nu(\delta^2-\gamma^2)\big)$;\\
 $b_2\eqg\nu\varsigma-\tau\delta^2+\alpha\nu(\gamma^2-4\delta^2)$;
$b_3\eqg2\alpha\nu\delta^2$;
\STATE ${s} \eqg \tfrac{b_1}{b_3} - \tfrac{b_2^2}{3b_3^2}$; ${t} \eqg \tfrac{b_0}{b_3} - \tfrac{b_1b_2}{3b_3^2} + \tfrac{2b_2^3}{27b_3^3}$;
\STATE ${u} \eqg \left(\tfrac{{s}}{3}\right)^3 + \big(\tfrac{{t}}{2}\big)^2$; ${v}\eqg \sqrt[3]{-\tfrac{{t}}{2} + \sqrt{{u}}}$; ${w} \eqg \sqrt[3]{-\tfrac{{t}}{2} - \sqrt{{u}}}$; 
\STATE $\omega \eqg -\tfrac{1}{2} + \tfrac{1}{2} \sqrt{3}i$; $\omega_s  \eqg -\tfrac{1}{2} - \tfrac{1}{2} \sqrt{3}i$; \COMMENT{ $\omega$ is the cubic root of 1 and $\omega_s \eqg\omega^2$}
\STATE $\tilde{\beta}_1 \eqg {v} + {w}$; 
$\tilde{\beta}_2 \eqg \omega {v} + \omega_s {w}$; 
$\tilde{\beta}_3 \eqg \omega_s {v} + \omega {w}$;
\STATE $\beta_1 \eqg \tilde{\beta}_1 - \tfrac{b_2}{3b_3}$;
$\beta_2 \eqg \tilde{\beta}_2 - \tfrac{b_2}{3b_3}$;
$\beta_3 \eqg \tilde{\beta}_3 - \tfrac{b_2}{3b_3}$;
\STATE $\beta_{\text{s}}\eqg\beta_i$, $i\in\{1,2,3\}$, \tq $\beta_i\in\R^*_+$.
\ENDIF
\end{algorithmic}
\end{algorithm}
}
\subsection{ Intersection of an ellipsoid with a hyperplane}\label{subsec_hyp}  
In this paragraph, the equality-type constraint \eqref{bound_HyperP0} is examined. This constraint on the state vector can be also viewed as a noiseless measurement, \aka pseudo-measurement
and results in the intersection of the state bounding ellipsoid with the hyperplane representing the measurement. This intersection 
leads to a degenerate ellipsoid, whose shape matrix loses one rank with each intersecting (not parallel and not containing) hyperplane. 

The theorem below gives the expression of thusly obtained ellipsoid. 
%
\stepcounter{lemm}
\begin{theo}[ellips./hyperplane inters.]\label{theo_ell_hyp_inters}
Let  $\cb\in~\Rn$, $P\in\Rnn$ SPSD, $\varsigma\in\R_+^*$, $\fb\in\Rn\text{--}\{\zero_n\}$ and $y\in\R$,
\begin{subequations}\label{ell_hyp_inters_eq}
\mathc
\begin{align}
\intertext{{if $y>\brho\vee y<-\urho$,}\hfill\normalfont{(case 1)}}
\EC(\cb ,\varsigma P)\cap\HC(\fb,y)&=\emptyset;\label{ell_hyp_empty}\\
\intertext{{else if $y=\brho=-\urho$},\hfill\normalfont{(case 2)}}
\EC(\cb ,\varsigma P)\cap\HC(\fb,y)&=\EC(\cb ,\varsigma P);\label{ell_hyp_ell}\\
\intertext{else if $y=-\urho$,\hfill{\normalfont{(case 3a)}}}
\EC(\cb,\varsigma P)\cap\HC(\fb,y)
&=\{\cb-\varsigma^{\frac{1}{2}}(\fb^TP\fb)^{-\frac{1}{2}} P\fb\};\hspace{-5mm}\label{cas3a_HP}\\
\intertext{else if $y=\brho$,\hfill{\normalfont{(case 3b)}}}
\EC(\cb,\varsigma P)\cap\HC(\fb,y)
&=\{\cb+\varsigma^{\frac{1}{2}}(\fb^TP\fb)^{-\frac{1}{2}} P\fb\};\hspace{-5mm}\label{cas3b_HP}\\
\intertext{{otherwise  $(\text{if } -\urho< y<\brho)$}\hfill\normalfont{(case 4)}}
\EC(\cb ,\varsigma P)\cap\HC(\fb,y)&={\EC}_{\HC}\eqg\EC({\cb}_{\HC},{\varsigma}_{\HC}{P}_{\HC}),\label{ell_hyp}\\
\intertext{where }{\cb}_{\HC}&\eqg\cb +\alpha\delta{P\fb},\label{eq_x}\\
{P}_{\HC}&\eqg P-{\alpha}{P\fb\fb^TP},\label{eq_P}\\ 
{\varsigma}_{\HC}&\eqg\varsigma-{\alpha}{\delta^2},\label{eq_sig}\\
\delta&\eqg y-\fb^T\cb \label{eq_delta}
\end{align}
and where $\alpha$, $\urho$  and $\brho$ are defined in \eqref{alpha_def}, \eqref{urho_def} and \eqref{brho_def} \resp.
\end{subequations}
\end{theo}
\debdemo \cf Appendix \ref{Appendix_theo_ell_hyp_inters}.
\findemo

\begin{figure}[ht]
\captionsetup[subfigure]{labelformat=empty}
\begin{subfigure}[b]{.32\textwidth}
  \centering
  \includegraphics[width=\linewidth]{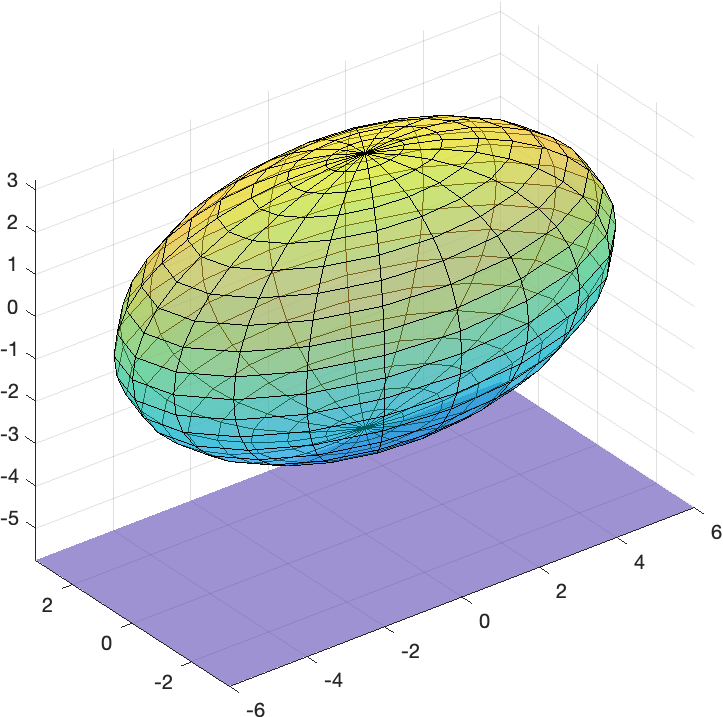}  
  \caption{{\tiny (1a): $y-\cb^T\fb>\sqrt{\varsigma\fb^TP\fb}$ }}
\end{subfigure}
\hfill
\begin{subfigure}[b]{.32\textwidth}
  \centering
  \includegraphics[width=\linewidth]{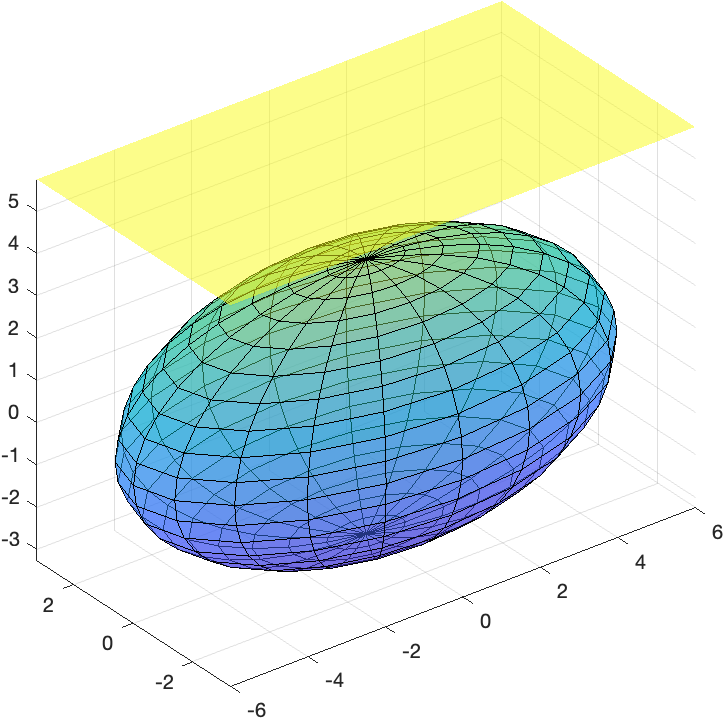}  
  \caption{{\tiny (1b): $y-\cb^T\fb<-\sqrt{\varsigma\fb^TP\fb}$} }
\end{subfigure}
\hfill
\begin{subfigure}[b]{.32\textwidth}
  \centering
  \includegraphics[width=\linewidth]{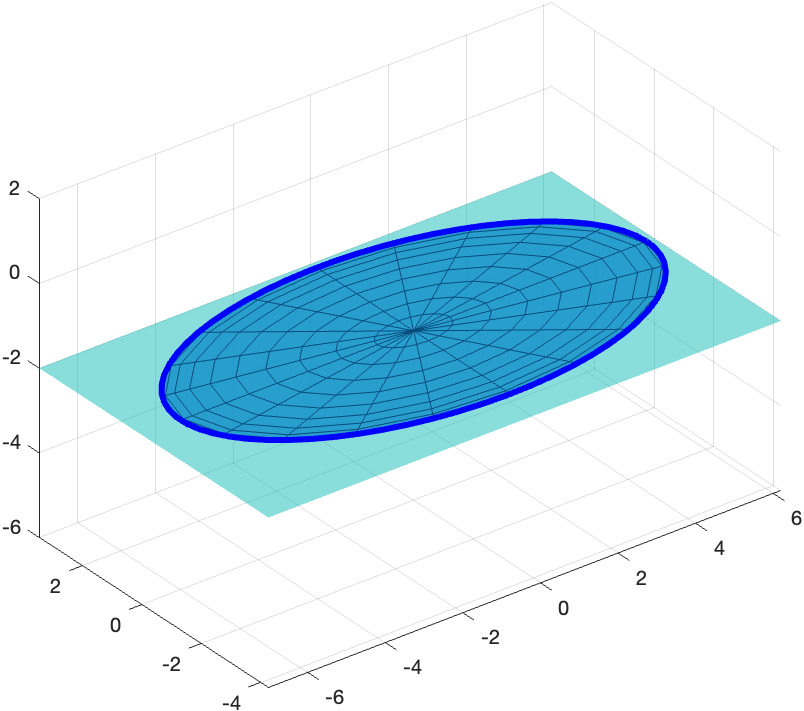}  
  \caption{{\tiny (2): \hspace{-2pt}${y-\cb^T\fb}\!=\!{\fb^TP\fb}\!=\!-{\fb^TP\fb}=0$}} 
\end{subfigure}

\begin{subfigure}[b]{.32\textwidth}
  \centering
  \includegraphics[width=\linewidth]{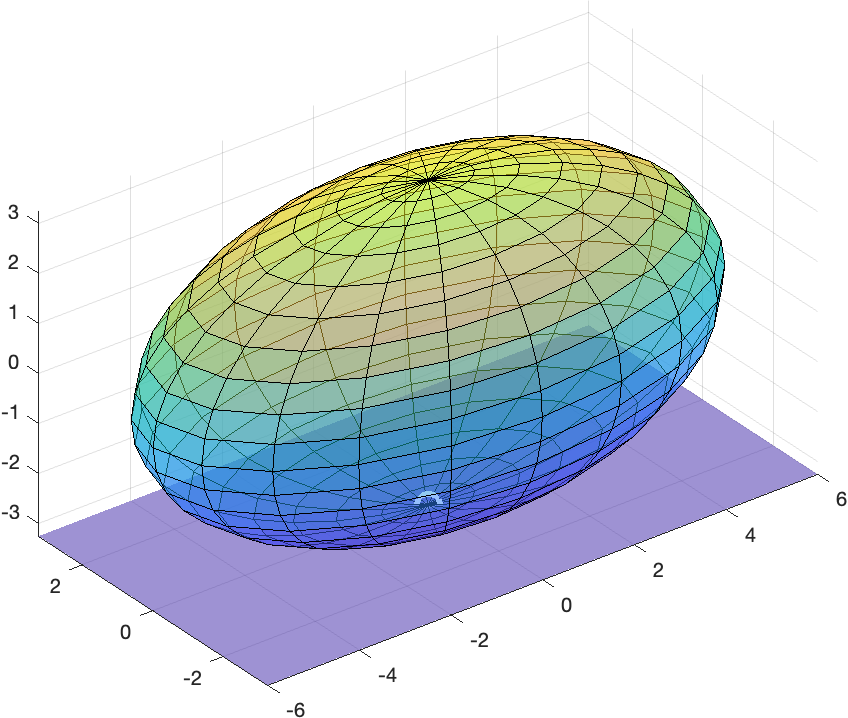}  
  \caption{{\tiny (3a): $y-\cb^T\fb=\sqrt{\varsigma\fb^TP\fb}$}} 
\end{subfigure}
\hfill
\begin{subfigure}[b]{.32\textwidth}
  \centering
  \includegraphics[width=\linewidth]{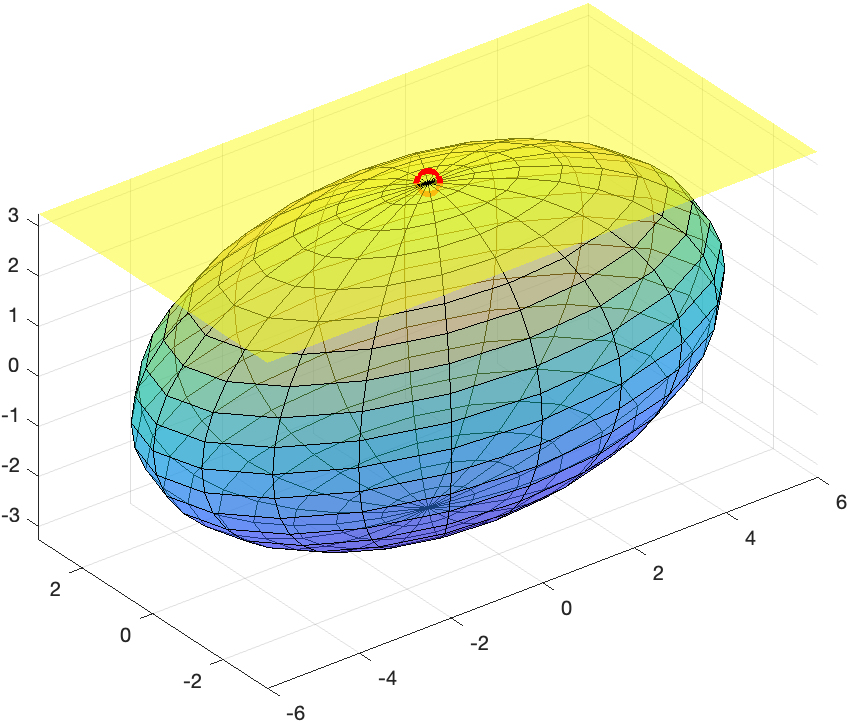}  
  \caption{{\tiny (3b): $y-\cb^T\fb=-\sqrt{\varsigma\fb^TP\fb}$}} 
\end{subfigure}
\hfill
\begin{subfigure}[b]{.32\textwidth}
  \centering
  \includegraphics[width=\linewidth]{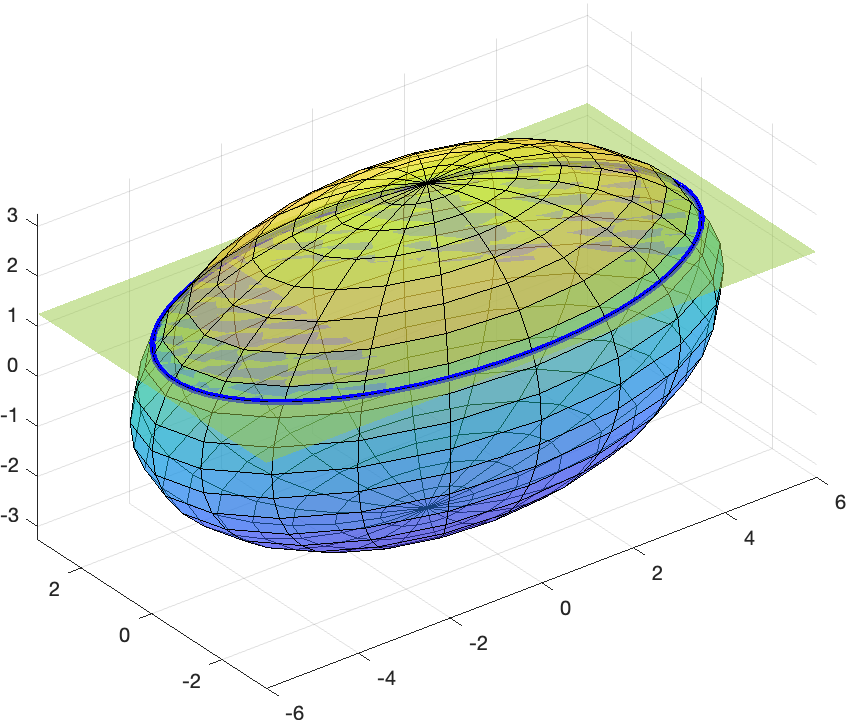} 
  \caption{{\tiny (4): 
  $\abs{y-\cb^T\fb}<\sqrt{\varsigma\fb^TP\fb}$}} 
 \label{ell_hs_inters_fig}
\end{subfigure}
\caption{Illustration of the four cases of Theorem \ref{theo_ell_hyp_inters}  for $n=3$}
\label{fig_ellips_hyp_inter}
\end{figure}

The figure \ref{fig_ellips_hyp_inter} is self-explanatory. It is plain to see that the intersection of a three-dimensional ellipsoid with an intersecting non parallel and non containing hyperplane is a degenerate ellipsoid (\cf the blue ellipse in fig. 4) and the good news is that there is no need to circumscribe it by an other one, as it is done in case of intersection with strip or halfspace.
Please note that the case 2 happens only when the ellipsoid $\EC(\cb,\varsigma P)$ is already degenerate and contained in the hyperplane $\HC(\fb,y)$, meaning that $y=\cb^T\fb$ and that $\fb\in\nul(P)$.

\subsection{The output update algorithm}\label{subsec_algo_meas_up}
Hereafter, the measurement update part of the state estimation algorithm is summarized.  All the variables are depending on the time step $k$ even if the subscript $k$ was skipped on some of them. 
\setcounter{topnumber}{0}
\begin{theo}[Correction stage]\label{theo_meas_update} 
\begin{subequations}\label{ellips_correc}
\mathl
{If $\xka\in\EC\kk\eqg\EC(\xkk,\sigkkk\Pkkk)$  satisfying \eqref{bounds}, 
 then }
\begin{align}
\forall\beta^{*}\in]0,1[,\ \xka\in\SC_k&\subseteq\EC_k\eqg\EC(\xk ,\sigk P_k),\text{  (\cf \eqref{SC_set}), where }\label{eq_thm_sig_min}
\end{align}
\mathl
\begin{align}
&\EC_k\eqg\EC\kk, &&\forall k\notin\KC \text{ (\ie\ $p_k=0$) and }\\
&\xk\eqg z_{p},\ \Pk\eqg\PPki_{p},\ \sigk\eqg\sigma_{p}, \  {q}_k\eqg \qqkj_{p},&&\forall k\in\KC \text{ and for $i\in\{1,\ldots,{p}\}$:}\label{eq_correc-x-P-sig}
\end{align}
\begin{align}
\PPki_{i}&\eqg\PPki_{i-1}-{\alphak{i}\betak{i}}{\phibki \phibki ^T},\label{eq_correc-P}\\ 
\xki&\eqg\xkiii+{\alphak{i}\betak{i}}\delki \phibki,\label{eq_correc-x}\\ 
\sigki&\eqg 
\begin{cases}
\sigkiii-{\alphak{i}\betak{i}^2\delkis},&\hspace{-48pt}\text{if }\big(\uyki=\byki \wedge -\urhoki\neq\brhoki\big)\vee \beta^{*}=\beta_{\varsigma},\\
\sigkiii+{\alphak{i}\betak{i}\bigg(\tfrac{\gammaki^2}{1-\betak{i}}-\delkis\bigg)},& \text{otherwise;}
\end{cases}\label{eq_correc-sig}\\
\PPki_{0}&\eqg\Pkk, \xkz\eqg\xkk\text{ and }\sigkz\eqg\sigkkk, \label{eq_correc0}\\
\intertext{
with $\qqkj_{0} \eqg q\kk$ computed at the prediction stage, in \eqref{eq_q-mu-vol-pred} (line \ref{Algo_Predic_VolMin_q} of Algorithm \ref{Algo_Predic_VolMin}) and ${q}_k\eqg\rank(P_k)\eqg \qqkj_{m}$;}  
\alphak{i}&\eqg 
\begin{cases}
\thetaki^{-1} , &\text{if }  \lamki  \neq 0,\\ 
0, &\text{otherwise};\end{cases}\label{eq_correc-alph}\\
\betak{i}&\eqg \begin{cases}
1,&\text{if }\uyki=\byki\wedge-\urhoki\neq\brhoki, (i\in\hscrk),\\ 
\beta^{*}, &\text{else if }-\urhoki<\uyki\vee\byki<\brhoki, (i\in\dscrk\cup\gscrk)\\ 
 0,&\text{otherwise;}  \end{cases}\label{eq_correc-bet}\\
  %
   \delki&\eqg \gammak{i}\yki-\tfrac{1}{2}(\brhoki-\urhoki), 
   \quad \yki\eqg \tfrac{1}{2\gammak{i}}({\byki+\uyki}),\label{eq_correc-del}\\
   \gammaki&\eqg  \tfrac{1}{2}(\byki-\uyki),\label{eq_correc-gam}\\
   \thetaki&\eqg \fbki^T\phibki, \label{eq_correc-the}\\
\phibki &\eqg {\PPki_{i-1}}\fbki,\label{eq_correc-phi}\\
\brhoki&\eqg \lamki+\fbki^T\xkiii 
\text{ and }
\urhoki \eqg
2\lamki-\brhoki,\label{eq_correc-brho-urho}\\
 \lamki&\eqg(\sigkiii\thetaki)^{\frac{1}{2}};\label{eq_correc-lam}\\
 \byki&\eqg \min(\bar{y}\ki,\brhoki)\text{ and }
\uyki\eqg \max(\uyki,-\urhoki),\ i\in\{1,\ldots,{p}\}.\label{by_uy_redef}
\intertext{and, if $\EC\kk$ is computed according to Thms \ref{theo_time_update} and \ref{theo_mu-vol_k} by \eqref{predic_eq}-\eqref{mu-vol_k}, then}
\qqkj_{i}& \big(\eqg\rank(\PPki_{i})\big) \eqg
\begin{cases}
\qqkj_{i-1}-1,&\text{if }\uyki=\byki \wedge -\urhoki\neq\brhoki \wedge \alphak{i}\neq 0,\hspace{-6pt}\\
\qqkj_{i-1},& \text{otherwise.}
\end{cases} \label{eq_correc-q}
\end{align}
\end{subequations}
Furthermore, if $\beta^{*}=\beta_{\varsigma}$, 
then $\d\max_{\xb\in\SC_k}{\Vscr}_{P_k}(\xb-\xk)$ (\cf \eqref{Lyap_def_sig0}) is minimized;\\
if $\beta^{*}=\beta_{\text{v}}$, 
then $\EC_k$ has a minimum pseudo-volume;\\
and if $\beta^{*}=\beta_{\text{s}}$, 
then $\EC_k$ has a minimum SSAL; 
where $\beta_{\varsigma}$,  $\beta_{\text{v}}$ and $\beta_{\text{s}}$, are given \resp.\ by Thm \ref{theo_beta_tr_min} - Algo. \ref{Algo_beta_TrMin}, Thm \ref{theo_beta_vol_min} - Algo. \ref{Algo_beta_VolMin} and
\eqref{beta_opt_sig_min} - Algo. \ref{Algo_beta_SigMin}; where $\alpha\eqg\alphak{i}$, $\gamma\eqg\gammaki$, $\delta\eqg\delki$, $\varsigma\eqg\sigkiii$, $\phib\eqg\phibki$, $P\eqg\PPki_{i-1}$ and ${q}\eqg \qqkj_{i-1}$. 
\end{theo}
{\debdemo 
\cf Appendix \ref{Appendix_theo_meas_update}. 
\findemo 
}
\renewcommand{\byki}{\bar{y}_{k_i}}
\renewcommand{\uyki}{\underb{y}_{k_i}}

The correction stage described in Thm \ref{theo_meas_update} is resumed in Algorithm \ref{Algo_Correc} and $\EC_k$ is computed from $\EC\kk$.
\begin{rema}
The strip reduction at eq. \eqref{by_uy_redef} (line \ref{algo_by_uy_redef} of Algo. \ref{Algo_Correc}) significantly lowers the resulting ellipsoid's size, when one of the two hyperplanes bounding the strip $\SC\ki$ is outside the ellipsoid $\EC\kii\eqg \EC(\xkiii,\sigkiii \PPki_{i-1})$, according to the idea of \cite{Bel:90}.

Note also that the updating of the rank of the shape matrix $\PPki_{i}$, by \eqref{eq_correc-q},  is done only when the volume minimization criterion is chosen at the prediction step. 
\end{rema}

 Let \texttt{Predic\_VolMin}  be a boolean variable set to \guil{\texttt{True}} when choosing the pseudo-volume minimization, $\Vol(\EC\kki)$, at the prediction stage and to \guil{\texttt{False}} when minimizing the SSAL of $\EC\kki$. 
 And let \texttt{Correc\_$\beta$}~$\in\{0,1,2\}$
 set to 0, 1 or 2 when choosing the minimization of $\Vol(\EC\ki)$, the SSAL of $\EC\ki$ or $\sigki$, \resp., during the correction stage. 

\newpage

{%
\begin{algorithm}[H]
\caption{Computation of the corrected ellipsoid}
\begin{algorithmic}[1]\label{Algo_Correc}
\REQUIRE $\x$, $ P$, $\varsigma$, ${q}$ (if \texttt{Predic\_VolMin}),
$\fb_{1},\ldots,\fb_{p}$, $\byb$, $\underb\yb$, 
\texttt{Predic\_VolMin},  \texttt{Correc\_$\beta$} ,  $p$  
\ENSURE $\x$,  $P$, $\varsigma$, $q$ (if \texttt{Predic\_VolMin})
\IF{$p\neq 0$}
\FOR{$i=1,\ldots p$}
\STATE  $\phib \eqg P\fb_i$; $\theta\eqg \fb_i^T\phib$;  \COMMENT{\cf  \eqref{eq_correc-phi}}; 
\IF{$\theta\neq 0$}
\STATE  $\alpha\eqg \theta^{-1}$;  $\eta\eqg(\varsigma\theta)^{\frac{1}{2}}$;  \COMMENT{\cf \eqref{eq_correc-alph}, \eqref{eq_correc-phi}}; 
\STATE $\brho\eqg \eta+\fb_i^T\x $; $\urho \eqg 2\eta-\brho$;
 \COMMENT{\cf \eqref{eq_correc-brho-urho}}; 
\STATE\label{algo_by_uy_redef} $\by\eqg \min(\by_i,\brho)$;
$\uy\eqg \max(\underb{y}_i,-\urho)$;  \COMMENT{\cf \eqref{by_uy_redef}};
\STATE  $\delta\eqg   \tfrac{1}{2}({\by+\uy}-\brho+\urho)$;  \COMMENT{\cf \eqref{eq_correc-del}};
\STATE $\gamma\eqg  \frac{1}{2}(\by-\uy)$;  \COMMENT{\cf \eqref{eq_correc-gam}}; 
\IF{$\uy=\by\text{ and }-\urho\neq\brho$}
\STATE $\beta = 1$;
	\IF{\texttt{Predic\_VolMin} $=$ \texttt{True}}
	\STATE{$q\gets q-1$}
	\ENDIF
\ELSIF{$-\urho<\uy$ or $\by<\brho$}
	\IF{ \texttt{Correc\_$\beta$}  $=0$}
		\STATE \texttt{Algorithm}  \ref{Algo_beta_VolMin} ({\bf Input:} $q$, $\alpha$, $\gamma$, $\delta$, $\varsigma$;
		{\bf Output:} $\beta_{\text{v}}$)
		\STATE $\beta \eqg \beta_{\text{v}}$;
	\ELSIF{ \texttt{Correc\_$\beta$}  $=1$}
		\STATE \texttt{Algorithm}  \ref{Algo_beta_TrMin} ({\bf Input:} $P$, $\alpha$, $\gamma$, $\delta$, $\varsigma$, $\phib$; {\bf Output:} $\beta_{\text{s}}$)
		\STATE $\beta \eqg \beta_{\text{s}}$;
	\ELSE 
		\STATE \texttt{Algorithm}  \ref{Algo_beta_SigMin} ({\bf Input:} $\delta$, $\gamma$; {\bf Output:} $\beta_{\varsigma}$ )
		\STATE $\beta \eqg \beta_{\varsigma}$;
	\ENDIF
\ELSE 
\STATE $\beta = 0$;
\ENDIF
\STATE  $P\leftarrow P-{\alpha\beta}{\phib \phib^T}$;
 $\x\leftarrow \x+{\alpha\beta}\delta\phib $;
 $\varsigma\leftarrow \varsigma-{\alpha\beta^2\delta^2}$; \COMMENT{\cf \eqref{eq_correc-P}--\eqref{eq_correc-sig}};   
\ENDIF
\ENDFOR
\ENDIF
\end{algorithmic}
\end{algorithm}
}
\setcounter{topnumber}{2}

\begin{rema}
This algorithm is of low computational complexity~: $O(n^2)$. 
Indeed, all the operations are simple sums and products: they were optimized in this regard and are thence suitable for systems with a high dimensional state vector (big $n$), potentially many measurements (big ${p_k}$) and potentially many unknown inputs (big $m_k$). The intermediate variables $\alphaki$, $\thetaki$, $\lamki$, $\phibki$ were added on to perform redundant vector and matrix operations only once. Thereby noticing that $\fbki^T\xkiii=\tfrac{1}{2}(\brhoki-\urhoki)$ allows to determine $\delki\eqg\tfrac{1}{2}({\byki+\uyki})-\fbki^T\xkiii$ and $\urhoki\eqg\lamki-\fbki^T\xkiii$ using addition of scalars, in \eqref{eq_correc-del} and \eqref{eq_correc-brho-urho} \resp., rather than multiplication of possibly high dimensional vectors.
\end{rema}
 \begin{rema}
 In the case where $\hscrk\neq\emptyset$, the matrix $\PPki_{i}$ 
 loses rank  with each intersecting hyperplane $\HC\ki$, $i\in\hscrk$, thusly entailing the progressive flattening of the ellipsoid $\EC_{k_i}$. Depending on the rank of the matrix $R_k$ (of which no assumption is made), the rank of $P_{k+1/k}$ can be recovered at the time-update phase.
The value of $\rank(\PPki_{i})$ is needed at each $i=1\ldots,m_k$ for each time step $k$, whenever the volume of the ellipsoid $\EC_k$ is minimized. Therefore, keeping track of this parameter through simple relations, as \eqref{eq_q-mu-vol-pred} during the time update and \eqref{eq_correc-q} during the observation update, spares its recalculation at each step $i$, provided that $\rank(A_0P_0A_0^T)$ is given.
\end{rema}
\begin{rema}
Setting either $\alphak{i}=0$ or $\betak{i}=0$ results in freezing $\EC_{k_{i-1}}$, meaning that the corresponding measurements $\fbki,\uyki,\byki$ do not bring any useful information.
\end{rema}
 \begin{rema}
 The cases 1 of Thms  \ref{theo_ellips_HSpace_inter}, \ref{theo_ellips_strip_inter} and \ref{theo_ell_hyp_inters} are not explicitly treated in this theorem assuming that they can not occur since the intervening measurements are supposed to be consistent with the system model;
yet the case where the measurement $\{\fbki,\uyki,\byki\}$ is aberrant is implicitly considered,
setting again either $\alphak{i}=0$ or $\betak{i}=0$,  preventing  so the updating of the ellipsoid $\EC_{k_{i-1}}$.
 \end{rema}
\section{Algorithm properties and stability analysis}\label{sec_prop_stability}
\subsection{The overall state estimation algorithm}\label{subsec_algo}
The time prediction stage given by Thm \ref{theo_time_update} with either Thm \ref{theo_mu-vol_k} or Thm \ref{theo_mu-tr} on one hand and the measurement correction phase, given by Thm \ref{theo_meas_update}, on the other, are concatenated to form the hole state estimation algorithm presented in Algorithm \ref{Algo_glo}, where  $N$ is the number of samples. 
{%
\begin{algorithm}[H]
\caption{Computation of the ellipsoid $\EC(\xk,\sigk P_{k})$}
\begin{algorithmic}[1]\label{Algo_glo}
	\REQUIRE $\xz$, $P_0$, $\sigz$, $N$,  \texttt{Predic\_VolMin},  \texttt{Correc\_$\beta$} 
\ENSURE $\xk$, $P_{k}$, $\sigk$, $k\in\{1,\ldots, N\}$
\STATE $n\leftarrow$ size of $\xz$\;
\FOR{$k=0,1,\ldots, N-1$}
\LCOMMENT{\bf // Prediction //}
\IF{ \texttt{Predic\_VolMin}}
\STATE Algorithm \ref{Algo_Predic_VolMin} ({\bf Input:} $\xk$,  $P_k$, $\sigz$, $q_k$, $A_k$, $B_k$, $R_k$, $\taubk$, \texttt{Predic\_VolMin};
{\bf Output}   $\xkkp$,  $\Pkkp$, 
$q\kkp$)
\ELSE
\STATE Algorithm \ref{Algo_Predic_TrMin} ({\bf Input:} $\xk$,  $P_k$, $\sigz$, $A_k$, $B_k$, $R_k$, $\taubk$;\\
{\bf Output:}   $\xkkp$,  $\Pkkp$) 
\ENDIF
\LCOMMENT{\bf // Correction //}
\STATE $k\gets k+1$
\STATE Algorithm \ref{Algo_Correc} ({\bf Input:} $\xkk$, $\Pkk$, $\sigz$, $\fb\kij{1},\ldots,\fb\kij{p}$, $\bybk$, $\uybk$, \texttt{Predic\_VolMin}, \texttt{Correc\_$\beta$} , $q\kk$; {\bf Output:}  $\xk$, $P_{k}$, $\sigk$, $q_k$)
\STATE\label{Algo_normalization}  
$\Pk\leftarrow \frac{\sigk}{\sigz} \Pk$; ${\varsigma}_{k}\leftarrow\frac{\sigz\sigk }{{\varsigma}_{k-1}}$;
\ENDFOR
\end{algorithmic}
\end{algorithm}
}

\begin{rema}
For more numerical stability and in order to avoid the explosion of the matrix $P_k$, caused by the set summations at the prediction step, 
a normalization is made at the line \ref{Algo_normalization}: 
$P_k$ would thereby represent, by itself, the shape matrix of the ellipsoid $\EC_k$ up to a constant factor $\sigz^{-1}$ and it is kept track of the evolution of ${\varsigma}_{k}$, since $\sigz$ is used, instead of $\sigk$ and $\sigkkk$, at the inputs of Algo. \ref{Algo_Predic_VolMin} and \ref{Algo_Correc} \resp.
\end{rema}
The proposed algorithm is designed in such a way as to fulfill the requirements \ref{requirement1} - \ref{requirement3}, expressed in the $\mathsection$\ref{sec_prob_form} and this is what will be shown in this section. 
\subsection{Algorithm properties}
In this paragraph, some algorithm's properties are shown, while the stability will be examined in the next.

\begin{theo}\label{theo_prop}
Consider the system \eqref{system0} subject to \eqref{bounds} and its state estimation algorithm given by Thms \ref{theo_time_update} and \ref{theo_mu-vol_k} or \ref{theo_mu-tr} on one hand and Thm \ref{theo_meas_update}, on the other.
  \begin{subequations}\label{Lyap_sig0}
\begin{enumerate}
  \item\label{theo_prop_E0_Ek} If $\xza\in\EC(\xz,\sigz P_0)$, then $\forall \bm{\beta}\in[0,1[^{p_k}$, $\forall \mub\in]0,+\infty[^{m}$, $\xka\in\EC(\xk,\sigk  P_k)$, $\forall k\in\N^*$;
  \item\label{theo_prop_acceptable} The vector $\xk$ is \emph{acceptable}, \ie, it  satisfies \eqref{boundb_HSpace0}--\eqref{bound_Strip0}: 
  \begin{align}
  \forall k\in\KC,\ \xk\in\bigcap_{i\in\gscrk}\GC\ki\cap\bigcap_{i\in\dscrk}\DC\ki\cap\bigcap_{i\in\hscrk}\HC\ki,  
  \end{align}
  where $\KC\eqg\{k\in\N|{p_k}\neq 0\}$.
  \item\label{theo_prop_lim_sig}  $\sigk$, defined in \eqref{eq_correc-sig}, satisfies,        
  \mathc
  \begin{align}
   \sigk&=\max_{\xbk\in\SC_k}\Vscr_{{P}_k}(\xka-\xk), \ \forall \beta^*\in]0,1[,
  \label{Lyap_def}
  \end{align}
where  $\SC_k$  and $\Vscr_{{P}}$ are given by   \eqref{SC_set} and \eqref{Lyap_def_sig0} \resp.

Moreover, if  in \eqref{eq_correc-bet}, $\beta^*=\beta_{\varsigma}$, 
given by \eqref{beta_opt_sig_min} - Algo. \ref{Algo_beta_SigMin}, then the sequence $\left\{\sigk\right\}_{k\in\N^*}$ is decreasing and convergent on $\R_+$.   
\end{enumerate}
   \end{subequations}
\end{theo}
{\debdemo
The proof of this lemma is detailed in the Appendix \ref{Appendix_theo_prop}.
\findemo
}
\subsection{Stability analysis}
The stability requirement \ref{requirement3} exploits the Input-to-State stability concept: \linebreak roughly speaking, for an ISS system, 
inputs that are bounded, \guil{eventually small}, or convergent, should lead to the state vector with the respective property; and the $\zero$-input system 
should be globally stable. 
We shall now recall more formal definitions and results about the ISS concept. Before doing so, let us recall some comparison functions, widely used in stability analysis. A continuous function
$\psi_1:\R_+\rightarrow\R_+$ is called positive definite if it satisfies $\psi_1(0)=0$ and $\psi_1(t)>0$, $\forall t>0$. A positive definite function is of class $\mathscr{K}$ if it is strictly increasing and of class $\kf_{\infty}$ if it is of class $\mathscr{K}$ and unbounded. A continuous function $\psi_2:\R_+\rightarrow\R_+$ is of class  $\mathscr{L}$ if $\psi_2(t)$ is strictly decreasing to 0 as $t\rightarrow \infty$ and a continuous function $\psi_3:\R_+\times\R_+\rightarrow\R_+$ is of class $\mathscr{KL}$ if it is of class $\mathscr{K}$ in the first argument and of class $\mathscr{L}$ in the second argument. 
\begin{defi}[\bf based on \cite{Jia:01}]\label{defi_ISS}
The system 
\mathl
\begin{subequations}\label{ISS_sys}
\begin{align}
\zb_{k+1}&=f_k\big(\zb_k,\ub_k\big),\\ 
\text{where } \qquad f(\zero,\zero)&=\zero, 
\end{align}
\end{subequations}
is globally input-to-state stable (ISS), if there exists a $\mathscr{KL}$-function $\phi$ and a $\mathscr{K}$-function $\psi$ such that, for each bounded input sequence $\ub_{[0,k]}\eqg\left\{\ub_0,\ldots,\ub_k\right\}$ and each $\zbz\in~\Rn$,
\mathc
\begin{align*}
\norm{\zb(k,\zbz,\ub_{[0,k-1]})}\leq\phi(\norm{\zbz},k)+\psi({\!\sup_{i\in\{1,\ldots, k-1\}}\!\norm{\ub_{i}}}),
\end{align*}
where $\zb(k,\zbz,\ub_{[0,k-1]})$ is the trajectory of the system \eqref{ISS_sys}, for the initial state $\zbz\in\Rn$ and the input sequence $\ub_{[0,k-1]}$.
\end{defi}
\begin{defi}[\bf\cite{Jia:01}]\label{defi_ISS_Lyap}\emph{
A continuous function $\Vscr: \Rn\rightarrow\R_+$ is an ISS-Lyapunov function for the system \eqref{ISS_sys}, if both conditions 1 et 2 are met:
\begin{enumerate}
\item there exists $\kf_{\infty}$-functions $\psi_1$ and $\psi_2$ such that for all $\zb\in\Rn$, 
\begin{align}
\psi_1(\norm{\zb})\leq\Vscr(\zb)\leq \psi_2(\norm{\zb});\label{ISS_cond1}
\end{align}
\item   there exists a {$\kf_{\infty}$-~\!function} $\psi_3$ and a  $\mathscr{K}$-function $\chi$ such that for all $k\in\N^*$, $\zb\in\Rn$ and all $\ub\in\Rm$,
\begin{align}
\Vscr\big(f_k(\zb,\ub)\big)-\Vscr(\zb)\leq -\psi_3(\norm{\zb})+\chi(\norm{\ub}).\label{ISS_cond2}
\end{align}
\end{enumerate}
  }
\end{defi}
\begin{lemm}[\bf\cite{Jia:01}]\label{lemm_ISS}
    The system \eqref{ISS_sys} 
    is ISS, if it admits a continuous ISS-~\!Lyapunov function. 
\end{lemm}
To prove the ISS stability, we will be using the candidate Lyapunov function defined in \eqref{Lyap_def}. For this purpose, the shape matrix $P_k$ should be bounded above and below.
Before studying the boundedness of $P_k$, 
we need to recall  the uniform controllability and observability notions: 
\begin{defi}[uniform observability and controllability]\label{defi_contr}
Consider \linebreak time-varying matrices $\breve{F}_{k}\in\R^{n\times p }$, ${R}_k$ and invertible ${A}_{k}\in\Rnn$. 
The pair $\{{A}_{k},\breve{F}_{k}^T\}$ 
is \emph{uniformly observable}, if there exist positive constants $\underb{\varrho}_{o}$ and $\bar{\varrho}_{o}$ and a positive integer $h$, such that, for all $k\geq h$, 
\mathc
\begin{align}
\underb{\varrho}_{o}I_n\leq {\OC}_{k,k-h} \leq \bar{\varrho}_{o}I_n,
\end{align}
where ${\OC}_{k+l,k}$ is an observability gramian of length $l$:
\mathc
\begin{align}
{\OC}_{k+l,k}&\eqg\sum_{i=k}^{k+l}{\Phi}_{i,k}^T\breve{F}_{i}\breve{F}_{i}^T{\Phi}_{i,k}.\label{gramian_obs}\\
\text{where \quad}{\Phi}_{k+l,k}&\eqg{A}_{k+l-1}\ldots{A}_{k}, \text{ with } {\Phi}_{k,k+l}={\Phi}_{k+l,k}^{-1}.\label{transition_matrix}
\end{align}
The pair $\{{A}_k,{R}_k\}$ is \emph{uniformly controllable}, if there exist positive constants ${\varrho}_3$ and ${\varrho}_4$ and a positive integer ${h}$, such that, for all $k\geq h$, 
\mathc
\begin{align}
{\varrho}_3I_n\leq {\CC}_{k,k-{h}} \leq {\varrho}_4I_n,
\end{align}
where  ${\CC}_{k+l,k}$ is a controllability gramian of length $h$:
\begin{align}
{\CC}_{k+l,k}&\eqg\sum_{i=k}^{k+l-1}{\Phi}_{k,i+1}{R}_i{R}_i^T{\Phi}_{k,i+1}^T.\label{gramian_con}
\end{align}
%
\end{defi}
It is needless to say that it is difficult to ensure the full rank for the matrix sum ${\OC}_{k,k-{h}}$ 
on a time window of constant length $h$ when dealing with sporadic measurements. The system \eqref{system0}-\eqref{bounds} can therefore not be uniformly observable.
For this purpose, let us introduce the new  observability criterion for systems with {sporadic measurements}, by allowing the observability gramian, used in \emph{uniform observability} 
to have a variable length $\spo_{k}(h)$ instead of the fixed one $h$:
\begin{defi}[sporadic observability]\label{defi_obs_spora} 
%
The  pair $\{{A}_{k},\breve{F}_{k}^T\}$  
is said \emph{sporadically observable}, if there exist positive constants $\underb{\varrho}_{c}$ and $\bar{\varrho}_{c}$ and a positive integer $h$, such that, for all $k\geq \spo_{k}(h)$,
\mathc
\begin{align*}
\underb{\varrho}_{c}I_n\leq {\OC}_{k,k-\spo_{k}(h)} \leq \bar{\varrho}_{c}I_n,
\end{align*}
where ${\OC}_{k+l,k}$ is the observability gramian given in \eqref{gramian_obs}
and ${\Phi}_{k+l,k}$, defined in \eqref{transition_matrix}, is the state transition matrix associated to ${A}_k\in\Rnn$, which is assumed to be \emph{invertible};  
$\spo_{k}(h)\in\N$  is \tq 
\mathc
\begin{align}
\card\big(\{i\in\breve{\KC}| k-\spo_{k}(h)\leq i \leq k\}\big)=h\label{card}
\end{align}
where $\card(\SC)$ stands for the cardinality (number of elements) of the set $\SC$.
\end{defi}
%
\begin{lemm}\label{Lem_Bounds_Pbar}
Consider the system \eqref{system0} subject to \eqref{bounds} and the matrix $P_k$ computed in line with either (\eqref{P_pred}-\eqref{P_pred1} and \eqref{mu-vol_k}) or \eqref{predic_eq_glo} on one hand and \eqref{ellips_correc}, 
on the other.
Let $\breve{H}_{k}\in~\R^{n\times{q}_k}$ 
whose columns form an orthonormal basis for $\range \big(P_k\big)$, 
where ${q}_k\eqg\rank(P_k)$ and let 
$\breve{P}_{k}\eqg \breve{H}_{k}P_k\breve{H}_{k}^T$. 
If the pair $\{A_{k},\breve{F}_{k}^T\}$ is \emph{sporadically observable} and $\{A_{k},R_k\}$ is uniformly controllable, 
then there exist positive finite numbers $\underb{\varrho}_{k}$ and $\bar{\varrho}_{k}$, \tq for all $k\geq \spo_{k}(l)$,
\mathc
\begin{align}
\underb{\varrho}_{k}I_{{q}_k}&\leq \breve{P}_k \leq \bar{\varrho}_{k}I_{{q}_k},\label{Bounds_Pbar}
\end{align}
$\spo_{k}$ being given in \eqref{card} of Definition \ref{defi_obs_spora}.
\end{lemm}
{\debdemo
The proof of this lemma is detailed in Appendix \ref{Appendix_Lem_Bounds_Pbar}.
\findemo
}

The following theorem shows the stability of the estimation algorithm according to the choice of the optimisation criterion made for $\betaki$ in  \eqref{eq_correc-bet}.
\begin{theo}\label{theo_ISS} Consider the system \eqref{system0} subject to \eqref{bounds} and its state estimation algorithm given by either Thms \ref{theo_time_update} and \ref{theo_mu-vol_k} or \ref{theo_mu-tr}, on one hand and \ref{theo_meas_update}, on the other. 
Let 
\begin{subequations}\label{bF_bp_KC_def}
\mathl
\begin{align}
\brdscrk&\eqg\bgscrk\cup\ugscrk\cup\dscrk=\pscrk-\hscrk,\text{ \cf \eqref{bounds},}\label{bdscrk_def}\\
\breve{F}_{k}&
\eqg[\fbki]_{i\in\brdscrk}\in\R^{n\times\breve{p} },\ k\in\breve{\KC}; \text{ setting }\breve{F}_{k}\breve{F}_{k}^T\eqg 0_{n, n},\ \forall k\notin\breve{\KC},\label{bF_def}\\
\breve{p}&\eqg\breve{p}_{k} \eqg\card(\brdscrk)\text{ and }\breve{\KC}=\{k\in\N^*|\bpk \neq 0\}.\label{bp_KC_def}
\end{align}
For each time step $k\in\N^*$ and each measurement $i\in\brdscrk$, consider {$\beta^*$ the value of $\betaki$ defined in \eqref{eq_correc-bet}.}
\renewcommand{\bpk}{\breve{p}}
If  $A_{k}$ is invertible and the pairs $\{A_{k},\breve{F}_{k}^T\}$ 
and $\{A_{k},R_k\}$  are \emph{sporadically observable}\footnote{\cf Definition \ref{defi_obs_spora}.}  and uniformly  controllable\footnote{\cf Definition \ref{defi_contr}.} 
\resp., then
\begin{enumerate}
 \item\label{theo_prop_vol_bounded} the volume of $\EC_k$ and all its axes lengths are bounded, $\forall \beta^*\in]0,1[$, in \eqref{eq_correc-bet}; 
\item\label{theo_prop_ISS} {If $\beta^*=\beta_{\varsigma}$ given in \eqref{beta_opt_sig_min} of Thm \ref{theo_beta_sig_min} or Algo.~\ref{Algo_beta_SigMin}, then \linebreak $\Vscr_{{P}_k}:\Rn\rightarrow\R_+$, $\xb\mapsto~\!\Vscr_{{P}_k}(\xb)\eqg\xb^TP_k^\dag\xb$ 
 is an ISS-Lyapunov function\footnote{\cf Definition \ref{defi_ISS_Lyap}} for the estimation error $\dxbk\eqg\xka-\xk$, which is ISS\footnote{\cf Lemma \ref{lemm_ISS}}.}
\item\label{theo_prop_vol_dec} If $\beta^*=\beta_{\text{v}}$ given by Thm \ref{theo_beta_vol_min} and Algo~\ref{Algo_beta_VolMin}, then $\tfrac{1}{{v}_{k}}\Vol\big(\EC_k\big)$ is nonincreasing and $\Vol\big(\EC_k\big)\leq {v}_{k}\Vol\big(\EC_0\big)$.
\item\label{theo_prop_ssal_dec} If $\beta^*=\beta_{\text{s}}$ given by Thm \ref{theo_beta_tr_min} and Algo.~\ref{Algo_beta_TrMin}, then $\tfrac{1}{{s}_{k}}\ssal\big(\EC_k\big)$ is nonincreasing and  $\ssal\big(\EC_k\big)\leq {{s}_{k}}\ssal\big(\EC_0\big)$;
\mathc
where 
\begin{align}
{v}_{k}&\eqg{\prod_{j=0}^{k-1}\Det{\breve{A}_jP_jP_j^\dag}^2\Det{\Qinv_j^\dag\Qinv_j+\big(\Qinv_j^\dag+I_n-\Qinv_j^\dag \Qinv_j\big)W_j}},\ {v}_{0}=1;\\
{s}_{k}&\eqg {\prod_{j=0}^{k-1}\Big(\tr\big(\breve{A}_j\breve{A}_j^T\big)+\tfrac{\tr(W_j)}{\tr(P_j)}\Big)},\ {s}_{0}=1;\\
\Qinv_k&\eqg \breve{A}_kP_k\breve{A}_k^T;\label{barQ}\\
\breve{A}_k&\eqg \prod_{i=1}^m\sqrt{1+\muki{i}} A_k =\sqrt{\chiki{1}}A_k, \ \muki{i}\in]0,1[; 
\label{barA} \\ 
W_k&\eqg \tfrac{1}{\sigk} R_k\diag\big(\tfrac{\chiki{i}}{{\muki{i}}}\big)_{i=1}^mR_k^T, \text{ where }\chiki{i} = \prod_{j=i}^m(1+\muki{j}).\label{W} 
\end{align}
\item\label{theo_prop_beta_sig_Lyap} 
Now, let $R_k=0_{n, m}$, $\forall k\in\N$. If $\beta^*=\beta_{\varsigma}$ 
 $\Vscr_{{P}_k}$ is a nonincreasing Lyapunov function for the estimation error which is Lyapunov stable.
\item\label{theo_prop_beta_ell_dec} Furthermore, if $\norm{A_k}\leq 1$, $\forall k\in\N$, then
\begin{enumerate}
\item\label{theo_prop_beta_ell_dec_sig} all the axes lengths of $\EC_k$ are nonincreasing, if $\beta^*=\beta_{\varsigma}$;
\item\label{theo_prop_beta_ell_dec_vol} $\Vol\big(\EC_k\big)$ is nonincreasing, if $\beta^*=\beta_{\text{v}}$;
\item\label{theo_prop_beta_ell_dec_ssal} $\ssal\big(\EC_k\big)$ is  nonincreasing, if $\beta^*=\beta_{\text{s}}$.
\end{enumerate}
\end{enumerate}
\end{subequations}
\end{theo}
{\debdemo
The proof is detailed in the Appendix \ref{Appendix_theo_ISS}.
\findemo
}
\begin{rema}
$H_k\eqg\left[\left.\breve{H}_{k}\right\vert \tilde{H}_{k}\right]\in\Rnn$, intervening in point \ref{theo_prop_ISS} of the proof, is a unitary matrix which rotates $P_k$ into a basis where it has two-bloc-diagonal form ($H_k$ can be obtained by QR decomposition of $P_k^{\frac{1}{2}}$ or by SVD of $P_k$).  $H_k^T\dxk=~[\dxbki{1}^T\vert\dxbki{2}^T]^T$ 
gives the components of  the state estimation error vector, $\dxbk\eqg\xka-\xk$, in this new rotated basis, where the ${q}_k$ first components are ISS (point \ref{theo_prop_ISS} of Thm \ref{theo_prop}) and the last $ \bqk$ ones are zero, 
meaning that the corresponding estimations are equal to their true values, in this rotated basis.
\end{rema}
\appendix
\section{Proofs of the results of section \ref{sec_time_update}}
\subsection{Proof of Lemma \ref{lem_ell_zono_sum}}
\label{Appendix_lem_ell_zono_sum}
The set containing every possible value of $\bxb\eqg A\xb+\wb $ can be schematized by $\bxb\in A\EC\oplus\ZC\subset\bEC(\mu) $; where  
$A\EC\eqg\EC(A\cb,\sig APA^{T})$ is the  image of the ellipsoid $\EC\eqg\EC(\cb,\sig P)$ by the endomorphism of matrix $A$; and $\bEC(\mu)\eqg\EC\big(\bcb,\sig\bP(\mu)\big)$ is the outer-bounding ellipsoid of the Minkowski sum (\cf $\mathsection$.\ref{subsec_notations}.\ref{Minkowski})
of $A\EC$ and the one-dimensional zonotope, \ie, interval segment, that is also a one-dimensional ellipsoid: $\ZC\eqg\ZC(\ub,\rb)=\EC(\ub,\rb\rb^T)$.
%
Now, the Minkowski sum of two ellipsoids $\EC(\cb_{1},P_{1})$  and $\EC(\cb_{2},P_{2})$ is not an ellipsoid, in general, yet can be bounded by a parametrized ellipsoid \cite{Che:94}: 
\begin{subequations}
\mathc
\begin{align}
\EC(\cb,P(\mu))&\supset\EC(\cb_{1},P_{1})\oplus\EC(\cb_{2},P_{2}), \ \forall\mu\in\R_+^*, \\
\text{ where }\cb&=\cb_{1}+\cb_{2}\text{ and }P(\mu)=(1+\mu)P_{1}+(1+\sfrac{1}{\mu})P_{2}.
\end{align}
\end{subequations}
Setting $P_1\eqg \sig APA^T$, $\cb_1=A\cb$, $P_2\eqg\rb\rb^T$ and $\cb_2=\ub$ completes the proof.
\subsection{Proof of Proposition \ref{propo_pseudo-determinant}}
\label{Appendix_propo_pseudo-determinant}
\mathl
\setdefaultleftmargin{0pt}{}{}{}{}{}
\begin{enumerate}
\item Let $\underb{q}\eqg n-{q}$. 
There exists a unitary matrix 
$U$, \tq $U^TQU=\Big[{{\Sigma}^2\atop 0_{ \bqk, {q}}} 
{0_{{q},\bqk}\atop 0_{\bqk,\bqk}}\Big]$, where $\Sigma\in~\!\Rqq$ is a diagonal matrix with nonzero singular values of ${Q}$ on its diagonal. 
Let 
$\bar{\rb}\eqg \sqrt{a}{\Sigma}^{-1}\big[I_q\ \zero_{q,\bqk }\big]U^T\rb\in~\!\Rq$ and 
$\underb{\rb}\eqg \sqrt{a}\big[\zero_{\bqk ,q}\ I_{\bqk }\big]U^T\rb\in\R^{\bqk }$.
\begin{align}
Q+a\rb\rb^T
%
&=U\!\left[{\Sigma\hphantom{_{ \bqk, {q}}} \atop0_{ \bqk, {q}}}{0_{q,\bqk-1}\atop0_{\bqk,\bqk-1}}{\Sigma\bar{\rb}\atop\underb{\rb}}\right]\!
\left[{\Sigma\hphantom{_{ \bqk, {q}}} \atop 0_{ \bqk, {q}}}{0_{q,\bqk-1}\atop0_{\bqk,\bqk-1}}{\Sigma\bar{\rb}\atop\underb{\rb}}\right]^T\!U^T\nn\\
&=U\!\left[{\Sigma\hphantom{_{ \bqk, {q}}} \atop0_{ \bqk, {q}}}{\Sigma\bar{\rb}\atop\underb{\rb}}\right]\!
\left[{\Sigma\hphantom{_{ \bqk, {q}}} \atop0_{ \bqk, {q}}}{\Sigma\bar{\rb}\atop\underb{\rb}}\right]^T\!U^T,\!\!\! \label{fac_Q}
\end{align}
$\d\rank(Q+a\rb\rb^T) = \rank\left(\left[{\Sigma\hphantom{_{ \bqk, {q}}} \atop 0_{ \bqk , {q}}}{0_{{q},\bqk }\atop I_{\bqk }}\right]\left[{I_q \atop0_{\bqk ,q}}{\bar{\rb}\atop\underb{\rb}}\right]\right)=\rank\left(\left[{I_q \atop0_{\bqk ,q}}{\bar{\rb}\atop\underb{\rb}}\right]\right)$.
Then if $\vb\eqg(I_n-QQ^\dag)\rb\neq\zero_n$, which is the projection of $\rb$ onto $\nul({Q})$, meaning that $\underb{\rb}\neq\zero_{\bqk }$, implying that $\rank\big({Q_+}\big)=\rank(Q+a\rb\rb^T)=~\!q+1$; otherwise $\rank\big({Q_+}\big)=~\!q$.
\item From \eqref{fac_Q},
\begin{align}
Q+a\rb\rb^T&=U \left[{\Sigma \atop 0_{ \bqk , {q}}}{0_{{q},\bqk }\atop I_{\bqk }}\right]
\left[{I_{q}+\bar{\rb}\bar{\rb}^T \atop \underb{\rb}\bar{\rb}^T}{\bar{\rb}\underb{\rb}^T\atop\underb{\rb}\underb{\rb}^T}\right]
\left[{\Sigma \atop0_{ \bqk , {q}}}{0_{{q},\bqk }\atop I_{\bqk }}\right]^TU^T\nn\\
\Det{Q+a\rb\rb^T}&=\abs{\Sigma}^2\Det{
\left[\begin{smallmatrix}
I_{q} &0_{q,\bqk } &\zero_q\\
0_{\bqk ,q} &I_{\bqk }&\zero_{\bqk }\\
\bar{\rb}^T& \underb{\rb}^T&1
\end{smallmatrix}\right]
\left[\begin{smallmatrix}
I_{q}+\bar{\rb}\bar{\rb}^T &\bar{\rb}\underb{\rb}^T  &\bar{\rb}\\
\underb{\rb}\bar{\rb}^T&\underb{\rb}\underb{\rb}^T&\underb{\rb}\\
\zero_{q}^T& \zero_{\bqk }^T&1
\end{smallmatrix}\right]
\left[\begin{smallmatrix}
I_{q} &0_{q,\bqk } &\zero_q\\
0_{\bqk ,q} &I_{\bqk }&\zero_{\bqk }\\
-\bar{\rb}^T& -\underb{\rb}^T&1
\end{smallmatrix}\right]
}\nn
\end{align}
\begin{align}
&=\abs{\Sigma}^2\Det{\left[\begin{smallmatrix}
I_{q} &0_{q,\bqk } &\bar{\rb}\\
0_{\bqk ,q} &0_{\bqk,\bqk }&\underb{\rb}\\
\zero_q^T& -\underb{\rb}^T&\bar{\rb}^T\bar{\rb}+\underb{\rb}^T\underb{\rb}+1
\end{smallmatrix}\right]
}=
\begin{cases}
\abs{\Sigma}^2\big(\bar{\rb}^T\bar{\rb}+1\big),&\text{ if }\underb{\rb}= \zero_{\bqk }\\
\abs{\Sigma}^2\underb{\rb}^T\underb{\rb},&\text{ otherwise.}
\end{cases}
\end{align}
Now, noticing that $\d\Det{b Q_+}= 
b^{q_+}\Det{Q_+}$ (the product of the $q_+$ nonzero singular values), that $\abs{\Sigma}^2=\Det{Q}$ and since $\underb{\rb}^T\underb{\rb}=a\vb^T\vb$ and $\bar{\rb}^T\bar{\rb}=a\rb^TQ^\dag\rb$, the proof of \eqref{eq_pseudo-det_rec} is completed.

\end{enumerate}

\subsection{Proof of Theorem \ref{theo_mu-vol}}
\label{Appendix_theo_mu-vol}
The pseudo-volume of an ellipsoid being proportional to the pseudo-determinant of its shape matrix and $\sig$ being considered as constant \wrt $\mu$,
\mathl
\begin{align}
\mu_{\text{v}}\eqg\arg\min_{\mu\in\R_+^*}\Det{\bP(\mu)}=\arg\min_{\mu\in\R_+^*}\log\Det{Q}.\nn
\end{align}
According to Proposition \ref{propo_pseudo-determinant}, where $b\eqg (1+\mu)$ and $a\eqg \tfrac{1}{\varsigma\mu}$, two cases are distinguished. 
\begin{description}
\item[Case 1:] the projection of the vector $\rb$ onto the nullspace of ${Q}$, embodied by $\vb$ is a zero vector, in this case $\rank\big(\bP(\mu)\big)=\rank\big(Q_+)=\rank({Q})=q$ and
\begin{align}
\bP(\mu)&\eqg(1+\mu)\left({Q}+\tfrac{1}{\sig\mu}\rb\rb^{T}\right)\Rightarrow
\Det{\bP(\mu)}=\Det{Q}(1+\mu)^{q}\big(1+\tfrac{h}{\mu}\big)\nn
\end{align}
\begin{align}
\tfrac{\partial}{\partial \mu}\log\Det{\bP(\mu)}=0&\eq\tfrac{{q}}{1+\mu}-\tfrac{h}{\mu(\mu+h)}=0
\eq \ {q}\mu^2+h({q}-1)\mu-h=0.\label{eq_quad_vol_min}
\end{align}
$\mu_{\text{v}}$ in \eqref{mu-vol}, is the only positive solution to the quadratic equation \eqref{eq_quad_vol_min}.
\item[Case 2:] the projection of $\rb$ onto the nullspace of ${Q}$ is nonzero, thus implying the rank incrementation of the shape matrix: $\rank\big(\bP(\mu)\big)=q+1$ and
\begin{align}
\Det{\bP(\mu)}=\Det{Q}\norm{\vb}^2\tfrac{(1+\mu)^{q+1}}{\varsigma\mu}\eq\tfrac{\partial}{\partial \mu}\log\Det{\bP(\mu)}=0&\eq\tfrac{{q}+1}{1+\mu}-\tfrac{1}{\mu}=0,\nn
\end{align}
this results in $\mu = \tfrac{1}{q}$ of \eqref{mu-vol}.
\end{description}



\subsection{Proof of Theorem \ref{theo_time_update}}
\label{Appendix_theo_time_update}
Conforming to \eqref{state_eq0}, the set containing every possible value of $\xka$ can be schematized by 
\mathc
\begin{align}
A_k\EC_k\oplus\ZC(B_k\taubk,R_k)=\EC\kkpj{0}\oplus\ZC(\zero_n,R_k).
\end{align}
%
Now, the zonotope $\ZC(\zero_n,\Rk)$, where $\Rk=\big[\r_{k_1}\cdots\r_{k_m}\big]$, can be viewed as the sum of $m$ degenerate ellipsoids \cite{kur:14}: 
$$\d\ZC(\zero_n,\Rk)=\mink_{i=1}^m\ZC(\zero_n,\rbki )=\mink_{i=1}^m\EC(\zero_n,\rbki \rbki ^T).$$
%
 Applying Lemma \ref{lem_ell_zono_sum} sequentially to 
\mathc
\begin{align}
\d\EC\kkp\supset\EC\kkpj{0}\oplus\left(\mink_{i=1}^m\EC(\zero_n,\rbki\rbki^T)\right),
\end{align} 
eventuates in \eqref{predic_eq}. 
\subsection{Proof of Theorem \ref{theo_mu-tr}}
\label{Appendix_theo_mu-tr}
Since the SSAL of an ellipsoid is the trace of its shape matrix, \linebreak
$\d\mu_{{\text{s}}_{i}}=\arg\min_{\mu_i}\ssal(\EC\kkpj{i})=~\!\arg\min_{\mu_i}\tr(\Qki_{i})$;
\eqref{mu-tr_k} is the direct corollary of  Theorem \ref{theo_mu-tr}.  
As for \eqref{predic_eq_glo}, it is a direct consequence of the result \cite{Dur:01} saying that the minimum trace ellipsoid containing the Minkowski sum of $m+1$ ellipsoids is : 
\mathc
\begin{subequations}\label{pf_theo_mu-vol-tr_1}
\begin{align}
\EC(\cb,P)&\eqg\oplus_{i=0}^m\EC(\cb_i,P_i),\\ 
\intertext{where }
\cb\eqg \sum_{i=0}^m \cb_i \text{ and }
P&\eqg \Big(\sum_{i=0}^m\sqrt{\tr(P_i)}\Big) \Big(\sum_{i=0}^m\big(\sqrt{\tr(P_i)}\big)^{-1}{P_i}\Big).
\end{align}
\end{subequations}
Then, after noticing that 
\mathc
\begin{align}
\tr(\rki\rkit)=\rkit\rki=\norm{\rki}^2\\
\intertext{and that}
\sum_{i=1}^m\left({\sqrt{\tr(\rki\rkit)}}\right)^{-1}{\rbki\rbki^{T}}={M}_k
\end{align}
\eqref{pf_theo_mu-vol-tr_1} applied to $\d\EC\kkpj{0}\oplus\big(\oplus_{i=1}^m\EC(\zero_n,\rki\rkit\big)$, leads clearly to \eqref{predic_eq_glo}.
It is also stated in \cite{Dur:01} that such an ellipsoid is the same that the one obtained sequentially in \eqref{mu-tr}.
\section{Proofs of the results of section \ref{sec_measurement_update}}
\subsection{Proof of Theorem \ref{theo_ellips_HSpace_inter}}\label{Appendix_theo_ellips_HSpace_inter}
{
\nopagebreak
\begin{defi}
The signed distance from a set $\SC\subset\Rn$ to a vector $\x\in\Rn$ is $\d\zeta(\SC,\x)\eqg\max_{\norm{\ub}=1}\ub^T\x-\rho_\SC(\ub)$.
\end{defi}
\begin{propo}[\cite{Kur:06}]\label{ellips_Hplane_dist}
The  signed distance from an ellipsoid  to a hyperplane is given by:
\begin{align}
\zeta\big(\EC(\cb,P),\HC(\db,a)\big)\eqg\norm{\db}^{-1}\Big(\abs{a-\cb^T\db}-\sqrt{\db^TP\db}\Big).
\end{align}
\end{propo}
Let $\EC\eqg\EC(\cb,\varsigma P)$. The signed distance from $\EC$ to $\HC(\fb,\by)$ is 
\begin{align}
\zeta\eqg\norm{\fb}^{-1}\left(\abs{\by-\cb^T\fb}-\sqrt{\varsigma\fb^TP\fb}\right).
\end{align}
\begin{itemize}
\item $\zeta\geq 0$ means that $\HC(\fb,\by)$ does not intersect $\EC$ in more than one point: 
\begin{enumerate}
\item if $\cb^T\fb>\by$, then $\EC\subset~\GC(-\fb,-\by)$ and $\EC\cap\GC(\fb,\by)=\emptyset\eq$ \eqref{cas1a}; 
\item if $\cb^T\fb\leq \by$, then $\EC\subset\GC(\fb,\by)\Rightarrow\EC\cap\GC(\fb,\by)=\EC\eq$ \eqref{cas1b};
\item if $\cb^T\fb-\by=\sqrt{\varsigma\fb^TP\fb}$, then $\HC(\fb,\by)$ is tangent to $\EC$
 and   \linebreak
 $\EC\cap\GC(\fb,\by)=\EC\cap\HC(\fb,\by)=\{{\cb}_{\HC}\}$, where ${\cb}_{\HC}$ is given in \eqref{cas3b_HP} and can be calculated using \eqref{eq_x},  letting $\delta\leftarrow \by-\cb^T\fb$, with
 $\delta^2={\varsigma\fb^TP\fb}$, $\fb\leftarrow\fb$, $P\leftarrow\varsigma^{-1} P$.
\end{enumerate}
\item If $\zeta\leq0$, then $\HC(\fb,\by)$ intersects $\EC$ and
$\HC(-\fb,\rho(-\fb))$ is the ellipsoid's supporting hyperplane of normal vector $-\fb$ which is contained in $\GC(\fb,\by)$.
Indeed, 
\begin{align}
\x\in\HC(-\fb,\rho(-\fb))&\eq\x^T\fb-\cb^T\fb=-\sqrt{\varsigma\fb^TP\fb}\leq \by-\cb^T\fb\nn\\
&\Rightarrow \x^T\fb\leq \by\eq\x\in\GC(\fb,\by).\\
%
\intertext{Thence, $\GC\big({-\fb},\rho(-\fb)\big)$ is its supporting halfspace and $\EC\subset\GC\big(-\fb,\rho(-\fb)\big)$.
Therefore, }
\EC\cap\GC(\fb,\by)&=\Big(\GC\big({-\fb},\rho(-\fb)\big)\cap\GC(\fb,\by)\Big)\cap\EC\\
\intertext{$\abs{\by-\cb^T\fb}<\sqrt{\varsigma\fb^TP\fb}$ means that $0<\by+\rho(-\fb)<2\sqrt{\varsigma\fb^TP\fb}$, 
entailing, on one hand, }
\GC(\fb,\by)&=\big\{\x\big|\x^T\fb\leq \by\big\}=\bigg\{\x\bigg|\frac{2}{\by+\rho(-\fb)}\x^T\fb\leq\frac{2\by}{\by+\rho(-\fb)}\bigg\}\nn\\ &=\GC(\gamma^{-1}\fb,{y}+1)\\
\text{and }
\GC(-\fb,\rho(-\fb))&=\bigg\{\x\bigg|-\frac{2}{\by+\rho(-\fb)}\x^T\fb\leq\frac{2\rho(-\fb)}{\by+\rho(-\fb)}\bigg\}\nn\\
&=\GC(-\gamma^{-1}\fb,-{y}+1),\\
\intertext{on the other hand. Finally, the proof \eqref{case4_Hspace}--\eqref{gamma_y_HSpace_def} is achieved thusly:}
\EC\cap\GC(\fb,\by)&=\EC\cap\Big(\GC\big(\gamma^{-1}\fb,{y}+1\big)\cap\GC\big(-\gamma^{-1}\fb,-{y}+1\big)\Big)\\
&=\EC\cap\DC\big(\gamma^{-1}\fb,{y}\big).\tag*{\ding{113}}
\end{align}
\end{itemize}
%

}
\subsection{Proof of Theorem \ref{theo_ellips_strip_inter}}
\label{Appendix_theo_ellips_strip_inter}
{
Let $\EC\eqg\EC(\cb ,\varsigma P)$.
The signed distance from the ellipsoid $\EC$ to each of the two hyperplanes $\HC(\fb,y\mp1)$, bounding the strip $\DC(\fb,y)$, is 
\begin{align}
\zeta\eqg\norm{\fb}^{-1}\big(\big| y\mp 1-\fb^T\cb\big|-\sqrt{\varsigma\fb^TP\fb}\big).
\end{align}
When $\zeta>0$, the ellipsoid doesn't intersect any of both hyperplanes meaning either that the ellipsoid is located outside the strip, in which case (case 1), the intersection is empty or that  it is situated between them \ie, contained in the strip (case 2). 
In the case 3, the interior of the ellipsoid is outside the strip touching it in only one point and the case 3 of Thm \ref{theo_ellips_HSpace_inter} is then applicable: $\DC(\fb,y)\cap\EC=\GC(\fb,y+1)\cap\EC$ (case 3.a) and $\DC(\fb,y)\cap\EC=\GC(-\fb,-y+1)\cap\EC$ (case 3.b).
In the case 4,  where  $\zeta\leq~0$, the intersection is not empty. It is then possible to introduce the following lemma, based on the results of \cite{Fog:82} and \cite{Tan:97}:
\begin{lemm}\label{lemm_Tan:97}
$\forall y\in\R$, $\bm{c}\in\Rn$, $\fb\in\R^{n}$, $\varsigma\in~\R_+^*$ and SPD $P\in\Rnn$, if $\DC(\fb,y)\cap\EC(\bm{c},\varsigma P)\neq\emptyset$, then 
\begin{subequations}\label{lemm_Tan:97_eq}
\begin{align}
\forall\omega\in\R_+^*,&\quad\EC(\tilde{\bm{c}}(\omega),\tilde{\varsigma}(\omega)\tilde{P}\big(\omega)\big)\supset\DC(\fb,y)\cap\EC(\bm{c},\varsigma P),\label{ellips_family_FH}\\
\intertext{where }   \tilde{P}(\omega)& \eqg  P-\tfrac{\omega}{\omega  \alpha + {1}} P {\fb}{\fb}^TP,\quad \\    
   \tilde{\bm{c}}(\omega)   & \eqg  \bm{c}+\tfrac{\omega}{\omega  \alpha + {1}}\delta P {\fb}= \bm{c}+\omega  \tilde{P}(\omega) ^{-1}{\fb}\delta,\quad   \\       
   \tilde{\varsigma}(\omega) & \eqg  \varsigma+\omega(1-\tfrac{\alpha}{\omega+ \alpha}\delta^2), \\
   \delta&\eqg y-{\fb}^T\bm{c} \text{ and  $\alpha$ is given in \eqref{alpha_def}}.
   \end{align}
\end{subequations}
\end{lemm}
This lemma is precisely the mono-output case of the \guil{observation update} part of 
Thm 1, \cite{Bec:08}. 
\eqref{P_ellips_strip_inter}, \eqref{c_ellips_strip_inter} and \eqref{sig_ellips_strip_inter} are obtained by setting \linebreak $\omega\eqg{\alpha\beta}({1-\beta})^{-1}$, thus $\beta={\omega}({\alpha+\omega})^{-1}$. 
%
%
But before applying the lemma above, it is suitable to reduce the strip $\DC(\fb,y)$ in case where one of the two hyperplanes does not intersect the ellipsoid $\EC$, \ie, when either $y+1>\brho$ or $y-1<-\urho$, by translating the aforementioned hyperplane so that it becomes tangent to the ellipsoid, as proposed in \cite{Bel:90}. The new strip so obtained is $\DC(\gamma^{-1}\fb,\breve{y})$, where $\gamma$ and $\breve{y}\eqg y$ are given in \eqref{gamma_y_HSpace_def} and result from applying (case 4) of Thm \ref{theo_ellips_HSpace_inter} to $\EC\cap\GC(-\fb,-y+1)$  and to $\EC\cap \GC(\fb,y+1)$.
 }
\subsection{Proof of Theorem \ref{theo_beta_sig_min}}
\label{Appendix_theo_beta_sig_min}
Applying the generalization of the Sherman-Morrison formula  to the pseudo-inverse of the matrix  \eqref{P_ellips_strip_inter} (\cf Corollary 3.5 \cite{Xu:17}), we can write 
\mathc
\begin{align}
{P}_{\DC}(\beta)^\dag=P^\dag+\tfrac{\alpha\beta}{1-\beta}{P^\dag P\fb\fb^TPP^\dag}.\label{Pbrevedag}
\end{align}
Now, noticing that $(PP^\dag)^T=P^\dag P$ and 
recalling that, $\forall\ub\in\range(P)$, 
$\exists\vb\in\Rn$, $\ub=P\vb$, so
$PP^\dag\ub=PP^\dag P\vb=P\vb=~\!\ub$, 
in particular for $\ub\eqg\x-{\cb}_{\DC}(\beta)$, $\forall\x\in~\!\EC(\cb,\varsigma P)$, since ${\cb}_{\DC}(\beta)\in\EC(\cb,\varsigma P)$;
replacing \eqref{Pbrevedag} in \eqref{Lyap_def_sig0} leads then to
\begin{subequations}\label{Lyap_sig_max}
\mathl
\begin{align}
{\Vscr}_{{P}_{\DC}(\beta)}\big(\xb-{\cb}_{\DC}(\beta)\big)&\eqg\big(\xb-{\cb}_{\DC}(\beta)\big)^T\Big(P^\dag+\tfrac{\alpha\beta}{1-\beta}{P^\dag P\fb\fb^TPP^\dag}\Big)\big(\x-{\cb}_{\DC}(\beta)\big)\nn\\
&=\big(\xb-{\cb}_{\DC}(\beta)\big)^T\Big(P^\dag+\tfrac{\alpha\beta}{1-\beta}{\fb\fb^T}\Big)\big(\x-{\cb}_{\DC}(\beta)\big).\label{Lyap_def_sig1} 
\end{align}
Inserting  \eqref{c_ellips_strip_inter} in \eqref{Lyap_def_sig1}, we can show, by the mean of some standard algebraic manipulations, that\footnote{${\Vscr}_{{P}_{\DC}(\beta)}$ is optimized on $\DC(\breve{\fb},\breve{y})\cap\EC(\cb,\varsigma P)$, it is then obvious that $\x\in\range (P)$, since $\xb\in\EC(\cb,\varsigma P)$.}, $\forall\xb\in\Rn$, \tq $(\xb-\cb)\in\range(P)$ (since $\xb\in\EC(\cb,\varsigma P)$),
\begin{align}
{\Vscr}_{{P}_{\DC}(\beta)}\big(\x-{\cb}_{\DC}(\beta)\big)&= \tfrac{\alpha\beta\gamma^2}{1-\beta}(\breve{y}-\breve{\fb}^T\x)^2-{\alpha\beta\delta^2}+(\x-\cb)^TP^\dag(\x-\cb)\nn\\
{\Vscr}_{{P}_{\DC}(\beta)}\big(\x-{\cb}_{\DC}(\beta)\big)-{\Vscr}_{{P}}(\x-\cb)&=\tfrac{\alpha\beta\gamma^2}{1-\beta}(\breve{y}-\breve{\fb}^T\x)^2-{\alpha\beta\delta^2}.
\end{align}
{Since $\d\max_{\bm{x}\in\DC(\breve{\fb},\breve{y})}(\breve{y}-\breve{\fb}^T\x)^2=1$ 
and $\d\max_{\bm{x}\in\EC(\cb,\varsigma P)}{\Vscr}_{{P}}(\x-\cb)=\varsigma$, it comes that}
\begin{align}
{\Vscr}_{{P}_{\DC}(\beta)}\big(\x-{\cb}_{\DC}(\beta)\big)-{\Vscr}_{{P}}(\x-\cb)&\leq\tfrac{\alpha\beta\gamma^2}{1-\beta}-{\alpha\beta\delta^2}={\varsigma}_{\DC}(\beta)-\varsigma;\nn\\
\ \max_{\bm{x}\in\DC(\breve{\fb},\breve{y})\cap\EC(\cb,\varsigma P)}{\Vscr}_{{P}_{\DC}(\beta)}\big(\x-{\cb}_{\DC}(\beta)\big)&=\tfrac{\alpha\beta\gamma^2}{1-\beta}-\alpha\beta{\delta^2}+\varsigma={\varsigma}_{\DC}(\beta).
\end{align}
\end{subequations}
The optimal value of $\beta$ is obtained by zeroing the derivative of ${\varsigma}_{\DC}$: 
\mathc
\begin{align}
\dfrac{\dif{\varsigma}_{\DC}}{\dif\beta}(\beta_{\varsigma})=0\eq\gamma^2\left(1-\beta_{\varsigma}\right)^{-2}-\delta^2=0
\eq\beta_{\varsigma}=1-\gamma\abs{\delta}^{-1}. 
\end{align}
Since $\beta_{\varsigma}\geq 0$, this solution is conditioned by $\abs{\delta}>~\gamma$; 
if $\abs{\delta}\leq~\gamma$,  the solution to the above minimization problem would be $\beta_{\varsigma}=0$. 

Next, let's prove \eqref{sig_decrease}. If $\abs{\delta}>~\!\gamma$, replacing $\beta\eqg\beta_{\varsigma}=1-\gamma\abs{\delta}^{-1}$ in the expression of ${\cb}_{\DC}(\beta)$ in  \eqref{sig_ellips_strip_inter}
\begin{align}
{\varsigma}_{\DC}(\beta_{\varsigma})&=\varsigma+\tfrac{\alpha\beta_{\varsigma}\gamma^2}{1-\beta_{\varsigma}}-\alpha\beta_{\varsigma}{\delta^2} \nn\\
{\varsigma}_{\DC}(\beta_{\varsigma}) & \eqg  \varsigma-{\alpha\beta_{\varsigma}}\left(\delta^2-\gamma^2({1-\beta_{\varsigma}})^{-1}\right), \text{ noticing that } {1-\beta_{\varsigma}}=\gamma\abs{\delta}^{-1}, \text{ if } \abs{\delta }>\gamma,\nn\\
{\varsigma}_{\DC}(\beta_{\varsigma}) &=  \varsigma-{\alpha\beta_{\varsigma}}\left(\delta^2-\gamma\abs{\delta}\right) 
= \varsigma-{\alpha\beta_{\varsigma}}\abs{\delta}^2\left(1-\gamma\abs{\delta}^{-1}\right)
=\varsigma-{\alpha\beta_{\varsigma}}^2{\delta}^2.
\label{sig_thm_pf}
\end{align}
Now, as defined in \eqref{alpha_def}, $\alpha>0$, implying \eqref{sig_decrease}.

As for \eqref{sig_vol_ssal_decrease}, it is clear from \eqref{P_ellips_strip_inter}, that ${P}_{\DC}(\beta_{\varsigma})-P=- {\alpha\beta} P\fb\fb^T P\leq 0$. This means that all the eigenvalues of ${P}_{\DC}(\beta_{\varsigma})$ are less than or equal to those of $P$. Moreover if $\beta_{\varsigma}\neq 0$ (whenever $\abs{\delta}>~\!\gamma$), some eigenvalues of ${P}_{\DC}(\beta_{\varsigma})$ are necessarily less than those of $P$ implying that their sum and their product obey to the strict inequality and since ${\varsigma}_{\DC}(\beta_{\varsigma})<\varsigma$,  it follows that ${\varsigma}_{\DC}(\beta_{\varsigma}){P}_{\DC}(\beta_{\varsigma})-\varsigma P<0$. Consequently, $\Det{{\varsigma}_{\DC}(\beta_{\varsigma}){P}_{\DC}(\beta_{\varsigma})}<\Det{\varsigma P}$ and 
$\tr\big({\varsigma}_{\DC}(\beta_{\varsigma}){P}_{\DC}(\beta_{\varsigma})\big)<\tr\big(\varsigma P\big)$. 

\subsection{Proof of Theorem \ref{theo_beta_vol_min}}
\label{Appendix_theo_beta_vol_min}
There exists an orthogonal matrix $U=[\bar{U} \ \underb{U}]\in\Rnn$, $\bar{U}\in\Rnq$, where $q\eqg\rank(P)$, \tq 
$U^TPU=\bdiag\big(\Sigma,\zero_{n-q,n-q}\big)$ and $\bar{U}\fb=\bar{\fb}$ and 
\begin{align}
{P}_{\DC}(\beta)&=P-\alpha\beta P\fb\fb^TP=U\bdiag\big(\Sigma-\alpha\beta\Sigma \bar{\fb}\bar{\fb}^T\Sigma,\zero_{n-q,n-q}\big)U^T\nn\\
\Det{{P}_{\DC}(\beta)}&=\Det{P-\alpha\beta P\fb\fb^TP}=\abs{\Sigma-\alpha\beta\Sigma \bar{\fb}\bar{\fb}^T\Sigma};\nn
\end{align}
then, 1) using the following proposition:
\begin{propo}[\cite{Mey:00}]\label{propo_det_one-rank-up}
If $Q\in\Rnn$ is nonsigular and $\ub$, $\vb\in\Rn$, then
$$\abs{Q+\ub\vb^T}=\big(1+\vb^TQ^{-1}\ub\big)\abs{Q}.$$
\end{propo}
2) $\abs{\Sigma}=\Det{P}$,  3) $\bar{\fb}^T\Sigma\bar{\fb}=\fb^TP\fb$ and 4) $\alpha\eqg\big(\fb^TP\fb\big)^{-1}$, we have
\begin{align}
\Det{{P}_{\DC}(\beta)}&=\abs{\Sigma}\big(1-\alpha\beta\bar{\fb}^T\Sigma\bar{\fb}\big)=\Det{P}\big(1-\alpha\beta\fb^TP\fb\big)=\Det{P}\big(1-\beta\big),
\nn\\
\tfrac{\partial}{\partial\beta}\Det{{\varsigma}_{\DC}(\beta){P}_{\DC}(\beta)}=0 &\eq \Det{P}\tfrac{\partial}{\partial\beta}{\varsigma}_{\DC}(\beta)^{{q}}\big(1-\beta\big)=0\eq {q}(1-\beta)\tfrac{\partial}{\partial\beta}{\varsigma }_{\DC}(\beta)-{\varsigma }_{\DC}(\beta)=0\nn\\
&\eq {q}\alpha\Big(\tfrac{\gamma^2}{1-\beta}-\delta^2(1-\beta)\Big)-\varsigma-{\alpha\beta}\left(\tfrac{\gamma^2}{{1-\beta}}-\delta ^2\right)=0\nn\\
&\eq a_2\beta^2+a_1\beta+a_0=0.\nn\\
&\eq\beta=\beta_{\textit{v}}=
  \begin{cases}
  \tfrac{-a_1-\sqrt{a_1^2-4a_0a_2}}{2a_2}, &\text{if } a_0<0,\hspace{-10pt}\\ 
      0, &\text{otherwise};\
   \end{cases}\label{beta_vol}
\end{align}
%
$\beta_{\text{v}}$ is then the unique solution in $[0,1[$ of the above quadratic equation. Indeed, it can be proven that the discriminant is always positive: 
\begin{align}
\Delta=a_1^2-4a_0a_2=\big(\varsigma-\alpha(\gamma-\delta^2)\big)^2+4({q}^2-1)\alpha^2\delta^2\gamma>0.\nn
\end{align}
Finally, considering \eqref{sig_vol_ssal_decrease} and the fact that $\Vol\big({\EC}_{\DC}(\beta_{\text{v}})\big)\leq\Vol\big({\EC}_{\DC}(\beta)\big)$, $\forall\beta\in~\!]0,1[$, it becomes obvious that $\Vol\big({\EC}_{\DC}(\beta_{\text{v}})\big)\leq\Vol\big({\EC}_{\DC}(\beta_{\varsigma})\big)<\Vol\big(\EC(\cb,\varsigma P)\big)$.
\subsection{Proof of Theorem \ref{theo_beta_tr_min}}
\label{Appendix_theo_beta_tr_min}
\begin{subequations}\label{eq-beta-opt-tr-demo}
\begin{align}
\tr\big({\varsigma}_{\DC}(\beta){P}_{\DC}(\beta)\big)={\varsigma}_{\DC}(\beta)\Big(\tr(P)-\alpha\beta\tr\big(P\fb\fb^TP\big)\Big)={\varsigma}_{\DC}(\beta)\Big(\tr(P)-\alpha\beta\fb^TP^2\fb\Big).\nn
\end{align}
Deriving the above expression \wrt $\beta$ and zeroing it eventuates in the following cubic equation:
\begin{align}
b_3\beta^3+b_2\beta^2+b_1\beta+b_0=0 \eq \beta^3+b\beta^2+c\beta+d&=0,\label{eq_cubic}
\end{align}
where $b=\frac{b_2}{b_3}$,  $c=\frac{b_1}{b_3}$ and  $d=\frac{b_0}{b_3}$.
The three explicit solutions of \eqref{eq_cubic}, $\beta_i$, $i\in\{1,2,3\}$, can be obtained applying the \guil{First course in the theory of equations} of L. E Dickson \cite{Dic:52} 
reproduced (copied/pasted) in \cf Appendix \ref{Appendix_cubic_eq}:
\mathl
\begin{align}
\beta_{s_1} &\eqg \tilde{\beta}_1 - \tfrac{b_2}{3b_3}, \qquad 
\beta_{s_2} \eqg \tilde{\beta}_2 - \tfrac{b_2}{3b_3}, \qquad
\beta_{s_3}\eqg \tilde{\beta}_3 - \tfrac{b_2}{3b_3}; \label{eq_cub_4}\\ 
\text{where }
b_0&\eqg\nu\varsigma-\tau(\delta^2-\gamma^2),\text{ where } \tau\eqg\tr(P)\text{ and }\nu\eqg\fb^TP^2\fb,
\label{beta_tr_b0}\\
b_1&\eqg2\big(\tau\delta^2-\nu\varsigma+\alpha\nu(\delta^2-\gamma^2)\big),  \label{beta_tr_b1}\\
b_2&\eqg\nu\varsigma-\tau\delta^2+\alpha\nu(\gamma^2-4\delta^2), \label{beta_tr_b2}\\
b_3&\eqg2\alpha\nu\delta^2; \quad\label{beta_tr_b3}\\
\tilde{\beta}_1 &\eqg {v} + {w},\qquad
\tilde{\beta}_2 \eqg \omega {v} + \omega^2 {w},\qquad
\tilde{\beta}_3 \eqg \omega^2 {v} + \omega {w};
\label{eq_cub_10}\\
\text{where }\omega   &\eqg -\tfrac{1}{2} + \tfrac{1}{2} \sqrt{3}i \text{ is the cubic root of unity, }
\omega^2 \eqg -\tfrac{1}{2} - \tfrac{1}{2} \sqrt{3}i,
\label{eq_cub_8}\\
{v}&\eqg \sqrt[3]{-\tfrac{{t}}{2} + \sqrt{{u}}},\qquad
{w} \eqg \sqrt[3]{-\tfrac{{t}}{2} - \sqrt{{u}}}, \qquad {u} \eqg \left(\tfrac{{s}}{3}\right)^3 + \big(\tfrac{{t}}{2}\big)^2, \label{eq_cub_9_7}\\
{s} &\eqg \tfrac{b_1}{b_3} - \tfrac{b_2^2}{3b_3^2} \text{ and }
{t} \eqg \tfrac{b_0}{b_3} - \tfrac{b_1b_2}{3b_3^2} + \tfrac{2b_2^3}{27b_3^3}.
\label{eq_cub_3}
\end{align}
\end{subequations}
The optimal solution, $\beta_{\text{s}}$, is the unique real positive value among $\beta_{s_i}$, $i\in~\!\{1,2,3\}$.
Using the discriminant $\Delta$ (\cf Appendix \ref{discriminant}), it can be proven that the condition $b_0<0$ corresponds to the existence of a unique positive solution to the equation \eqref{eq_cubic} (\cf \cite{Dic:52,Fog:82,Del:94} ).

\nid Finally, considering \eqref{sig_vol_ssal_decrease} and the fact that $\ssal\big({\EC}_{\DC}(\beta_{\text{s}})\big)\leq\ssal\big({\EC}_{\DC}(\beta)\big)$, $\forall\beta\in~\!]0,1[$, it becomes obvious that $\ssal\big({\EC}_{\DC}(\beta_{\text{s}})\big)\leq\ssal\big({\EC}_{\DC}(\beta_{\varsigma})\big)<\ssal\big(\EC(\cb,\varsigma P)\big)$.
\def\qnul{\bqk}
\subsection{Proof of Theorem \ref{theo_ell_hyp_inters}
}\label{Appendix_theo_ell_hyp_inters}
{
To start with, recall that an affine map 
  $\Fd:~\Rn\rightarrow\Rn$, $\x\mapsto L\x+\bm{a}$ turns an ellipsoid $\EC(\cb,P)$ into another one $\EC(L\cb+\bm{a},L^TPL)$ and the hyperplane $\HC(\fb,y)$ into $\HC(L^{\dag}\fb,y+\fb^TL^{\dag}\bm{a})$. %
Throughout this proof, we'll be changing coordinate systems but dealing with one and the same hyperplane $\HC\eqg\HC(\fb,y)$ and one and the same ellipsoid $\EC\eqg\EC(\cb,\varsigma P)$. 
Consider the vector $\fb\in\Rn-\{\zero_n\}$.\linebreak Two cases (different from those of the Thm) will be distinguished depending on whether $\fb\in{\nul }(P)$ (1) or $\fb\notin{\nul }(P)$ (2).

1. $\fb\in{\nul }(P)$. This means that the matrix $P$ is SPSD and singular (having at least one zero eigenvalue) and  $\fb^TP\fb=0$.  In this case $\EC\subset\HC'$, where $\HC'\eqg\{\x\in\Rn|\fb^T\x=\cb\}$ is the hyperplane of normal vector $\fb$ and containing the center $\cb$ of $\EC$. If $\cb\notin\HC$, \ie, $\fb^T\cb\neq y$ (corresponding to case 1 of the Thm, with $\fb^TP\fb=0$),  $\EC$ is a subset of the hyperplane $\HC'$ parallel to $\HC$ and $\EC\cap\HC=\emptyset$, as in \eqref{ell_hyp_empty}. Otherwise (case 2), $\HC'=\HC$, meaning that $\EC\subset\HC$ and $\EC\cap\HC=\EC$, as in \eqref{ell_hyp_ell}. 

2. Consider now $\fb\notin{\nul }(P)$ and let  ${q}\eqg \rank(P)\le n$. We shall  define the affine transformation  
that maps the unit hypersphere or ball  into the ellipsoid $\EC(\cb,\varsigma P)$: 
\begin{align}
\BC_2^n\xrightarrow{\Fd_1}\EC(\cb,\varsigma P),\text{ \ie, }\Fd_1:\bxb\mapsto\x=\racine{(\varsigma P)}(\bxb+\cb).
\end{align}
Now consider its 
inverse transform $\Fd_1^\dag$ that maps the ellipsoid into a possibly degenerate
unit ball: $\d\EC(\cb,\varsigma P)\xrightarrow{\Fd_1^\dag}~\EC({\cb_{1}},{P_{1}})$ , where
 \begin{align}
 {\cb_{1}}=\zero_n \text{ and } {P_{1}}=\bar{I}_{n,{q}} \text{ where } \bar{I}_{n,{q}}\eqg
 \left[{I_{{q}} \atop 0_{\bqk, \bqk} } {0_{ \bqk, \bqk} \atop 0_{ \bqk, \bqk}}\right],\  \bqk\eqg n-{q}
 \label{bc_bP}
\end{align}
and $\d\HC(\fb,y)\xrightarrow{\Fd_1^\dag}~\HC({\fb_{1}},{y_{1}})$, \ie, $\bxb\in\HC\eq\bxb^T{\fb_{1}}={y_{1}}$. In the new coordinates system transformed thusly, the unit normal vector to the hyperplane $\HC$ and its minimum signed distance from origin are \resp.  
\begin{align}
{\fb_{1}}&\eqg\frac{\racine{P}\fb}{\sqrt{\fb^TP\fb}}, \text{ with } \norm{{\fb_{1}}}=1
\text{ and }{y_{1}}\eqg\frac{y-\fb^T\cb}{\sqrt{\varsigma \fb^TP\fb}}.\label{bf_by}
\end{align}
 
 \nid Let $\ib\eqg\ib_{1}\eqg[1\, 0\ldots0]^T\in\Rn$  (\cf $\S$~\ref{subsec_notations} \ref{identity}) the first vector of the identity matrix and 
\begin{align}
H\eqg I_n-\tfrac{2}{\norm{{\fb_{1}}-\ib}^2}({\fb_{1}}-\ib)({\fb_{1}}-\ib)^T\label{Householder}
\end{align}
is the Householder symmetric ($H=H^T$) and unitary ($HH^T=I_n$) matrix that transforms ${\fb_{1}}$ into $\ib$: 
$H{\fb_{1}}=\ib\eq{\fb_{1}}=H^T\ib=\hb_1$.
 
\nid Next, let 
$\Fd_2:~\bxb~\mapsto~\bbxb~=~H(\bxb~-{y_{1}}{\fb_{1}})$ that transforms the former (second) coordinate system into the third one, in which the considered hyperplane is orthogonal to $\ib$ and contains the origin: $\bbxb\in\HC\eq\bbxb^T\ib=0$, \ie, 
\begin{align}
\HC({\fb_{1}},{y_{1}})\xrightarrow{\Fd_2}\HC({\fb_{2}},{y_{2}})\text{ where }{\fb_{2}}\eqg\ib\text{ and }{y_{2}}\eqg 0.\label{bbf_bby}
\end{align}
The (possibly degenerate) unit ball $\EC({\cb_{1}},{P_{1}})$ is transformed, by $\Fd_2$, into the (possibly degenerate) hypersphere $\EC({\cb_{2}},{P_{2}})$, where 
\begin{align}
{\cb_{2}}\eqg H{\cb_{1}}-{y_{1}} H{\fb_{1}}=-{y_{1}} \ib\text{ and }{P_{2}}\eqg H{P_{1}} H^T=\bar{I}_{n,{q}}\label{bbc_bbP}
\end{align}

\nid Now, the distance between the center of the ellipsoid $\EC({\cb_{2}},{P_{2}})$ and the hyperplane $\HC({\fb_{2}},{y_{2}})$, $\abs{{\cb_{2}}^T{\fb_{2}}-{y_{2}}}=\abs{-{y_{1}}\ib^T\ib-0}$ $=\abs{{y_{1}}}$ is compared to the projection of the radius of the former onto the normal vector to the latter:
\begin{align}
\sqrt{{\fb_{2}}^T{P_{2}}{\fb_{2}}}=\sqrt{\ib^T\bar{I}_{n,{q}}\ib}=1.
\end{align}
If $\abs{{y_{1}}}>1$ (case 1 with $\fb^TP\fb\neq0$), then $\EC\cap\HC=\emptyset$. Otherwise (case 4),
the possibly degenerate hypersphere 
resulting from the intersection of 
$\EC(-{y_{1}}\ib,\bar{I}_{n,{q}})$ and the hyperplane $\HC(\ib,0)$ is $\EC({\cb_{3}},{P_{3}})$ where 
\begin{subequations}\label{bbcc_bbcP}
\begin{align}
{\cb_{3}}&\eqg{\cb_{2}}+\ib^T{\cb_{2}}\ib= -{y_{1}}\ib+(\ib^T\ib){y_{1}}\ib=\zero_n,\label{bbcc}\\
{P_{3}}&\eqg\big(1-(\ib^T{\cb_{2}})^2\big)\big(\bar{I}_{n,{q}}-\ib\ib^T\big)=\big(1-{y_{1}^2}\big)\big(\bar{I}_{n,{q}}-\ib\ib^T\big)\label{bbcP}.
\end{align}
\end{subequations}

This ellipsoid is expressed in the third coordinate system. Well, we have to find its expression in the orignal one and for this purpose, the inverse former transformations will be applied in reverse order: $\EC({\cb_{3}},{P_{3}})\xrightarrow{\Fd_1\circ\Fd_2^{-1}}~\EC({\cb}_{\HC},{\varsigma}_{\HC}{P}_{\HC})$.
To start with, we'll apply the inverse transformation $\Fd_2$ to the spheroid:\linebreak $\EC({\cb_{3}},{P_{3}})\xrightarrow{\Fd_2^{-1}}\EC({\cb_{3}'},{P_{3}'})$, to obtain 
\begin{subequations}\label{bcc_bcP}
\begin{align}
{\cb_{3}'}&\eqg H^T{\cb_{3}}+{y_{1}}{\fb_{1}}={y_{1}}{\fb_{1}}\\
{P_{3}'}&\eqg H^T{P_{3}} H=\big(1-{y_{1}^2}\big)\big(H^T\bar{I}_{n,{q}}H-H^T\ib\ib^TH\big)=\big(1-{y_{1}^2}\big)\big(\bar{I}_{n,{q}}-{\fb_{1}}{\fb_{1}}^T\big).\nn
\end{align}
\end{subequations}

\nid Then, applying $\Fd_1$: 
$\EC({\cb_{3}'},{P_{3}'})\xrightarrow{\Fd_1}~\EC({\cb}_{\HC},{\varsigma}_{\HC}{P}_{\HC})$, yields to 
\begin{subequations}\label{cc_cP}
\begin{align}
{\cb}_{\HC}&\eqg \racine{(\varsigma P)}{\cb_{3}'}+\cb=\cb+\racine{(\varsigma P)}{y_{1}}{\fb_{1}}\\
{\varsigma}_{\HC}{P}_{\HC}&=\racineT{(\varsigma P)}{P_{3}'}\racine{(\varsigma P)}=\varsigma\big(1-{y_{1}^2}\big)\big(P-\racine{P}{\fb_{1}}{\fb_{1}}^T\racine{P}\big).
\end{align}
\end{subequations}
Lastly, choosing ${\varsigma}_{\HC}\eqg\varsigma\big(1-{y_{1}^2}\big)$ and ${P}_{\HC}\eqg P-\racine{P}{\fb_{1}}{\fb_{1}}^T\racine{P}$ and replacing afterwards ${y_{1}}$, ${\fb_{1}}$, ${\cb}_{\HC}$, ${P}_{\HC}$ and ${\varsigma}_{\HC}$ by their respective expressions, \eqref{bf_by} and \eqref{cc_cP},  we get to \eqref{eq_P}$-$\eqref{eq_sig}.
The two last cases of the theorem can be combined in a single case where  $ -\urho\leq y\leq\brho$ and the case 3 would be the particular case where the ellipsoid resulting from the intersection of the ellipsoid $\EC(\cb,\varsigma P)$ with each of its (tangent) support hyperplanes of vector $\fb$, $\HC(\fb,-\urho)$ and $\HC(\fb,\brho)$, reduces to a single point: its center given by \eqref{cas3a_HP} and \eqref{cas3b_HP} \resp.
  }
\subsection{Proof of Theorem \ref{theo_meas_update}}
\label{Appendix_theo_meas_update}
This theorem is a direct application of Thms \ref{theo_ellips_HSpace_inter}, \ref{theo_ellips_strip_inter}, \ref{theo_beta_vol_min}, \ref{theo_beta_tr_min},\ref{theo_beta_sig_min} and \ref{theo_ell_hyp_inters} to $\SC_k$ given in \eqref{SC_set}.
In particular, \eqref{eq_correc-del} in case $\beta^{*}=\beta_{\varsigma}$, 
is obtained 
considering \eqref{sig_decrease}.
Now if $\abs{\delta }\leq \gamma$ (when $\beta^{*}=\beta_{\varsigma}$), if $a_0\geq 0$ (when $\beta^{*}=\beta_{\text{v}}$) or   if $b_0\geq 0$ (when $\beta^{*}=\beta_{\text{s}}$); setting $\beta=0$, \eqref{sig_thm_pf} is still equivalent to \eqref{sig_ellips_strip_inter}; and when it comes to the intersection with a hyperplane, replacing $\beta=1$ in \eqref{sig_thm_pf} leads to \eqref{eq_sig}.

As for \eqref{eq_correc-q}, 
it is obtained considering, from \eqref{eq_correc-P}, that  
\begin{align}
\PPki_{i}\eqg\PPki_{i-1}^{\frac{1}{2}}\big(I_n-\betak{i}{\brfbki\brfbki^\dag}\big)\PPki_{i-1}^{\frac{1}{2}}\nn
\end{align}
where $\brfbki=\PPki_{i-1}^{\frac{1}{2}}\fbki$ and $I_n-{\brfbki\brfbki^\dag}$ is the orthogonal projector onto the kernel of $\brfbki$, implying that
$\rank(\PPki_{i})=\rank(\PPki_{i-1})-1$ whenever $\betak{i}=1$ and $\rank(\PPki_{i})=\rank(\PPki_{i-1})$, otherwise.
\section{Proofs of the results of section \ref{sec_prop_stability}}
\subsection{Proof of Theorem \ref{theo_prop}}
\label{Appendix_theo_prop}
{
\begin{enumerate}
\item This point is satisfied by construction: from \eqref{SC_set}, Thms \ref{theo_time_update} and \ref{theo_meas_update}, \setcounter{Thmnbr}{\value{enumi}}
\mathl
\begin{align}
\text{we have }\xza\in\EC_0&\Rightarrow\left(\x_1\in\EC_{1/0}\right)\land\bigg(\x_1\in\big(\PC_1\cap\ZC_1\cap\bigcap_{i\in\hscr{1}}\HC_{1_i}\big)\bigg)\nn\\
&\Rightarrow\x_1\in\SC_{1}\Rightarrow\x_1\in\EC_{1}\Rightarrow\cdots
\Rightarrow\xkka\in\EC\kkk\nn\\
&\Rightarrow\left(\xka\in\EC\kk\right)\land\bigg(\xka\in\PC_k\cap\ZC_k\cap\bigcap_{i\in\hscrk}\HC_{k_j}\bigg)\nn\\
&\Rightarrow\xka\in\SC_{k}\Rightarrow\xka\in\EC_{k}, \ \forall k\in\N^*.
\end{align}
%
%
\item 
This point is also granted by construction. To check it, consider \linebreak $\xkz\eqg\xkk$.  From \eqref{eq_correc-x} of Thm \ref{theo_meas_update},
\mathc
\begin{align*}
\fbk{i}^T\xki=\fbk{i}^T\xkiii+{\alphak{i}\betak{i}}\delki\fbk{i}^T\phibki .
\end{align*}
If $\fbk{i}^T\phibki =0$ or $\abs{\delki}\leq \gamki$, it means that $\xki$ is already in $\DC\ki$ or $\HC\ki$. 
Else, 
\begin{align}
\fbk{i}^T\xki=\fbk{i}^T\xkiii+\betak{i}\delki.\label{pr_acc1}
\end{align}
Now, if $i\in\hscrk$, $\beta=1$ according to \eqref{eq_correc-bet}; then inserting \eqref{eq_correc-del} in \eqref{pr_acc1}, results in $\fbk{i}^T\xki=\yki$ meaning that $\xki\in\HC\ki$; otherwise, $\beta=~\!1-~\!\gammaki\abs{\delki}^{-1}$ and  $\fbk{i}^T\xki-\yki=-1$, if $\delki<-\gammaki$ and $\fbk{i}^T\xki-\yki=1$, if $\delki>\gammaki$; this means that $\xki\in\DC\ki$. Combining these results for $i\in\gscrk\cup\dscrk$, leads to $\xk\in\SC_k$, where $\SC_k$ is defined  in \eqref{SC_set} and considering \eqref{bounds}, the proof of  the point \ref{theo_prop_acceptable} is achieved.
\item 
\eqref{Lyap_def} is clearly established considering \eqref{Lyap_def_sig0}. 
Now, if $\beta^*\eqg\beta_{\varsigma}$, using \eqref{eq_correc-sig}, $\sigki-\sigkiii=-{\alphak{i}\betak{i}^2\delkis}$. Since $\alphak{i}$, defined in \eqref{eq_correc-alph}, is a quadratic form when it is nonzero, it is obvious that  $\sigki-\sigkiii\leq~\!0$. From \eqref{eq_correc-x-P-sig} and \eqref{eq_correc0},
$\sigk\eqg\sigma_{p}$ and  $\sigkkk\eqd\sigkz$, then 
\begin{align}
\sigk-\sigkkk=\sigma_{p}-\sigkz=-\sum_{i=0}^{p_k}{\alphak{i}\betak{i}^2\delkis}\leq0.\label{sigma_dec_demo}
\end{align}
The sequence $\big(\sigk\big)_{k\in\N}$ is decreasing, bounded above by $\sigz$ and hence convergent. 
\end{enumerate}
 }
\subsection{Proof of Lemma \ref{Lem_Bounds_Pbar}}\label{Appendix_Lem_Bounds_Pbar}
\renewcommand{\bpk}{\breve{p}}
\begin{manualtheorem}{5.3}
Consider the system \eqref{system0} subject to \eqref{bounds} and the matrix $P_k$ computed in line with either (\eqref{P_pred}-\eqref{P_pred1} and \eqref{mu-vol_k}) or \eqref{predic_eq_glo} on one hand and \eqref{ellips_correc}, 
on the other.
Let $\breve{H}_{k}\in~\R^{n\times{q}_k}$ 
whose columns form an orthonormal basis for $\range \big(P_k\big)$, 
where ${q}_k\eqg\rank(P_k)$ and let 
$\breve{P}_{k}\eqg \breve{H}_{k}P_k\breve{H}_{k}^T$. 
If the pair $\{A_{k},\breve{F}_{k}^T\}$ is \emph{sporadically observable} and $\{A_{k},R_k\}$ is uniformly controllable, 
then there exist positive finite numbers $\underb{\varrho}_{k}$ and $\bar{\varrho}_{k}$, \tq for all $k\geq \spo_{k}(l)$,
\mathc
\begin{align}
\underb{\varrho}_{k}I_{{q}_k}&\leq \breve{P}_k \leq \bar{\varrho}_{k}I_{{q}_k},\label{Bounds_Pbar}
\end{align}
$\spo_{k}$ being given in \eqref{card} of Definition \ref{defi_obs_spora}.
\end{manualtheorem}
The proof of this lemma will be carried out in two phases. We'll be ultimately using the observability and controllability properties of the Kalman filter in $\S$~\ref{sec_obs_contr}. For this purpose, we have to start by showing the analogy of the latter with the proposed algorithm. 
\subsubsection{Kalman filter analogy}\label{sec_KF}
To begin with, 
consider  the following  linear time-varying stochastic system with some bounded matrix $\breve{A}_k\in\Rnn$:
\mathl
\begin{subequations}\label{kf_sys}
\begin{align}
\xibk&=\breve{A}\kkk\xibkk+B\kkk\taubkk+\twbkk,\  k\in\N^*\\
\brybk&=\breve{F}_{k}^T\xibk+\vk,\ \forall k\in\breve{\KC},\text{ (\cf\eqref{bp_KC_def})}\label{kf_y_def}\\
\text{where \quad} \twbk&\sim\NC(\zero_m,W_k)\\
  \vk&\sim\NC(\zero_{\bpk },V_k),\quad\text{ where $\bpk$ is defined in\eqref{bp_KC_def},}
\end{align}
\end{subequations}%
where $\xibk\in\Rn$ is the unknown state vector, $\brybk\eqg[\yki]_{i\in\brdscrk}\!\in\R^{\bpk }$, $\ybk\eqg\tfrac{1}{2}(\bybk-~\!\uybk)$ and $\breve{F}_{k}\in~\R^{n\times\bpk }$, defined in \eqref{bF_bp_KC_def}, are the output vector and the observation matrix \resp.; 
$B\kkk\taubkk$ is the known input intervening in \eqref{state_eq0};
and $\twbk$ and $\vk$ are gaussian centered noise vectors of covariance matrices $W_k$ and $V_k$ resp. 
Now consider the Kalman filter, designed for the system~\eqref{kf_sys}:\hspace{-4pt}
\mathc
\begin{subequations}\label{kf}
\begin{align}
 \xibhk   & =  \xibhkk+K_{k}\delk   \label{kf_xest}        \\
   \breve{P}_{k} & = (I_n-K_{k}\breve{F}_{k}^T)\breve{P}\kk    \label{kf_Pest}\\
   \delk & \eqg  \brybk-\breve{F}_{k}^T\xibhkk  \label{kf_innovation}\\
    K_{k} & \eqg   \begin{cases}
   \breve{P}\kk \breve{F}_{k}(\breve{F}_{k}^T\breve{P}_{k-1}\breve{F}_{k} + V_k)^{-1},&\text{if } k\in\breve{\KC}\\
      0_{n,\bpk }, &\text{otherwise};
      \end{cases}
 \label{kf_gain}\\
    \xibhkk &=\breve{A}\kkk\xibhkkk+B\kkk\taubkk\label{kf_xpred}\\
    \breve{P}\kk&=\breve{A}\kkk \breve{P}\kkk\breve{A}\kkk^T+W\kkk.\label{kf_Ppred}
\end{align}
\end{subequations}
The time prediction stage, \eqref{predic_eq}, 
of Thm \ref{theo_time_update} 
can be seen as the prediction stage of the Kalman filter \eqref{kf_xpred}-\eqref{kf_Ppred} and the measurement correction stage  \eqref{ellips_correc}, given in Thm \ref{theo_meas_update} is nothing else than \eqref{kf_xest}-\eqref{kf_gain}. This is stated in Proposition \ref{algo_eq_kf}. 
Forasmuch as the Kalman filter undergoes numerical stability issues 
when the system \eqref{kf_sys} is subject to equality constraints (the matrix $\breve{F}_{k}^TP_{k-1}\breve{F}_{k} + V_k$ in the Kalman gain, \eqref{kf_gain}, becoming ill-conditioned), \eqref{bound_HyperP0} are not considered.
\renewcommand{\omki}{\omega_{i}}
\begin{propo}\label{algo_eq_kf}
If $\xk$ is computed in line with 
\eqref{predic_eq}  of Thm \ref{theo_time_update}, for any $\mub
\in~\!\!]0,\:+\infty[^m$
and \eqref{ellips_correc} of Thm \ref{theo_meas_update}, 
for any value of $\betak{i}=\beta^*\in]0,1[$, ${i\in\gscrk\cap\dscrk}$ 
and if $\xibhk$ is the Kalman estimator \eqref{kf} designed for the system \eqref{kf_sys},  such that $\breve{F}_k$ is given by \eqref{bF_def}, 
\mathc
\begin{subequations}\label{kf_A_W_V_def}
\begin{align}
\quad\breve{A}_k&\eqg\prod_{j=1}^m\sqrt{1+\muki{j}}A_k=\sqrt{\chi_{1}} A_k, \text{ \cf \eqref{barA},}
\label{kf_A} \\ 
W_k&\eqg \tfrac{1}{\sigk} R_k\diag\big(\tfrac{\chi_{i}}{{\muki{i}}}\big)_{i\in\{1,\ldots,m\}}R_k^T, \text{ where }\chi_{i} = \prod_{j=i}^m(1+\muki{j}),
\label{kf_W}\\ 
V_k&\eqg \diag\big(\tfrac{1}{\omki}\big)_{i\in\gscrk\cap\dscrk},\text{ where }  \omki\eqg\tfrac{\alphak{i}\betak{i}}{1-\betak{i}}, \label{kf_V}\\
\sigk &\eqg \sigkkk+{\alphak{i}\betak{i}\bigg(\tfrac{\gammaki^2}{1-\betak{i}}-\delkis\bigg)},\ \text{ for a fixed $\sigkz$;} \label{kf_sig}
\end{align}
 \end{subequations}
 \mathl
%
and if $\xz=\xibh_0$ and $\breve{P}_0=P_0$, then
\mathc
\begin{align}
\forall     k\in\N^*,\ \xk= \xibhk \text{ and } \breve{P}_k=P_k.
\end{align}
\end{propo}
\debdemo 
Replacing $\betak{i}=\tfrac{{\omki}}{\omki+\alphak{i}}$ and $\alphak{i}$ from \eqref{eq_correc-alph} in \eqref{eq_correc-P}, the latter can be rewritten 
\mathc
\begin{align}
\PPki_{i}&=\PPki_{i-1}-{\PPki_{i-1}\fbki \big(\fbki ^T\PPki_{i-1}\fbki +\tfrac{1}{\omki}\big)^{-1}\fbki ^T\PPki_{i-1}}.\nn
\end{align}
Then, using the inversion lemma, 
it comes that 
\begin{align}
P_k^{-1}={\PPki_{\bpk }^{-1}}={\PPki_{0}^{-1}}+\sum_{i=1}^{\bpk }\omki{\fbki \fbki ^T}. 
\nn
\end{align}
Recalling that $P_k={\PPki_{\bpk }}$ and that $\PPki_{0}=\Pkk$ and noticing that  
\begin{align}
\sum_{i=1}^{\bpk }{\omki\fbki \fbki ^T}&=\breve{F}_{k}^TV_k\breve{F}_{k},\nn\\
\intertext{we have}
P_k^{-1}&=\Pkk^{-1}+\breve{F}_{k}^TV_k^{-1}\breve{F}_{k}.\label{pr_acc3}
\end{align}
Applying the inversion lemma again to \eqref{pr_acc3}, the algorithm \eqref{ellips_correc} can be rewritten as \eqref{kf_y_def}, \eqref{kf} and \eqref{kf_A_W_V_def}.
Finally, 
using $P\kkp$ defined in \eqref{predic_eq_glo}
and considering \eqref{kf_A} and \eqref{kf_W},
we obtain \eqref{kf_xpred}-\eqref{kf_Ppred}.
Now, repeatedly using \eqref{P_pred1} for $i=m,\: m-1,\ldots,0$ in \eqref{P_pred} with \eqref{P_pred0}, produces
\begin{align}
P\kkp&\eqg\!\prod_{i=1}^m(1+\muki{i})A_kP_kA_k^T+\!\sum_{i=1}^m\prod_{j=i}^m\tfrac{1+\muki{j}}{\muki{j}\sigk}\r_{k_i}\r_{k_i}^{T} 
\label{Ppred_algo_eq_kf}
\end{align}
{then, considering \eqref{kf_A},  \eqref{kf_W} and the fact  that }
\begin{align}
\sum_{i=1}^m\prod_{j=i}^m\tfrac{1+\muki{j}}{\muki{j}\sigk}\r_{k_i}\r_{k_i}^{T}&=W_k,\nn
\end{align}
we obtain \eqref{kf_xpred}-\eqref{kf_Ppred} thus completing the proof.
\findemo 

\subsubsection{Boundedness of the shape matrix $P_k$}\label{sec_obs_contr}
%
\begin{propo}[\cite{Son:95,Bag:09}]\label{prop_P_bound}
Consider the time-varying system \eqref{kf_sys} and let $\breve{\KC}=~\!\N^*$ (\cf\eqref{bp_KC_def}). If the matrix pairs  $\{\breve{A}_k,W_k^{\frac{1}{2}}\}$ 
and $\{\breve{A}_{k},V_k^{-\frac{1}{2}}\breve{F}_{k}^T\}$ are uniformly  controllable and observable \resp., the estimation covariance matrix of the Kalman filter \eqref{kf}, designed for the system  \eqref{kf_sys}, satisfies the following inequalities, for all $k\geq l$:
\mathc
\begin{align}
(\breve{\OC}_{k,k-l}+\breve{\CC}_{k,k-l}^{-1})^{-1}\leq \breve{P}_k \leq \breve{\OC}_{k,k-l}^{-1}+\breve{\CC}_{k,k-l}.\nn
\end{align}
\end{propo}
\begin{propo}\label{prop_obs_con_unif}
The pairs $\{\breve{A}_k,W_k^{\frac{1}{2}}\}$ 
and $\{\breve{A}_{k},V_k^{-\frac{1}{2}}\breve{F}_{k}^T\}$ are uniformly controllable and observable \resp., \ssi  $\{{A}_k,{R}_k\}$ and $\{{A}_{k},\breve{F}_{k}^T\}$ have the respective properties.
\end{propo}
\debdemo 
Since $\lambda_k\eqg\prod_{j=1}^m\sqrt{1+\muki{j}}$, $W_k^{\frac{1}{2}}$ and $V_k^{-\frac{1}{2}}$, given in \eqref{kf_A} and \eqref{kf_W}, are all bounded and positive  (\resp. SPD), the observability and controllability gramians, associated to the matrices  $\breve{A}_k\eqg \lambda_k A_k$, $W_k^{\frac{1}{2}}$ 
and $\breve{F}_{k}V_k^{-\frac{1}{2}}$ are
 \begin{subequations}\label{gramian_con_obs}
 \mathl
\begin{align}
\breve{\CC}_{k+l,k}&\eqg\sum_{i=k}^{k+l-1}\breve{\lambda}_{i+1,k}^{-2}{\Phi}_{k,i+1}W_i^T{\Phi}_{k,i+1}^T\label{gramian_con_}\\
\breve{\OC}_{k+l,k}&\eqg\sum_{i=k}^{k+l}\breve{\lambda}_{i,k}^{2}{\Phi}_{i,k}^T\breve{F}_{i}V_i^{-1}\breve{F}_{i}^T{\Phi}_{i,k}\label{gramian_obs_}\\
\text{where }\qquad \breve{\lambda}_{k+l,k}&\eqg{\lambda}_{k+l-1}\ldots{\lambda}_{k},\quad\text{ $\lambda_k$ is defined in \eqref{kf_A};}\label{blambda_def}
\end{align}
 \end{subequations}
$\breve{\CC}_{k,k-l}$ and $\breve{\OC}_{k,k-l}$ are SPD bounded matrices  \ssi ${\OC}_{k,k-l}$ and ${\CC}_{k,k-l}$, given by \eqref{gramian_obs} and \eqref{gramian_con}, associated to $A_k$, $R_k$ and $\breve{F}_{k}$ are also bounded SPD matrices.
\findemo 

Considering Definition \ref{defi_obs_spora} for this kind of systems, the direct consequence of Propositions \ref{prop_P_bound} and \ref{prop_obs_con_unif} applied to the system with all \linebreak$i\in~\!\gscrk\cap\dscrk\cap\hscrk$ measurements \eqref{bounds} including equality constraints \eqref{bound_HyperP0} complete the proof of Lemma \ref{Lem_Bounds_Pbar}.

\subsection{Proof of Theorem \ref{theo_ISS}}
\label{Appendix_theo_ISS}
{
\nopagebreak\begin{description}[style=unboxed,leftmargin=0.5cm]
\item[\ref{theo_prop_vol_bounded}] Let $\bqk\eqg\tilde{q}_k\eqg  n-{q}_k=n-\rank(n)$
and $\tilde{H}_{k}\in~\!\R^{n\times \bqk}$ 
whose columns form orthonormal basis for $\nul \big(P_k\big)$,  \tq $H_k\eqg\left[\left.\breve{H}_{k}\right\vert \tilde{H}_{k}\right]$ is a unitary matrix. Hence, we have 
$H_k^TP_kH_k=\bdiag(\breve{P}_k,0_{\bqk,\bqk})$. 
The $\EC_k$'s nonzero semi-axes lengths are the singular values of the matrix 
	$\sigk \breve{H}_{k}^TP_k\breve{H}_{k}=\sigk \breve{P}_k$. On one hand, it is shown, at  the point \ref{theo_prop_lim_sig} of Thm \ref{theo_prop} (\cf Appendix \ref{Appendix_theo_prop}, point \ref{theo_prop_lim_sig}), that the sequence $\big\{\sigk\big\}_{k\in\N}$ is decreasing, bounded above by $\sigz$ and convergent. On the other hand, as stated in \eqref{Bounds_Pbar} of Lemma \ref{Lem_Bounds_Pbar} (\cf Appendix \ref{Appendix_Lem_Bounds_Pbar}.), the singular values of $\breve{P}_k$ are bounded and so are the ellipsoid's axes lengths, as well as their product representing the ellipsoid's volume.
\item[\ref{theo_prop_ISS}] According to Definition \ref{defi_ISS_Lyap}, we need first to show that $\Vscr_k(\dxbk)$ is bounded despite of the deficient rank of the matrix $P_k$. For any possible value of  the true state vector $\xka$, we have
\begin{align}
\xka\in\EC(\xk,\sigk P_k)&\eq \dxk\eqg\xka-\xk\in\EC(\zero_n,\sigk P_k)\nn\\
&\eq\dxk=(\sigk P_k)^{\frac{1}{2}}\ubk, \ubk\in\BC_2^n,\text{ (\cf $\S$~\ref{subsec_notations}. \ref{unit_ball}).}
\end{align}
It means that $\dxk\in\range \big(P_k\big)$, 
which is a subspace of $\Rn$ of dimension ${q}\leq n$, where ${q}\eqg {q}_k\eqg \rank(P_k)$: 
\begin{align}
H_k^T\dxk &=\sigk^{\frac{1}{2}}H_k^TP_k^{\frac{1}{2}}H_kH_k^T\ubk
=\sigk^{\frac{1}{2}}\left[{\breve{P}_{k}^{\frac{1}{2}}\atop 0_{ \bqk, {q}}} {0_{ {q}, \bqk}\atop 0_{ \bqk, \bqk}}\right]H_k^T\ubk
=\left[{\sigk^{\frac{1}{2}}\breve{P}_k^{\frac{1}{2}}\bar{\ub}_k\atop \zero_{ \bqk}}\right], \nn
\end{align}
where  $\bar{\ub}_k\eqg \breve{H}_{k}^T\ubk\in\BC_2^{q}$, meaning that 
\begin{align}
\forall\xka\in\EC_k, \dxk=H_k^T[\dxbki{1}^T \zero_{ \bqk}^T]^T, \text{ where }
\dxbki{1}\eqg \breve{H}_{k}^T\dxbk \text{ and } \tilde{H}_{k}^T\dxbk =\zero_{ \bqk}.\nn
\end{align} 
%
Now we shall show that $\Vscr_{{P}_k}$ is an ISS-Lyapunov function for all possible values of $\dxk\in\range \big(P_k\big)$. 
First,  let $\Vscr_{k}\eqg\Vscr_{{P}_k}(\dxbk)$;
\mathl
\begin{align}
\Vscr_{k}&\eqg \dxbk^TH_k H_k^TP_k^\dag H_k H_k^T\dxbk
=\left[\dxbki{1}^T\ \dxbki{2}^T\right]\left[{\breve{P}_{k}^{-1}\atop 0_{ \bqk, {q}}} {0_{ {q}, \bqk}\atop 0_{ \bqk, \bqk}}\right]\!\!\left[\dxbki{1}^T\ \dxbki{2}^T\right]^T=\dxbki{1}^T \breve{P}_k^{-1} \dxbki{1}\nn
\end{align}
noticing that   $\norm{\dxbki{1}}=\norm{H_k^T\dxk}=\norm{\dxk}$ and by virtue of \eqref{Bounds_Pbar}, it can be deduced that 
\mathc
\begin{align}
\underb{\psi}_{k}(\norm{\dxk})&\leq\Vscr_{k}\leq\bar{\psi}_{k}(\norm{\dxk}), 
\end{align} 
where
$\underb{\psi}_{k}$ (\resp. $\bar{\psi}_{k}):\R_+\rightarrow\R_+$, 
$t \mapsto\underb{\psi}_{k}(t)=\underb{\varrho}_{k}^{-1}t^2$ (\resp. $\bar{\psi}_{k}(t)=\bar{\varrho}_{k}^{-1}t^2$), \cf \eqref{Bounds_Pbar},
are $\kf_{\infty}$ functions. Now, since ${v}_{k}$ and ${s}_{k}$ are bounded above and below, it comes out that $\Vscr_k$ satisfies the condition \eqref{ISS_cond1}.

Second, 
to prove that  $\Vscr_{k}$ meets the condition \eqref{ISS_cond2}, we shall begin with proving that $\Vscr_{k}-\Vscr\kk\leq 0$, where
\begin{subequations}
\begin{align}
\Vscr\kk&\eqg \dxbkk^T\Pkk^\dag\dxbkk=\Vscr\kz;
\label{Lypap_kk_def}\\
\Vscr\ki&\eqg \dxbki{i}^T\PPki_{i}^\dag\dxbki{i} \text{ and }\Vscr\kij{p}=\Vscr_{k} \label{Lypap_ki_def}\\
\dxbkk&\eqg\xka-\xkk=\xka-\xbki{0} \eqd \dxbki{0};\label{error_x_kk_def}
\end{align}
\end{subequations}
{From \eqref{Lyap_def} and \eqref{Lyap_sig_inequality}, we have }
\mathc
\begin{align}
\Vscr_k-\Vscr\kk= \sum_{i=1}^{p}\Vscr\ki-\Vscr\kij{i-1}\leq\sum_{i=1}^{p}\sigki-\sigkkki&=\sum_{i=1}^{p} {\alphak{i}\betak{i}\big(\tfrac{\gammaki^2}{1-\betak{i}}-\delkis\big)}\nn\\&=\sigk-\sigkkk\leq 0.
\label{theo_ISS_pf_1}
\end{align}
Thanks to point \ref{theo_prop_lim_sig} of Thm \ref{theo_prop}, 
\begin{align}
\Vscr_k-\Vscr\kk&\leq 0, \text{ if $\beta^*=\beta_{\varsigma}$ { given in \eqref{beta_opt_sig_min} /Algo. \ref{Algo_beta_SigMin},}}
\end{align}
Now, considering $P\kkp$ given by \eqref{predic_eq}, we have (\cf \eqref{Ppred_algo_eq_kf} of Proposition \ref{algo_eq_kf}) 
\mathl
 \begin{align}
\Pkk&\eqg\lambda\kkk^2A\kkk \Pkkk A\kkk^T+W\kkk, \text{ where } \lambda_k\eqg\prod_{i=1}^m\sqrt{1+\muki{i}};
\label{Ppred_algo_eq_kf_}
\end{align} 
and $W_k$ SPD, defined in \eqref{W} 
 \resp. and both are bounded.
  Basing on the same reasoning as done in Lemma 3 in \cite{She:18}, it can be
    shown that for any vectors $\xb,\yb\in\Rn$ and any matrices $A,B\in\Rnn$, 
     \mathc
    \begin{align}
(\xb+\yb)^T(A+B)^{\dag}(\xb+\yb)\leq \xb^TA^{\dag}\xb+\yb^TB^{\dag}\yb.
    \end{align}
On the other hand, $\forall\x\in\range(P)$, $\exists\ub\in\Rn$, $\xb=P\ub$; then recalling that $\forall X\in\Rnn$, $XX^\dag X=X$, $\forall A\in\Rnn$ of full rank:
   \mathl
\begin{align}
\xb^TA^T(APA^T)^\dag A\xb 
&=\ub^TA^{-1}APA^T(APA^T)^\dag APA^TA^{-T}\ub\nn\\
&=\ub^TA^{-1}(APA^T)A^{-T}\ub=\ub^TP\ub\nn\\
&=\ub^TPP^\dag P\ub=\xb^T P^\dag\xb.
\end{align}
  Therefore, since $\dxkkk\in\range(\Pkkk)$ 
    \mathl
    \begin{align}
    \Vscr\kk&\leq\tfrac{1}{\lambda_{k-1}^{2}}\dxkkk^TA\kkk^T\big(A\kkk\Pkkk A\kkk^T\big)^{\dag}A\kkk\dxkkk{+\bwbkk^T{W}_{k-1}^{\dag}\bwbkk}\nn\\
   &\leq\tfrac{1}{\lambda_{k-1}^{2}}\dxkkk^T\Pkkk^{\dag}\dxkkk
    {+\bwbkk^T{W}_{k-1}^{\dag}\bwbkk}\nn\\
  \Vscr\kk&\leq \tfrac{1}{\lambda_{k-1}^{2}}\Vscr\kkk+\norm{W_{k-1}^{\dag}}\norm{\bwbkk}^2.\label{theo_ISS_pf_2}
    \end{align}
    Now, from \eqref{theo_ISS_pf_1}, we have 
    \mathc
    \begin{align}
    \Vscr_k-\Vscr\kkk\leq\Vscr\kk-\Vscr\kkk;\label{theo_ISS_pf_3}
    \end{align}
 and consequently, \eqref{theo_ISS_pf_2} becomes 
    \begin{align}
    \Vscr_k-\Vscr\kkk\leq\Vscr\kk-\Vscr\kkk\leq-\phi_{k-1}\Vscr\kkk+\psi_{k-1}(\norm{\bwbkk}),\label{theo_ISS_pf_4}
    \end{align}
    \mathl
 where $\phi_k\eqg 1-\tfrac{1}{\lambda_{k}^{2}}>0 $, since $\lambda_k>1$,
 and 
$\psi_k:\R_+\rightarrow\R_+$, \linebreak
$t \mapsto\psi_k(t)=\norm{W_{k}^{\dag}}t^2$,
 is a $\kf-$ function.
This means that $\Vscr_k$ is an ISS-Lyapunov function for the system of state vector $\dxk$. Thus
applying Lemma \ref{lemm_ISS}  completes the proof of this point.
\item[\ref{theo_prop_vol_dec}] %
First, we shall establish some properties of the pseudo-determinant. For any $P\in\Rnn$ of rank $q$, 
any {SPSD} $R\in\Rnn$ and any full rank $A\in\Rnn$,
\mathl
\begin{align}
\Det{P}&=\abs{P+I_n-P^\dag P};\nn\\
{P+{R}}&=\big(P+I_n-P^\dag P\big)\big(P^\dag P+(P^\dag+I_n-P^\dag P){R}\big)\nn\\
\Det{P+{R}}&=\Det{P}\Det{P^\dag P+(P^\dag+I_n-P^\dag P){R}};\nn\\
APA^T&=APP^\dag PA^T=(APP^\dag) P(P^\dag PA^T)\Rightarrow\Det{APA^T}&=\Det{APP^\dag}^2\Det{P}.\nn
\end{align}
From Thm \ref{theo_time_update}, it is clear that $P\kkp=\breve{A}_k P_k \breve{A}_k^T+W_k=\Qinv_k+W_k$, thus
\begin{align}
\sigk^{q_k} \Det{P\kkp}&=\Det{\breve{A}_kP_kP^\dag_k}^2\Det{\Qinv_k^\dag \Qinv_k+(\Qinv^\dag_k+I_n-\Qinv^\dag_k \Qinv_k)W_k}\sigk^{q_k} \Det{P_k}\nn\\
&=\tfrac{{v}_{k+1}}{{v}_{k}}\sigk^{q_k} \Det{P_k}. 
\label{Det_kk_dec}
\end{align}
Now, let us recall that the rank of the matrix $P_k$ is varying with each time step $k$ and with each measurement $i$: $\rank(\PPki_{i})\eqd \qqkj_i$. From \eqref{vol_decrease} of Thm \ref{theo_beta_vol_min}, 
\begin{multline}
\sum_{i=1}^{p_k}\sigkj{i}^{\qqkj_{i}} \Det{\PPki_{i}}- \sigkj{i-1}^{\qqkj_{i-1}}\Det{\PPki_{i-1}}\leq 0\Rightarrow \sigk^{q_k} \Det{P_k}\leq \sigkkk^{q\kkk} \Det{\Pkk}.\label{Det_k_dec}
\end{multline}
Considering \eqref{Det_k_dec} and \eqref{Det_kk_dec},
\mathc
\begin{align}
\sigk^{q_k} \Det{P_k}&\leq \frac{{v}_{k}}{{v}_{k-1}}\sigkkk^{q\kkk} \Det{P\kkk}
\leq \frac{{v}_{k}}{{v}_{k-2}}\sigj{k-2}^{q_{k-2}} \Det{P_{k-2}}
\leq \tfrac{{v}_{k}}{{v}_{0}}\sigz^n \Det{P_0}={{v}_{k}}\sigz^n \Det{P_0}.\nn
\end{align}
\item[\ref{theo_prop_ssal_dec}] From \eqref{ssal_decrease} of Thm \ref{theo_beta_tr_min},
\begin{align}
	\sum_{i=1}^{p_k}\sigkj{i} \tr(\PPki_{i})- \sigkj{i-1}\tr(\PPki_{i-1})\leq 0
\Rightarrow \sigk\tr(P_k)\leq \sigkkk \tr(\Pkk)\label{tr_k_dec}
\end{align}
\begin{align}
\tr(P\kkp)&=\tr\big(\breve{A}_k P_k \breve{A}_k^T\big)+\tr(W_k)= \lambda_k^2\tr\big(P_k A_k A_k^T\big)+\tr(W_k)\nn\\
&\leq  \lambda_k^2\tr(P_k)\tr\big(A_k A_k^T\big)+\tr(W_k);\nn\\
\sigkp\tr(P\kp)& \leq \sigk \tr(P_k)\Big(\lambda_k^2\tr\big(A_k A_k^T\big)+\tfrac{\tr(W_k)}{\tr(P_k)}\Big)
=\tfrac{{s}_{k+1}}{{s}_{k}} \sigk \tr(P_k).\nn
\end{align}
{Hence}
\begin{align}
\sigk\tr(P_k)&\leq \tfrac{{s}_{k}}{{s}_{k-1}}\sigkkk \tr\big(P\kkk \big)\leq\ldots\leq\tfrac{{s}_{k}}{{s}_{0}} \sigz \tr\big(P_0 \big)={{s}_{k}}\sigz \tr\big(P_0 \big).
\label{tr_k_dec}
\end{align}

\item[\ref{theo_prop_beta_sig_Lyap}] The proof of this point is the direct consequence of the point \ref{theo_prop_ISS}: the fact that $R_k=0_{n, m}$ means that $\bwbk=\zero_n$ and the ISS stability of a system implies its Lyapunov stability with 0-input.
It can also be obtained by simply replacing  $\phi_k\eqg1-\tfrac{1}{\lambda_{k}^{2}}=0 $ in \eqref{theo_ISS_pf_2}-\eqref{theo_ISS_pf_4}, where it comes out that $ \Vscr_k-\Vscr\kkk\leq0$.
\item[\ref{theo_prop_beta_ell_dec}]
Now,  
because $R_k=0_{n,m}$, $\lambda_k=1$ and $W_k=0_{n,n}$, it is clear that, 
\mathl
\begin{align}
&P\kkp=A_k P_k A_k^T\leq \norm{A_k}^2P_k\leq P_k, \text{ as } \norm{A_k}\leq 1.\label{Pkk_dec}\\
\intertext{On the other hand,  $\forall\beta^*\in]0,1[$, }
&P\kp\eqg P\kkp-\sum_{i=1}^{p_k}{\alphak{i}\betak{i}}{\phibki \phibki ^T}\leq P\kkp.\label{Pk_dec}\\
&\text{the obvious consequence of \eqref{Pkk_dec} and \eqref{Pk_dec} is } 
P\kp\leq P_k, \ \forall\beta^*\in]0,1[. \label{P_dec}
\end{align}
and this means that all the eigenvalues of $P_k$ are nonincreasing.
\begin{description}
\item[\ref{theo_prop_beta_ell_dec_sig}] According to the point \ref{theo_prop_lim_sig} of Thm \ref{theo_prop_acceptable}, when $\beta^*=\beta_{\varsigma}$, 
$\sigkp\leq\sigk$ and by the use of \eqref{P_dec}, $\sigkp P\kp\leq\sigk P_k$. This complets the proof of this point, recalling that the eigenvalues of $\sigk P_k$ are the semi-axes' lengths of the ellipsoid $\EC_k$.
\item[\ref{theo_prop_beta_ell_dec_vol}] If  $\beta^*=\beta_{\text{v}}$, \eqref{Det_k_dec} and \eqref{P_dec} imply that $\sigkp^{q\kp} \Det{P\kp}\leq \sigk^{q_k} \Det{P_k}$.
\item[\ref{theo_prop_beta_ell_dec_ssal}] If  $\beta^*=\beta_{\text{s}}$, \eqref{tr_k_dec} and \eqref{P_dec} imply that $\sigkp \tr\big({P\kp}\big)\leq \sigk \tr({P_k})$.
\end{description}
\end{description}
    The cases where $k\notin\breve{\KC}$ can be viewed as measurements $i$ for which $\alphak{i}=0$ or $\betak{i}=0$.
}

\section{Cubic Equation}\label{Appendix_cubic_eq}
\subsection{Reduced Cubic Equation}
If, in the general cubic equation
\index{Cubic equation!reduced}%
\begin{align}
\beta^3 + b\beta^2 + c\beta + d = 0,
\label{1}
\end{align}
we set $\beta = \tilde{\beta}-b/3$, we obtain the \emph{reduced cubic equation}
\begin{align}
\tilde{\beta}^3 + {s}\tilde{\beta} + {t} = 0,
\label{2}
\end{align}
lacking the square of the unknown~$\tilde{\beta}$, where
\begin{align}
{s} = c - \frac{b^2}{3}, \qquad
{t} = d - \frac{bc}{3} + \frac{2b^3}{27}.
\label{3}
\end{align}
After finding the roots $\tilde{\beta}_1$, $\tilde{\beta}_2$, $\tilde{\beta}_3$ of~\eqref{2}, we shall know the roots of~\eqref{1}:
\begin{align}
\beta_1 = \tilde{\beta}_1 - \frac{b}{3}, \qquad
\beta_2 = \tilde{\beta}_2 - \frac{b}{3}, \qquad
\beta_3 = \tilde{\beta}_3 - \frac{b}{3}. 
\label{4}
\end{align}
\subsection{Algebraic Solution of the Reduced Cubic Equation}   We shall
employ the method which is essentially the same as that given by Vieta
in~1591. We make the substitution
\begin{align}
\tilde{\beta} = z - \frac{{s}}{3z}
\label{5}
\end{align}
in~\eqref{2} and obtain
\[
z^3 - \frac{{s}^3}{27z^3} + {t} = 0,
\]
since the terms in $z$ cancel, and likewise the terms in~$1/z$. Thus
\[
z^6 + {t}z^3 - \frac{{s}^3}{27} = 0.
\label{6}
\]
Solving this as a quadratic equation for~$z^3$, we obtain
\[
z^3 = -\frac{{t}}{2} ±\sqrt{{u}},\qquad
{u} = \left(\frac{{s}}{3}\right)^3 + \left(\frac{{t}}{2}\right)^2.
\label{7}
\]


Any number has three cube roots, two of which are the products
of the remaining one by the imaginary cube roots of unity:
\[
\omega   = -\tfrac{1}{2} + \tfrac{1}{2} \sqrt{3}i,\qquad
\omega^2 = -\tfrac{1}{2} - \tfrac{1}{2} \sqrt{3}i.
\label{8}
\]
We can choose particular cube roots
\begin{align}
{v}= \sqrt[3]{-\frac{{t}}{2} + \sqrt{{u}}},\qquad
{w} = \sqrt[3]{-\frac{{t}}{2} - \sqrt{{u}}},
\label{9}
\end{align}
such that ${v}{w} = -{s}/3$, since the product of the numbers under the cube
root radicals is equal to~$(-{s}/3)^3$. Hence the six values of~$z$ are
\begin{align*}
{v},\quad \omega {v},\quad \omega^2 {v},\quad
{w},\quad \omega{w},\quad \omega^2 {w}.
\end{align*}
These can be paired so that the product of the two in each pair is~$-{s}/3$: 
\[
{v}{w} = -\frac{{s}}{3},\qquad
\omega {v}\cdot\omega^2 {w} = -\frac{{s}}{3},\qquad
\omega^2 {v}\cdot\omega {w} = -\frac{{s}}{3}.
\]
Hence with any root~$z$ is paired a root equal to~$-{s}/(3z)$. By~\eqref{5}, the sum
of the two is a value of~$\tilde{\beta}$. Hence the \emph{three} values of~$\tilde{\beta}$ are
\begin{align}
\tilde{\beta}_1 = {v} + {w},\qquad
\tilde{\beta}_2 = \omega {v} + \omega^2 {w},\qquad
\tilde{\beta}_3 = \omega^2 {v} + \omega {w}.
\label{10}
\end{align}

It is easy to verify that these numbers are actually roots of~\eqref{2}. For
example, since $\omega^3 = 1$, the cube of $\tilde{\beta}_2$ is
\[
{v}^3 + {w}^3 + 3\omega {v}^2 {w} + 3\omega^2 {v}{w}^2
  = -{t} - {s}(\omega {v}+ \omega^2 {w}) = -{t} - {s}\tilde{\beta}_2,
\]
by~\eqref{9} and ${v}{w} = -{s}/3$.

The numbers~\eqref{10} are known as \emph{Cardan's formulas} for the roots of a
\index{Cardan's formulas}%
reduced cubic equation~\eqref{2}. The expression ${v} + {w}$ for a root was first
published by Cardan in his \textit{Ars Magna} of~1545, although he had obtained
it from Tartaglia under promise of secrecy.

\subsection{Discriminant}\label{discriminant} The product of the squares of the differences of
the roots of any equation in which the coefficient of the highest power of
the unknown is unity shall be called the \emph{discriminant} of the equation.
For the reduced cubic~\eqref{2}, the discriminant is
\index{Discriminant!of cubic}%
\begin{align}
(\tilde{\beta}_1 - \tilde{\beta}_2)^2 (\tilde{\beta}_1 - \tilde{\beta}_3)^2 (\tilde{\beta}_2 - \tilde{\beta}_3)^2 = -4{s}^3 - 27{t}^2,
\label{11}
\end{align}
\begin{rema}
\emph{The discriminant $\Delta$ of the general cubic~\eqref{1} is equal to the discriminant
of the corresponding reduced cubic~\eqref{2}.} For, by~\eqref{4},
\begin{align}
\beta_1 - \beta_2 = \tilde{\beta}_1 - \tilde{\beta}_2, \qquad
\beta_1 - \beta_3 = \tilde{\beta}_1 - \tilde{\beta}_3, \qquad
\beta_2 - \beta_3 = \tilde{\beta}_2 - \tilde{\beta}_3.
\end{align}

Inserting in~\eqref{11} the values of ${s}$ and~$q$ given by~\eqref{3}, we get
\begin{align}
\Delta = 18bcd - 4b^3 d + b^2 c^2 - 4c^3 - 27d^2.
\label{12}
\end{align}

It is sometimes convenient to employ a cubic equation
\begin{align}
a\beta^3 + b\beta^2 + c\beta +d = 0 \quad   (a \neq 0),
\label{13}
\end{align}
in which the coefficient of $\beta^3$ has not been made unity by division. The product~$\bar\Delta$
of the squares of the differences of its roots is evidently derived from~\eqref{12} by replacing
$b$, $c$, $d$ by $b/a$, $c/a$, $d/a$. Hence
\[
a^4 \bar\Delta = 18 abcd - 4b^3 d + b^2 c^2 - 4ac^3 - 27a^2 d^2.
\label{14}
\]
This expression (and not $P$ itself) is called the discriminant of~\eqref{13}.
\end{rema}

\subsection[Number of Real Roots of a Cubic]{Number of Real Roots of a Cubic Equation.}
\index{Cubic equation!number of real roots}
\index{Number!of roots}%
\begin{theo}
A cubic equation
with real coefficients has three distinct real roots if its discriminant~$\Delta$ is positive,
a single real root and two conjugate imaginary roots if $\Delta$ is negative, and at
least two equal real roots if $\Delta$ is zero.
\end{theo}


{%
\bibliographystyle{MyIEEEtran}
\bibliography{Biblio}
}
\end{document}